\documentclass[apj,square,numbers]{emulateapj}
\usepackage{graphicx}
\usepackage{latexsym}
\usepackage{epsfig}
\usepackage{color}
\usepackage[colorlinks, linkcolor=blue, citecolor=blue, urlcolor=blue]{hyperref}
\usepackage{apjfonts}

\usepackage{amsmath,amssymb,bm,color}
\usepackage{ulem}
\def\bea{\begin{eqnarray}}
\def\eea{\end{eqnarray}}
\def\be{\begin{equation}}
\def\ee{\end{equation}}

\newcommand{\de}{\mathrm d}

\newcommand{\odm}{{\Omega_\mathrm{DM}}}

\newcommand{\mdm}{{m_{\rm DM}}}

\newcommand{\sv}{{\langle\sigma_a v\rangle}}

\newcommand{\g}{$\gamma$}
\newcommand{\ns}{{\sc ns}}
\newcommand{\low}{{\sc low}}
\newcommand{\high}{{\sc high}}
\newcommand{\sigmav}{\langle\sigma_a v \rangle}
\newcommand{\Fermi}{{\sl Fermi}}

\shorttitle{Dark matter searches in the $\gamma$ rays with galaxy catalogues}
\shortauthors{Cuoco et al.}

\begin{document}
\title{Dark matter searches in the $\gamma$-ray extragalactic background via \\ cross-correlations with galaxy catalogues}

\author{Alessandro Cuoco$^{1,2}$}
\author{Jun-Qing Xia$^{3,4}$}
\author{Marco Regis$^{1,2}$}
\author{Enzo Branchini$^{5,6,7}$}
\author{Nicolao Fornengo$^{1,2}$}
\author{Matteo Viel$^{8,9}$}

\affiliation{$^1$ Dipartimento di Fisica, Universit\`{a} di Torino, via P. Giuria 1, I--10125 Torino, Italy}
\affiliation{$^2$ Istituto Nazionale di Fisica Nucleare, Sezione di Torino, via P. Giuria 1, I--10125 Torino, Italy}
\affiliation{$^3$ Department of Astronomy, Beijing Normal University, Beijing 100875, China}
\affiliation{$^4$ Key Laboratory of Particle Astrophysics, Institute of High Energy Physics, Chinese Academy of Science, P. O. Box 918-3, Beijing 100049, P. R. China.}
\affiliation{ $^5$ Dipartimento di Matematica e  Fisica, Universit\`a degli Studi ``Roma
Tre'', via della Vasca Navale 84, I-00146 Roma, Italy}
\affiliation{$^6$ INFN, Sezione di Roma Tre, via della Vasca Navale 84, I-00146 Roma, Italy}
\affiliation{$^7$ INAF Osservatorio Astronomico di Roma, Osservatorio Astronomico di Roma, Monte Porzio Catone, Italy}
\affiliation{$^8$ INAF Osservatorio Astronomico di Trieste, Via G. B. Tiepolo 11,
I-34141, Trieste, Italy}
\affiliation{$^9$ INFN, Sezione di Trieste, via Valerio 2, I-34127, Trieste, Italy}

\email{cuoco@to.infn.it, xiajq@ihep.ac.cn, regis@to.infn.it, branchin@fis.uniroma3.it, fornengo@to.infn.it,  viel@oats.inaf.it}


\begin{abstract}
We compare the measured angular cross-correlation between the \Fermi-LAT \g-ray sky and 
catalogues of extra-galactic objects with the expected signal induced by weakly interacting massive particle (WIMP) dark matter (DM).
We include a detailed description of the contribution of astrophysical \g-ray emitters such as blazars, 
misaligned AGN and star forming galaxies, and perform a global fit
to the measured cross-correlation.
Five catalogues  are considered: SDSS-DR6 quasars, 2MASS galaxies, NVSS radio galaxies, 
SDSS-DR8 Luminous Red Galaxies and SDSS-DR8 main galaxy sample.
To model the cross-correlation signal we use the halo occupation distribution formalism
to estimate the number of galaxies of a given catalogue in DM halos and their spatial correlation properties.
We discuss uncertainties in the predicted cross-correlation signal arising from the DM clustering and 
WIMP microscopic properties, which set the DM \g-ray emission.
The use of different catalogues probing objects at different redshifts reduces significantly, though not completely,  
the degeneracy among the different \g-ray components.
We find that the presence of a significant WIMP DM signal is allowed by the data but not significantly preferred by the fit, 
although this is mainly due to a degeneracy with the misaligned AGN component.
With modest substructure boost, the sensitivity of this method excludes thermal annihilation cross sections at 95\% level.
for WIMP masses up to few tens of GeV.
Constraining the low-redshift properties of astrophysical populations with future data will further improve the sensitivity to DM.
\end{abstract}


\maketitle

\section{Introduction}
\label{sec:intro}

The last few years have seen a tremendous improvement in our understanding of the \g-ray sky, mostly thanks to the observations
performed by the Large Area Telescope (LAT) on board of the \Fermi\ satellite \citep{2009ApJ...697.1071A}. Among the main issues that have been investigated, an important one is the understanding of the origin of the Isotropic Gamma-Ray Background (IGRB) \citep{2015IGRB,Fornasa:2015qua}, i.e. the fraction of the extra-galactic
 \g-ray background (EGB) that has not been resolved into individual sources.
The nature of the extragalactic emission is a recurrent issue which arises each time a new observational window of the electromagnetic spectrum
is opened on the Universe. A good example is the quest for the origin of the soft X-ray background, with the important difference that the latter has now been largely resolved (see, e.g.,  \cite{HM07}) whereas a significant fraction of the \g-ray flux is still diffuse, leaving large room for potential new discoveries.

The interest in the IGRB also stems from the consideration that the \g-ray band is a potential ``golden channel" for the indirect detection of particle
Dark Matter (DM). In fact,  among the conventional astrophysical sources that contribute to the IGRB there is the possibility that
a characteristic signal from DM annihilation or decay may also be present.
After its first detection and early attempts to shed light on the origin of the  IGRB (see e.g. \cite{1972ApJ...177..341K,1973ApJ...186L..99F,
1982A&A...105..164M,padovani,1996ApJ...464..600S,1998ApJ...494..523S,2004JCAP...04..006K,
2004ApJ...613..956S}), 
 a significant step forward has recently been possible thanks to  {\Fermi}-LAT that is resolving 
an ever growing number of sources \citep{2FGL,3FGL,AGNfermi},
most of which have been identified as  blazars, almost equally split into Flat Spectrum Radio Quasars (FSRQs) and BL~Lacs sub-classes. 

The properties of the resolved sources can be used to extrapolate their contribution to the 
IGRB~\citep{AjelloFSRQs, AjelloBLLs,dimauro2014a,DiMauro:2014wha,ajello2015IGRB}.
These population studies suggest that unresolved blazars account for only about 20\% of the unresolved IGRB integrated above 100 MeV, while they can be the dominant component above few GeV.
The remaining IGRB fraction is thought to be contributed by star-forming galaxies (SFGs) and misaligned AGN (mAGN), two types of sources that can contribute 10--50\% each to the extragalactic \g-ray emission \citep{inoue2011,Fermi:2012eba,DiMauro:2013xta}.
The contribution from additional potential sources like
the millisecond pulsars located in our Galaxy at high Galactic latitudes turned out to be small \citep{SG11,Calore:2014oga}.
The contribution from known astrophysical sources to the IGRB has thus significant uncertainties and leaves 
room for an additional contribution by more exotic sources like DM.

Additional constraints on the origin of the IGRB can be obtained by analyzing the angular correlation properties of its fluctuations. Refs.
 \cite{Fermi:2012eba,cuoco2012,Ackermann:2012uf} have confirmed the conclusions derived from the IGRB mean-intensity and source populations studies:
a blazar population that contributes, at energy below $\sim$10 GeV, about 20\% of the unresolved IGRB 
can account for
 the whole measured angular power,
thus providing an independent confirmation that the IGRB is not dominated by emission from blazars in the low energy part of the spectrum. 
Constraints on the DM contribution have been  derived in \cite{Ando:2013ff,Gomez-Vargas:2013cna,Fornasa:2012gu}.

The accuracy in the analysis of the IGRB and its fluctuations is limited by the presence of Galactic foregrounds and bright sources. If incorrectly subtracted they can induce spurious contributions to both the mean IGRB intensity and its anisotropies. An effective way of dealing with this problem and filter out contaminations is to cross-correlate the IGRB with maps of sources (observed in other wavelengths or by other means) that trace the same structures where the actual IGRB sources reside
but do not correlate with Galactic foregrounds. Basically, all catalogs of extragalactic objects at any redshift satisfy these conditions.
In the framework of the IGRB investigation, the cross-correlation strategy has been proposed in \cite{Cuoco:2007sh,2009MNRAS.400.2122A} 
and recently revisited in \cite{2014JCAP...10..061A,2014PhRvD..90b3514A}.
The measurement was pioneered  in  \cite{xia11} using the first 21 months of \Fermi\ data. In that case, no statistically significant signal was observed. 
The analysis has then been recently updated using 60-months \Fermi\ maps \citep{Xia:2014}. 
This time a significant (more than 3.5 $\sigma$ C.L.) cross-correlation signal has been detected. 
The signal is present on angular scales smaller than 1$^{\circ}$ in the cross-correlation between the diffuse \g-ray emission  cleaned by the 
Galactic foregrounds and four types of   Large Scale Structure (LSS) tracers:  radio galaxies in the NRAO VLA Sky Survey (NVSS) \citep{NVSS},  near infra-red selected galaxies in the
 Two Micron All Sky Survey (2MASS) \citep{2MASS} , optically selected galaxies in the Sloan Digital Sky Survey (SDSS)-DR8 catalog \citep{SDSSDR8}
 and quasi stellar objects (QSO) in   the SDSS-DR6 catalogs (QSO-DR6) \citep{SDSSDR6QSO}. 
 No significant correlation was observed  with Luminous Red Galaxies (LRG) also from SDSS. 
The analysis further confirms that blazars provide a minor contribution $<20\%$ to the IGRB as found 
in the IGRB mean-intensity and source populations studies, while
a mixture of SFGs and mAGNs can in principle contribute to the majority of the IGRB. 
 
 A promising, possibly more effective in the context of DM, way to apply the cross-correlation technique is to use weak gravitational lensing maps (cosmic shear) instead 
 of catalogs of LSS tracers
\citep{Camera:2012cj,Fornengo:2013rga,Fornengo:2014cya,Camera:2014rja,Shirasaki:2014noa}. 
 This alternative approach, originally proposed in   \citep{Camera:2012cj}, has the advantage of 
probing directly the matter distribution,
therefore avoiding  the so-called `biasing' issue, i.e., the fact that the mapping between the spatial distribution of extragalactic sources and that of the underlying mass density field is ill-known, and needs to be modelled.
Cross-correlation of \g-rays with cosmic shear will become available in the next years with the release of cosmic-shear maps 
from wide area surveys, like, e.g., the Dark Energy Survey (DES) \citep{DES} and, in the next decade, by the satellite-based  Euclid survey \citep{Euclid}. 
Finally, a similar technique, based on the cross-correlation of \g-rays with Cosmic Microwave Background (CMB) lensing maps, has 
been recently adopted in \cite{Fornengo:2014cya}, where an evidence of $3.2\sigma$ has been reported providing a
further direct evidence of the extragalactic origin of the IGRB,  and of a subdominant role of Galactic sources.

In this paper we investigate the implications 
of the recent measurement of a cross-correlation between \g-rays and LSS tracers by  \cite{Xia:2014} 
for both the DM and the main astrophysical contributors to the IGRB.
This work builds upon the results obtained by \cite{Regis:2015zka} in which we concentrated on the low-redshift 2MASS catalog as a tracer of the LSS in the local Universe, and we have assumed that 
DM-induced \g-rays provide the dominant source of cross-correlation for such a low redshift observations. 
That approach has been motivated by the fact that the DM contribution to the cross-correlation is dominated by \g-rays emission at low-redshift 
(see e.g. \cite{Fornengo:2013rga} or Appendix \ref{sec:windows}), which is where 2MASS galaxies mostly reside. In that analysis we found that the observed cross-correlation signal can indeed be explained by a DM emission, 
while its contribution to the total mean intensity is significantly below the IGRB intensity measured by \Fermi. 
This implies that the cross-correlation technique can be a powerful probe of the particle nature of DM, even when the DM contribution to the 
IGRB is subdominant, which is what we expect in a realistic scenario. In \citep{Regis:2015zka} 
we found that the cross-correlation signal can be explained by a DM particle with
mass in the tens to hundreds GeV range (depending on the \g-rays production channel) and, once the uncertainties in the DM distribution modeling is properly accounted for,  a ``thermal'' value for the annihilation cross section ($\sigmav = 3 \times 10^{-26}$ cm$^3$ s$^{-1}$), which is the most appealing case for a weakly interacting massive particle (WIMP) DM.
From the same analysis we have obtained upper bounds on the DM annihilation cross section and decay rate that turn out to be 
quite competitive with those obtained with different techniques, based either on local (Galactic halo, dwarf galaxies) or extra-galactic \g-ray emission.
We point out that those constraints are conservative precisely because the DM is assumed to be the only source of the  \g-ray signal.

In this 
follow-up
 paper we extend the study of \cite{Regis:2015zka} to the inclusion of astrophysical \g-ray emitters, and to the whole set of LSS-tracers catalogs. As it will be discussed in the next Sections, the redshift distributions of the  \g-ray signal is a fingerprint that characterises the contribution of different astrophysical sources and of the DM. For this reason, 
 the possibility to use  catalogues of objects whose distributions peak at different redshifts is an effective way to extract the information encoded in the \g-rays maps and remove degeneracies. To this aim, in addition to DM, we account here for contributions from blazars (of both BL Lac and FSRQs types), SFGs and mAGNs. 
Consequently, we do not limit our cross correlation study to the 2MASS catalog but consider
NVSS, SDSS-DR6 QSO, LRGs and SDSS-DR8 Main Galaxy samples as well.
The approach will be similar to \cite{Xia:2014} where, however, only astrophysical sources have been fitted to the observed correlations.
Here, beside including DM in the fit, we will use an improved description of the cross-correlation modeling between astrophysical sources
and LSS tracers based on the Halo Occupation Distribution (HOD) formalism. As for the DM, we shall use the halo model to trace
 its spatial distribution and predict its  cross-correlation with 
LSS  (see e.g., \cite{Cooray:2002dia,Ando:2013ff,Fornengo:2013rga}).


The paper is organized as follows. 
In Section \ref{sec:formalism} we present the theoretical estimate of the angular cross-correlation function and angular power spectra. 
Section~\ref{sec:analysis} describes the statistical techniques employed in the determination of the parameters
of the \g-rays emitters (DM and astrophysical sources) from the measured cross-correlation reported in  \cite{Xia:2014}. Section~\ref{sec:results} then shows our results, and finally Section \ref{sec:concl} summarizes our conclusions. The technical aspects of our theoretical modeling are presented in a set of three Appendices. Appendix \ref{sec:windows} introduces the modeling of the window functions of DM and astrophysical \g-rays sources and of catalogs of LSS tracers. Appendix \ref{sec:HOD} discusses the HOD of galaxies for the various catalogs. Appendix \ref{sec:3dps} describes  the derivation of three-dimensional (3D)  power spectra (PS). These are the ingredients used in Section \ref{sec:formalism}.

In this work we assume a fiducial flat $\Lambda$CDM model with the cosmological parameters derived by the {\it Planck} Collaboration in  \cite{Planck:2015xua}: matter density parameter $\Omega_{\rm m} = 0.31$, baryon density parameter
$\Omega_{\rm b} h^2 = 0.022$, reduced Hubble constant $h = 0.68$, rms matter fluctuations in a comoving sphere of 8 Mpc $\sigma_8=0.83$ and spectral index of primordial scalar perturbations $n_{\rm s} =0.96$.

\section{Formalism}
\label{sec:formalism}

To quantify the cross-correlation between \g-ray sources and the LSS tracers
in the various catalogues, we consider both the 2-point angular cross-correlation function 
(CCF) and its Legendre transform, i.e. the cross angular power spectrum (CAPS).
In the Limber approximation \citep{1953ApJ...117..134L,1992ApJ...388..272K,1998ApJ...498...26K}, the CAPS can be obtained by integrating the 3D-PS of cross-correlation $P_{\gamma g}(k,z)$:
\be
 C_\ell^{(\gamma g)}=\int \frac{d\chi}{\chi^2} W_{\gamma}(\chi)\, W_{g}(\chi)\,P_{\gamma g}\left(k=\ell/\chi,\chi\right)\;,
\label{eq:clfin}
\ee
where $\chi(z)$ denotes the radial comoving distance, $W(\chi)$ is the so-called window function that characterizes the distribution of objects and 
\g-ray emitters along the line of sight, $k$ is the modulus of the wavenumber and $\ell$ is the multipole.  
We relate the cosmological redshift $z$ to the radial comoving distances $\chi$ through the differential relation, valid in a flat cosmology,
$d\chi=c\,dz/H(z)$, where $H(z)$ is the expansion rate of the Universe. 

The indices \g\ and $g$ denote $\gamma$-ray emitters and extragalactic objects in different catalogs, respectively.
We consider five types of \g-ray sources: three different flavours of AGNs 
(BL Lacs, FSQRs, mAGN), SFGs and DM. We will consider both the case of annihilating and decaying DM particles.
For the LSS tracers, we consider five different catalogues: quasars in SDSS-DR6, 2MASS galaxies, 
NVSS radio sources, SDSS-DR8 Luminous Red Galaxies  and SDSS-DR8 ``main'' galaxies.

Denoting the density fields of an LSS tracer  with $f_g(\chi,\bm r)$,   
and that of the gamma ray emitter with $ f_{\gamma}(\chi,\bm r)$, where $\bm r$ indicates the position in comoving coordinates and $\chi$ labels time (given the one-to-one correspondence between time and distance), the cross-power spectrum 
is defined as:
\begin{equation}
 \langle \hat f_{\gamma} (z,\bm k) \hat f_{g}^\ast (z^\prime,\bm k^\prime) \rangle = (2\pi)^3 \delta^3 (\bm k + \bm k^\prime) P_{\gamma g} (k,z,z^\prime)\, ,
 \end{equation}
where $\hat f$ is the Fourier transform  
of $f(\chi(z),\bm r)/\langle f(\chi(z))\rangle$,  $\langle  \, . \, \rangle$ indicates the
average over the survey volume and the explicit dependence on $z$ and $z^\prime$ highlights the possibility that the two populations under study (\g-ray emitters and extragalactic LSS tracers) are located at two different redshifts.
From the Limber approximation one gets $\delta(z-z')$, so, in practice, only $P_{\gamma g}(k,z)$ is used.
The modeling of the various power spectra used in our analysis is derived in Appendix~\ref{sec:3dps}.
Objects in the catalogs are described in terms of their halo occupation distribution (HOD), which is discussed in Appendix~\ref{sec:HOD}.

The window function $W_{g}(z)$ appearing in Eq.~(\ref{eq:clfin}) weights the contribution of objects at different redshifts to the cross-correlation signal.
In the case of LSS tracers it coincides with the redshift distribution of the objects, $dN_g/dz$.
 More precisely, $W_{g}(z)\equiv H(z)/c\,dN_g/dz$ such that $\int \de\chi W_{g}(\chi)=1$ for a redshift distribution $dN_g/dz$ normalized to unity. 
 The expressions of $dN_g/dz$ for the different types of LSS tracers that we consider here are 
 the same as in  \cite{Xia:2014} (see also Appendix~\ref{sec:Wgal}).

For a \g-ray emitter the window function $W_\gamma(\chi)$ can be defined in term of the \g-ray  intensity integrated along the line-of-sight,
  $I_{f_{\gamma}}\!(\bm n)$  as function of the direction in the sky $\bm n$, which can be written as:
\be
 I_{f_{\gamma}}\! (\bm n) = \int d\chi\, \frac{f_{\gamma}(\chi,\bm r)}{\langle f_\gamma(\chi)\rangle} \, W_{\gamma}(\chi)
  \label{eq:int}
\ee
so that $\langle I_{f_\gamma} \rangle = \int d\chi\ W_\gamma(\chi)$.  
We will use a coordinate system centered on the observer so that $\bm r = \chi \, \bm n$. 
The expression of the density fields $f_{\gamma}$ and window functions $W_\gamma$ for the different classes of \g-ray sources  are provided in Appendix~\ref{sec:windows}.

\begin{figure*}[t]
\vspace{-4cm}
\includegraphics[width=10cm]{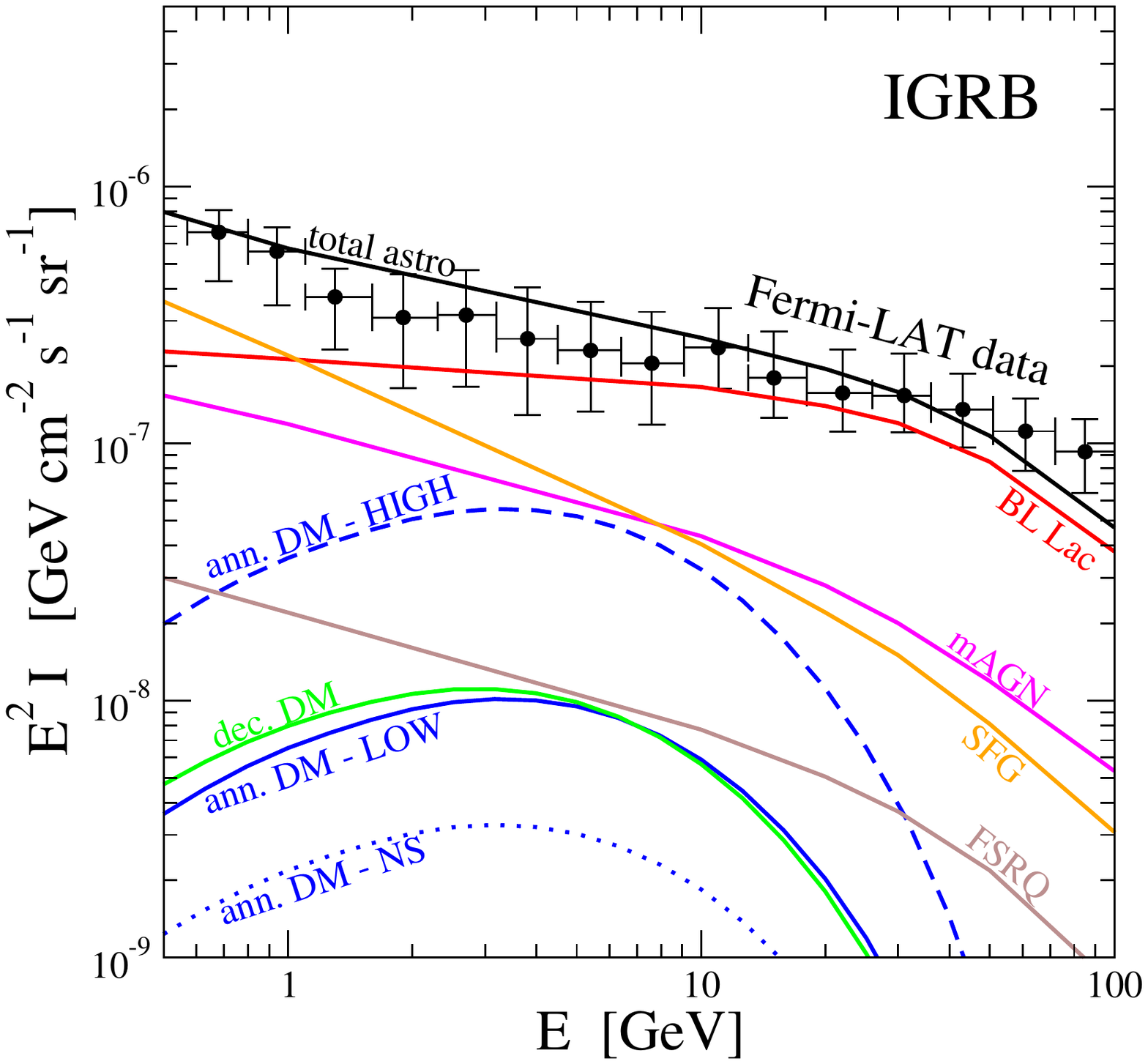}  \hspace{-1cm}
\includegraphics[width=10cm]{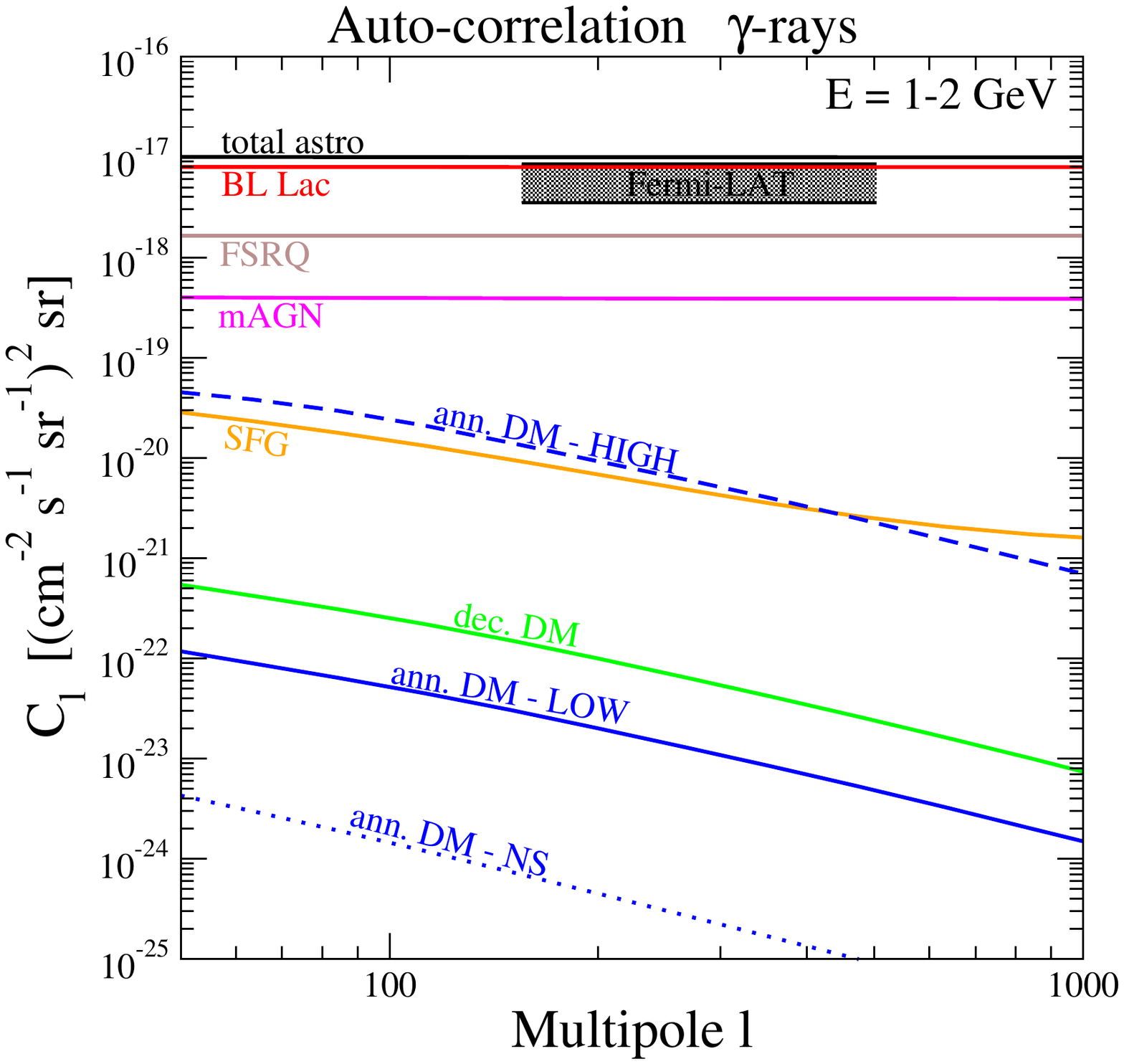}
\caption{Average gamma-rays intensity $I_\gamma$ as a function of photon energy ({\sl left}) and auto-correlation APS $C^{(\gamma\gamma)}_l$ in the (1--2) GeV energy band ({\sl right}) for the benchmark \g-ray models considered in this work. The black lines denote the total contribution arising from astrophysical sources (i.e. the sum of BL-Lac, mAGN, FSRQ and SFG emission). \Fermi-LAT data are shown as black points (left, from \cite{2015IGRB}, adding in quadrature systematic and statistical uncertainties) and a shaded region (right, from \cite{Ackermann:2012uf}, including the measurements with and without foreground cleaning). 
}
\label{fig:bench}
\end{figure*}

In the Appendices we also define our reference models for the astrophysical and DM \g-rays emitters, and for their cross-correlations with the LSS tracers. 
The mean  intensity  $I_\gamma = \langle I_{f_\gamma} \rangle $ as function of energy of the different \g-ray emitters for our reference models
is shown in the left panel of Fig.~\ref{fig:bench}.
The various curves in color indicate the contribution of each component, as indicated by the labels,
while the black line indicates the sum of all astrophysical contributions.
The predicted total energy spectrum matches the recent \Fermi-LAT measurements~\citep{2015IGRB} (solid dots with $1\sigma$ error bars).
Similarly, as shown in the right panel of Fig.~\ref{fig:bench}, we also verified that our reference model matches 
the observed  angular power spectrum of the diffuse extragalactic gamma ray background measured
in the 1--2 GeV energy band by the \Fermi-LAT (grey strip) \cite{Ackermann:2012uf}. 
The different curves in color  show the predicted angular power spectra of the various emitters that contribute to the total angular spectrum (solid black line).
The model angular power spectra for the various gamma ray emitters have been derived 
by using in Eq. (\ref{eq:clfin}) the power spectrum of the source $P_{\gamma \gamma}$ instead of the cross-spectrum 
$P_{\gamma g}$, and $W_{\gamma}^2(\chi)$ instead of the product $W_{g}W_{\gamma}$.
With respect to the more accurate procedure used in \cite{DiMauro:2014wha}, here we use the simplifying assumption that
all the sources of a given population have the same photon spectral index (see Appendix~\ref{sec:windows}).

In the next Section we will fit the theoretical predictions to the measured cross-correlations and will estimate
 the free parameters of the models. For the astrophysical components, the results will be in the form of {\it deviations} from reference models, which we adopt from the literature as updated benchmarks. We will therefore allow variations only in their normalization, plus a correction term as specified in the next Section.

A technical remark to take into account when comparing the model with the data is that
the experimental CAPS  determined from the data are {\it not} deconvolved from the effect   
of the point spread function (PSF) of the instrument and the effect of map pixelization.
To account for these effects we thus convolve our model prediction in Eq.~(\ref{eq:clfin}) with the PSF and pixelization 
using the same procedure described in  \cite{Xia:2014}.
Formally, this is implemented defining the new quantity  
directly comparable with the data as  $\tilde C_\ell^{(\gamma g)}=W_\ell^{B}\,C_\ell^{(\gamma g)}$
where  the effective beam window function $W_\ell^{B}$ parameterizes the PSF and pixelization effects
(see  \cite{Xia:2014} for more details).

Finally, in the following, we perform our analyses in terms of the cross correlation function  $CCF^{(\gamma g)}(\theta)$ rather than the cross-angular power spectra $C_\ell^{\gamma g}$. 
To obtain the CCF we perform a Legendre transformation on our CAPS as follows:
\begin{equation}
 CCF^{(\gamma g)}(\theta) = \sum_\ell\frac{2\ell+1}{4\pi}\tilde C_\ell^{\gamma g} P_\ell[\cos(\theta)] \;,
\label{eq:2point}
\end{equation}
where $\theta$ is the angular separation in the sky and $P_\ell$ are the Legendre polynomials.

\section{Statistical Analysis}
\label{sec:analysis}

In order to assess the possible presence of a DM signal in the IGRB,
and its robustness to the presence of astrophysical emitters,
we perform a statistical analysis fitting the observed cross-correlation data of  \cite{Xia:2014}
with a combination of both DM and astrophysical source models.
Specifically, we define  a $\chi^2$ statistic  from the data $D$, i.e., the
observed CCF between the \Fermi\ maps and the number of sources in catalogues \citep{Xia:2014}, and  $M$, i.e. the model CCF calculated for the different types of $\gamma$-ray emitters as introduced in the previous Section and detailed in the Appendices.
The $\chi^2$ is defined as:
\begin{equation}
   \chi^2 = \sum_{p=1}^5\,\sum_{n=1}^3\,\sum_{\theta_i\,\theta_j}   \left(D_{\theta_i}^{(p,n)} -M_{\theta_i}^{(p,n)}(\bm A) \right) \, \left[{\mathcal C}^{(p,n)}\right]^{-1}_{\theta_i\theta_j}  \left(D_{\theta_j}^{(p,n)} -M_{\theta_j}^{(p,n)}(\bm A) \right) \, ,
\label{eq:chi2}
\end{equation}
where the index $p$ runs over the five different catalogues of extragalactic sources 
 (2MASS, NVSS, SDSS-DR6 QSO, SDSS-DR8 Main Sample Galaxies and SDSS-DR8 Luminous Red Galaxies), 
the index $n$ runs over three \g-rays energy ranges ($E>0.5$ GeV, $E>1$ GeV and $E>10$ GeV), whereas 
the indices $\theta_i$ and $\theta_j$ run over 10 angular bins logarithmically spaced between $\theta= 0.1^\circ$ and $100^\circ$.
${\mathcal C}^{(p,n)}$ is the covariance matrix that quantifies the errors on the CCFs in each angular bin and the covariances among different bins, and $\bm A$ denotes the vector of free parameters which the CCF model $M$ depends upon (specified below). Both the covariance matrix 
${\mathcal C}^{(p,n)}$ and the measured CCFs $D_{\theta_i}^{(p,n)}$ are taken from  \cite{Xia:2014}. 
In Eq.~(\ref{eq:chi2}) the total $\chi^2$ is obtained by adding up the individual $\chi^2$ computed in three overlapping energy bands. 
There is, thus, in principle, a statistical dependence among the different energy  bands that should be accounted for. 
Nonetheless, such dependence is expected to be small since 
photon counts are heavily dominated by events near the lower end of each energy interval 
because of the steep IGRB energy spectrum  $\propto E^{-2.3}$ \citep{2015IGRB}.  
For this reason we will treat the CCFs estimated in the three energy intervals as 
statistically independent in the $\chi^2$ analysis.

For any given catalog of LSS tracers, energy band and angular bin  (i.e. for a given choice of $p$, $n$, and $\theta_i$) the 
theoretical CCF $M_{\theta_i}^{(p,n)}$ can be expressed as a sum of different contributions:
\begin{equation}
M_{\theta_i}^{(p,n)}= \sum_{\alpha=1}^5 A_{\alpha}  c_{\alpha}^{(p,n)} (\theta_i) +  A_{1h}^{(p)}  c_{1h}^{(n)} (\theta_i) \, .
\label{eq:model}
\end{equation}
The sum runs over the five different \g-ray emitters: BL Lacs, FSRQs, SFGs, mAGNs and DM. 
The terms $c_{\alpha}^{(p,n)} (\theta_i)$ denote the benchmark theoretical model CCFs 
described in  the Appendix  and $A_\alpha$ is a free normalization parameter that quantifies the individual contribution to the observed cross-correlation.
Values  $A_\alpha=1$  thus denote models equal to our benchmarks, while values $A_\alpha\neq1$ would correspond to deviations from
the benchmarks.

Besides the normalization of each component, we have introduced in the fit also a further free parameter,  $A_{1h}$,
which we dub {\it 1-halo correction-term}. This term is introduced as a correction
for possible inaccuracies in the modeling of the 1-halo contribution of the power spectrum (and thus mainly to the small scale cross-correlation signal)
of the \g-ray sources (hence its name),  as discussed after Eq. (\ref{eq:PSBd1}) in  Appendix \ref {sec:3dps}.
For simplicity we model it  as a constant term added in the CAPS, which
is  a good approximation of the 1-halo term itself for astrophysical components, except at very high multipoles $\ell>1000$,
which we do not considered in our analysis.
In real space, and taking into account the modulation introduced by the PSF of the instrument,
the  {\it 1-halo correction-term}  $M^{(p,n)}_{1h}(\theta_i)$ explicitly reads:
\begin{equation}
M^{(p,n)}_{1h}(\theta_i) = A_{1h}^{(p,n)}
 \sum_\ell\frac{2\ell+1}{4\pi} W_\ell^{B_n} P_\ell[\cos(\theta_i)] \, ,
\label{eq:1halo}
\end{equation}
where  $W_\ell^{B_n}$ is the (energy dependent) window function  of the PSF, also introduced in the previous section.
For definiteness we have assumed that $M^{(p,n)}_{1h}$ has the same energy dependence as the
IGRB spectrum $\propto E^{-2.3}$.  With this assumption we can separate the energy dependence from that on the source catalog
$A_{1h}^{(p,n)} =  A_{1h}^{p}\,f_n$ where $f_n= 2.46, \,1,\, 0.05$ for $E_n>0.5,\, 1,\, 10$ GeV (we take $E=1$ GeV as the normalization energy).  In this way, combining Eq.s (\ref{eq:model}) and (\ref{eq:1halo}) we have:
$c_{1h}^{(n)} =f_n\,\sum_\ell\frac{2\ell+1}{4\pi} W_\ell^{B_n} P_\ell$.
%
%
Notice that, in principle, each candidate \g-ray emitter has its own {\it 1-halo correction-term} for each catalog and energy band.
However they are degenerate and only their sum can be constrained. 
We have thus grouped them together so that in Eq. (\ref{eq:model})  the reported {\it 1-halo correction-term}
actually represents  the sum of the {\it 1-halo correction-terms} from all the components, for a given catalog and energy band.

All five values $A_{1h}^{p}$ are treated as free parameters in the analysis. 
Notice that they can be either positive or negative since we intend them
as possible correction to our benchmarks, and the natural expectation 
would thus be $A_{1h}^{p}$= 0 if the benchmarks are correct. 
The whole set of $A_\alpha$ and $A_{1h}^{p}$ coefficients (plus the additional parameter represented by the DM mass, upon which the DM signal depends) defines the parameter vector $\bm A$ of Eq. (\ref{eq:chi2}), which represents the full set of parameters over which our analysis is performed.

For what concerns the particle DM contribution, we consider both the case where \g-rays are produced 
through DM particle annihilation and the case of DM decay. 
The DM mass is varied from 10 GeV to 5 TeV and we will show the results for a DM which dominantly annihilates/decays into one
of the following \g-rays production channels: $b\bar b$, $\mu^+\mu^-$, $\tau^+\tau^-$ and $W^+W^-$.
For annihilating DM, the signal strongly depends on the clustering at small scales and, in particular, on the amount of substructures.
As discussed in Appendix \ref{sec:windows}, results will be shown for three DM substructure models: \high, \low\ and \ns. 
This will bracket the uncertainty on the reconstructed DM parameters arising from DM structure modeling. 
The \high\ scenario provides a more optimistic case with the largest boosting factor for the \g-ray annihilation flux.
The \ns\ scheme, where DM substructures are absent, provides a lower limit to to the annihilation signal and therefore
represent the most conservative scenario.
Finally,  the \low\  scheme represents an intermediate case which can be (currently) considered as the most realistic one and that 
we regard as our reference model.
Each one of these three scenarios predicts a different CCF, $c^{(p,n)}_{\alpha} (\theta_i)$ with $\alpha = DM$ in Eq. (\ref{eq:model}). 
Since the intensity of the DM signal is proportional to the DM annihilation cross section $\sigmav$ (or DM decay rate, $\Gamma_d$), 
we normalize our calculations to the reference values  $\sv_0=3\cdot10^{-26} {\rm cm^3 s^{-1}}$, i.e. the so-called ``thermal value" which correspond to a DM particle thermally produced in the early Universe which, alone, would account for the  observed DM relic abundance.
For decaying DM we normalize the models to a decay rate of
 $\Gamma_{d,0}=1.67\cdot10^{-28}{\rm s^{-1}}$, which is the decay rate which would
produce a DM signal equal to the one of an annihilating DM with thermal cross section in the \low\  substructure scheme (for DM masses around 100 GeV).
The parameter $A_{\alpha}$  for $\alpha = DM$ can be thus seen as the annihilation or decay rate in units of $\sv_0$ or $\Gamma_{d,0}$, respectively.

\begin{figure*}[t]
\centering
\includegraphics[width=0.9\textwidth]{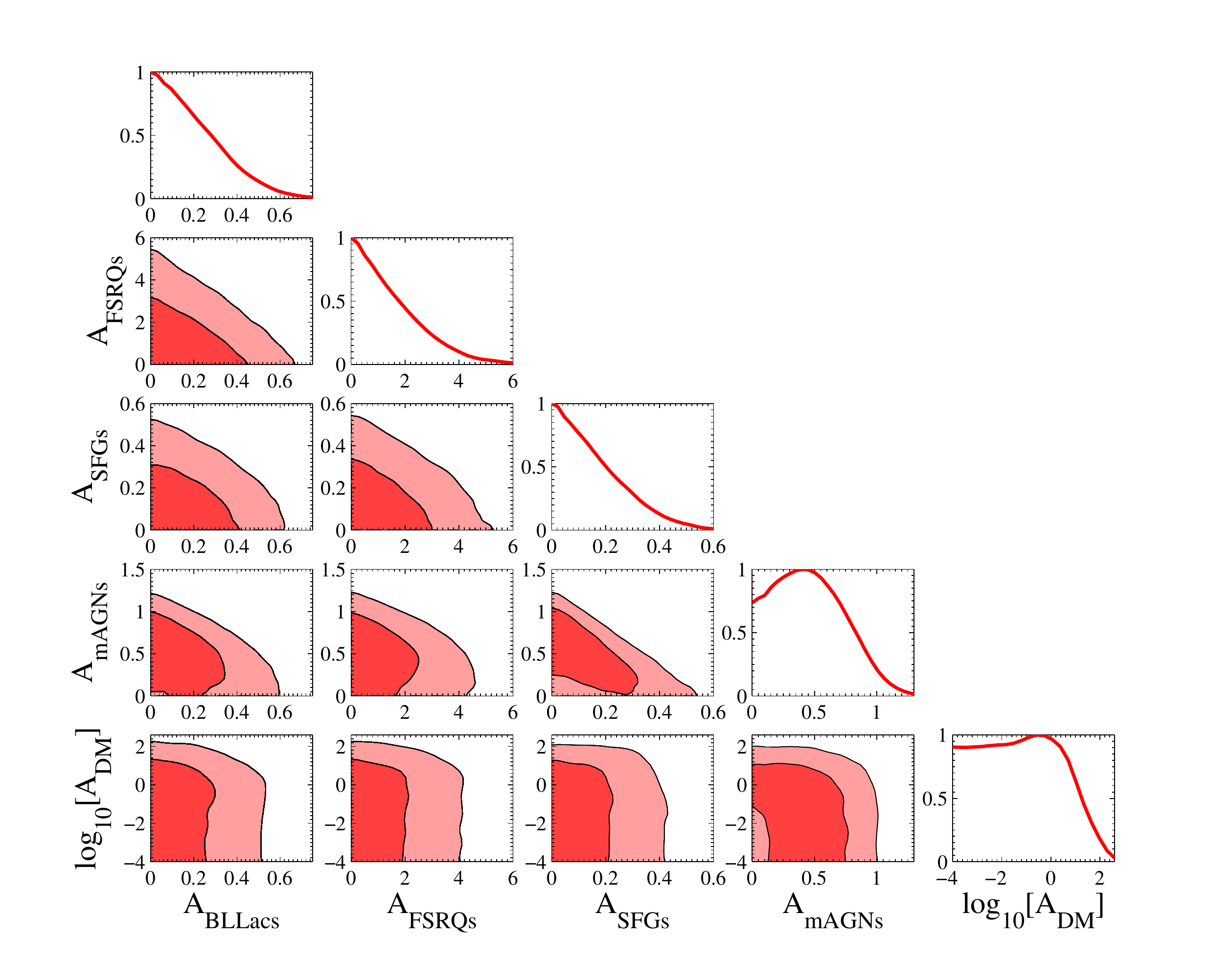}
\caption{Triangle plot of the parameters posterior distributions, for our reference fit setup, 
for the \low\ DM substructure scheme and for a DM particle annihilating into $b\bar b$.
The darker (innermost) and the lighter (outermost) areas denote the $1\sigma$ and 
$2\sigma$ credible regions, respectively, for each combination of parameters considered in the analysis.
The plots along the diagonal show the marginalized one-dimensional posterior distributions for each parameter. 
Notice that, for clarity, only the $5\times5$ sub-triangle plot with the $A_\alpha$ parameters is shown,
instead of the full $11\times 11$ full triangle plot, which includes also the dark matter mass $m_{DM}$ and the 1-halo correction $A_{1h}^{(k)}$ amplitudes.
}
\label{fig:tri_astrodm}
\end{figure*}

In summary, the global fit will be performed in a 11-dimensional parameter space, with the parameter vector given by 
\mbox{${\bm A} = (A_{DM},\,m_{DM},A_{BL Lac},A_{mAGN},A_{SFG},A_{FSRQ},A_{1h}^{k=1,2,3,4,5})$}.
All the parameters in the fit are linear, except $m_{DM}$ which enters non-linearly in the fit through
$c^{(k,n)}_{DM} (\theta_i)$.
Beside the above fit, we will also consider different configurations, namely different parameter vectors $\bm A$, to cross-check the robustness of the results.
In particular, we will consider the case where the $A_{1h}^k$ are set equal to zero. These additional analyses will be described in more details in the next Section.

In order to efficiently scan the multi-parameter space we adopt the Markov Chain Monte Carlo (MCMC) strategy
publicly available in the {\sl cosmomc} package \citep{Lewis:2002ah}.
We will use linear priors limited to positive values for  the normalization of the astrophysical 
components $A_{BL Lac},A_{mAGN},A_{SFG},A_{FSRQ}$, although we will also check log priors.
For the $A_{1h}$ parameters we allow for linear priors with negative values
since the {\it 1-halo correction-term} can either correct for over-estimation
or under-estimation of the small-scale cross-correlation.
Finally we will use a logarithmic prior for  $A_{DM}$ and $m_{DM}$ since,
 theoretically, the possible values of the DM mass and signal normalization can span several orders of magnitude.

Notice that in our $\chi^2$ analysis we consider only the cross-correlation signal 
and ignore the intensity and the auto-correlation of the IGRB.
These additional observational inputs will be used to perform an independent {\it a posteriori}  check on the validity of our results. 
While the total intensity $I_\gamma$ and the \g-rays autocorrelation $C_l^{(\gamma\gamma)}$ calculated 
from the derived best-fit configurations 
must not exceed the measured values (this will be a sanity check),  if they
 fall short of accounting for the data this might indicate either that the measured IGRB contains an unaccounted contribution 
which does not correlate with extragalactic tracers (possibly of Galactic origin) 
or that the modeling of the known components is imperfect/incomplete.
We will discuss more in detail these aspects in section~\ref{sanitycheck}  and in the conclusions.

\begin{figure*}[t]
\centering
\includegraphics[width=0.43\textwidth]{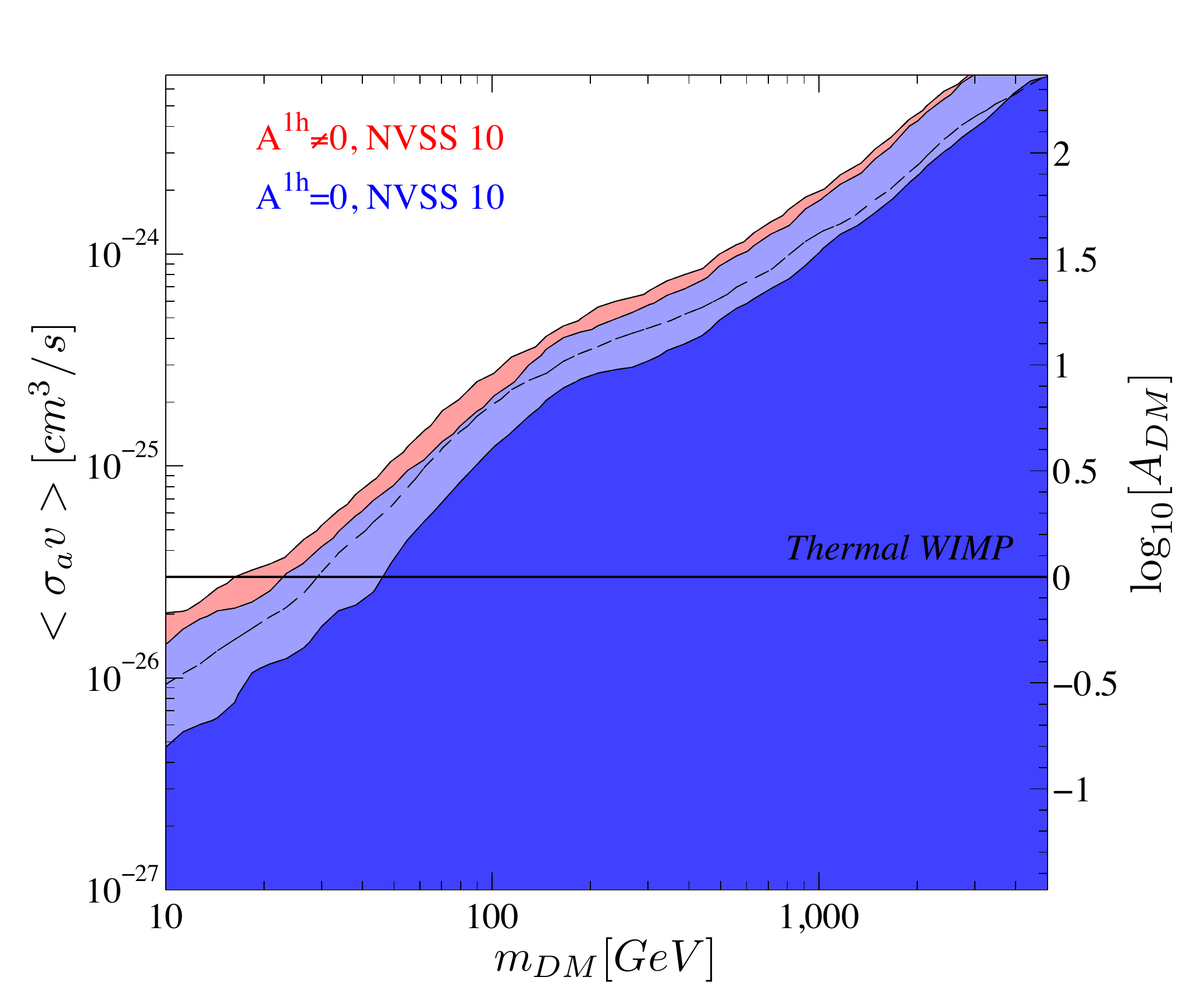}
\hspace{8mm}
\includegraphics[width=0.49\textwidth]{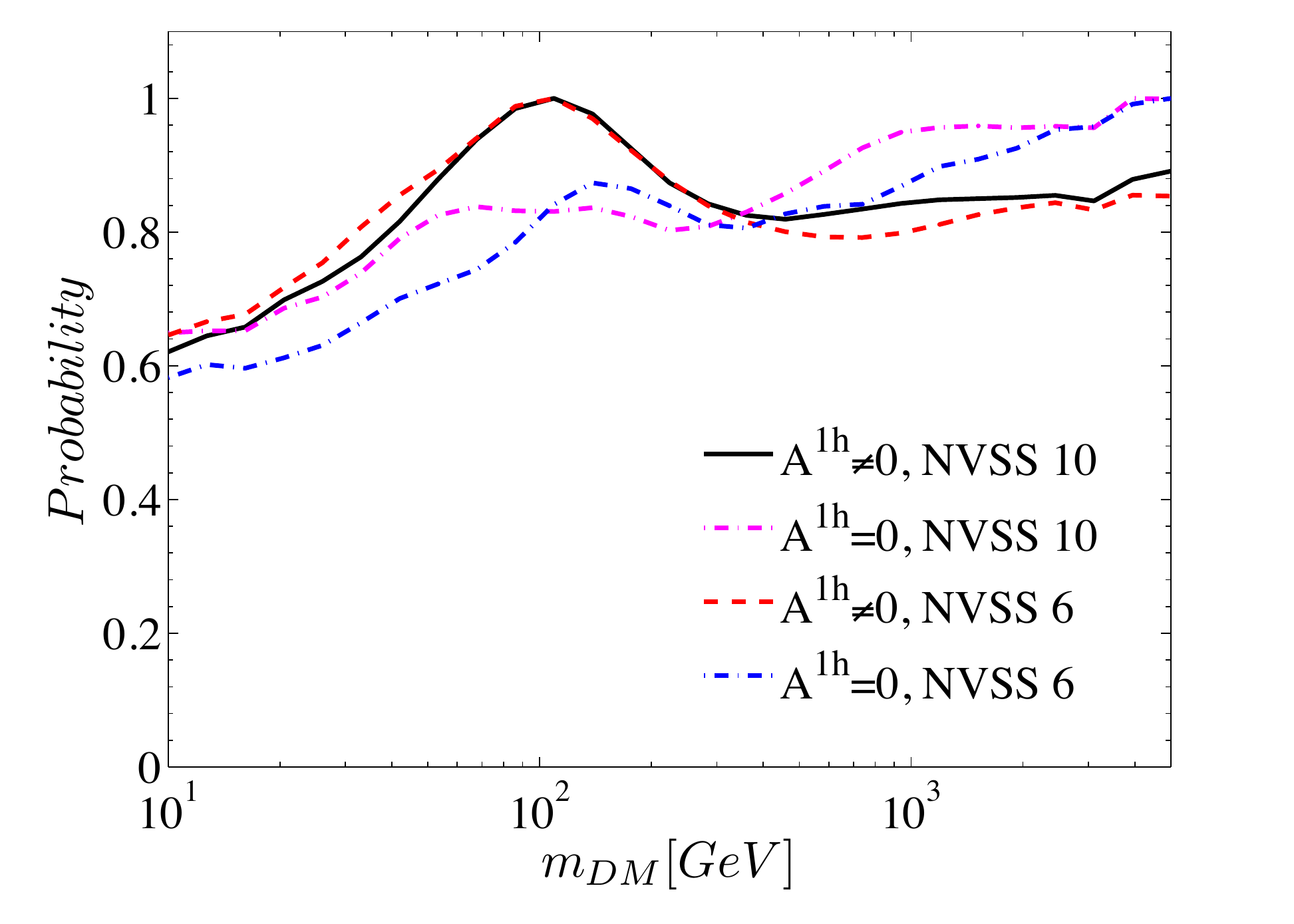}
\caption{{\sl Left}: $1\sigma$ and $2\sigma$  allowed credible regions for the annihilation rate $\sigmav$ versus the DM mass $m_{DM}$ in the NVSS-10 $A_{1h}^k=0$ (blue) and NVSS-10 $A_{1h}^k\neq0$ (red) fit setups. A DM particle annihilating into $b\bar b$ and the \low\ substructure scheme are assumed.
The lower 1$\sigma$ and $2\sigma$ contours of both cases extend down to $\sigmav=0$ (providing therefore only upper limits on the annihilation rate).
{\sl Right}: Marginalized 1D posterior probability for the DM mass, for the same DM annihilation channel ($b\bar b$) and substructure scheme (\low) as in the left panel.
The four lines refer to the four different fit setups described in the text, as labeled. }
\label{fig:DMtree}
\end{figure*}

\section{Results and Discussion}
\label{sec:results}

\begin{figure*}[t]
\centering
\includegraphics[width=0.9\textwidth]{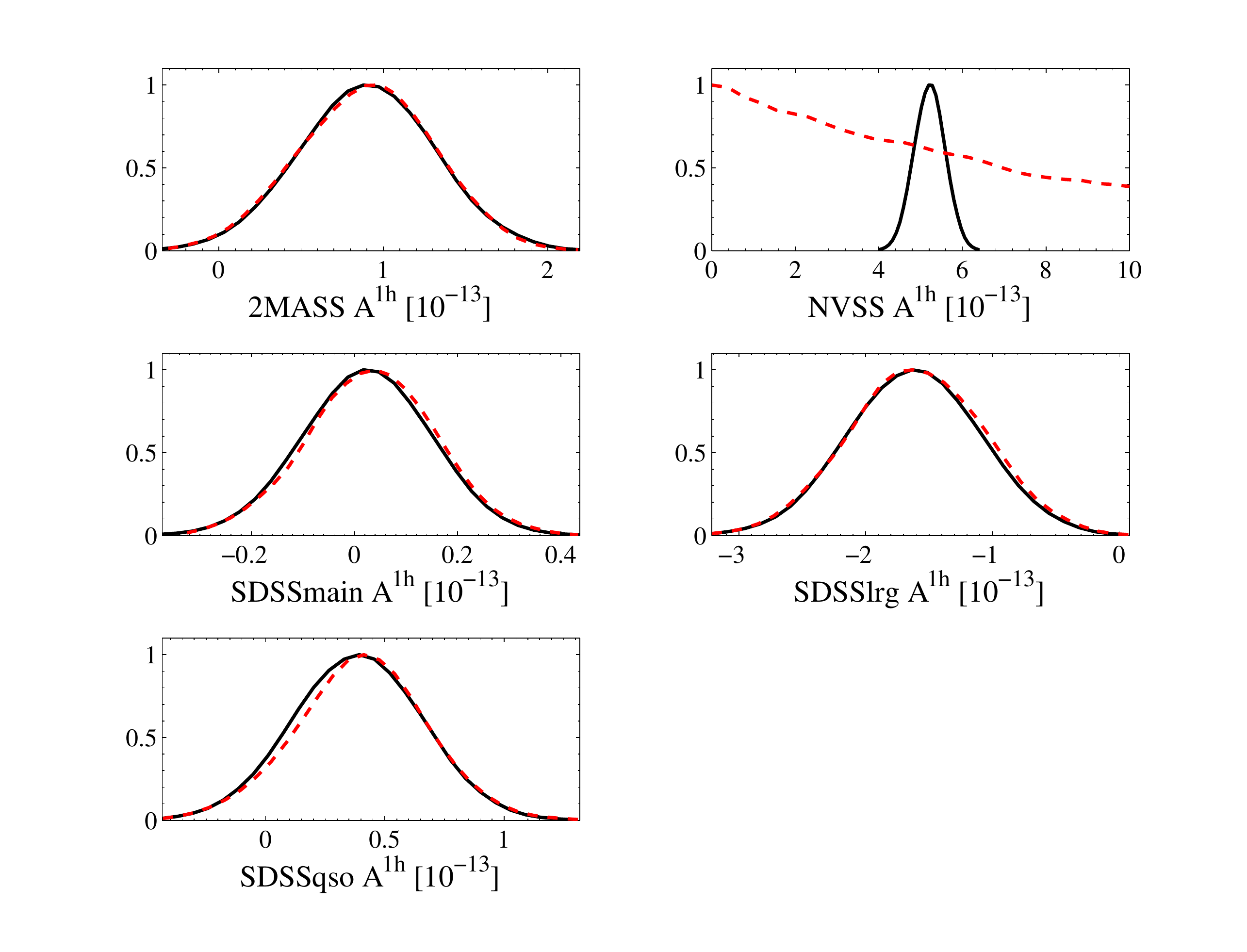}\\
\caption{Marginalized 1D posterior probabilities for the 1-halo correction terms $A_{1h}^{k}$, for the NVSS-10, $A_{1h}^k\neq0$ (black-solid) and NVSS-6, $A_{1h}^k\neq0$ (red-dashed) fits. They are in units of $10^{-13} {\rm cm^{-2}s^{-1}}$.}
\label{fig:c1hpost}
\end{figure*}

\begin{figure*}[ht!]
\begin{center}
%
\includegraphics[width=0.33\textwidth]{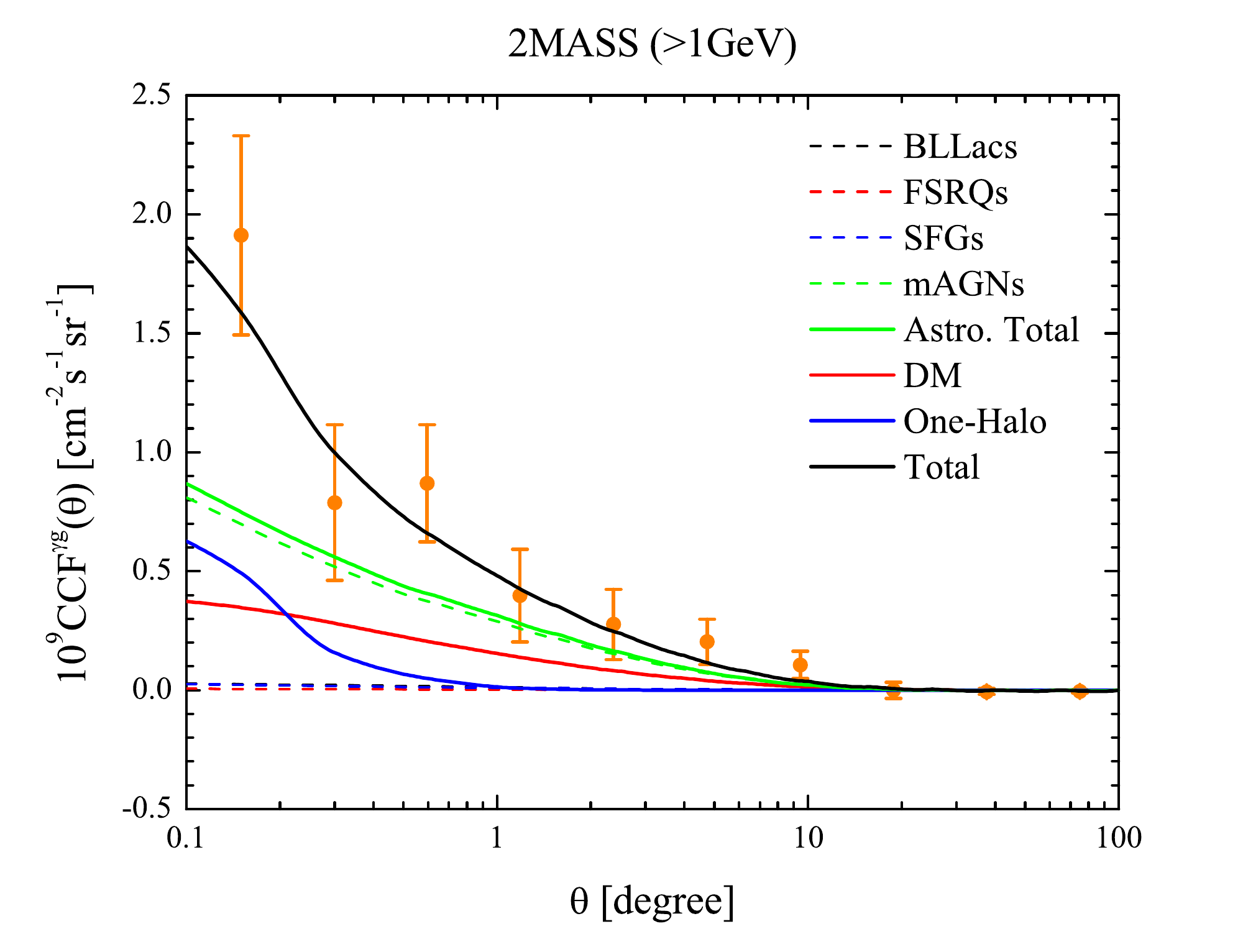} 
\hspace{-0.5cm}
\includegraphics[width=0.33 \textwidth]{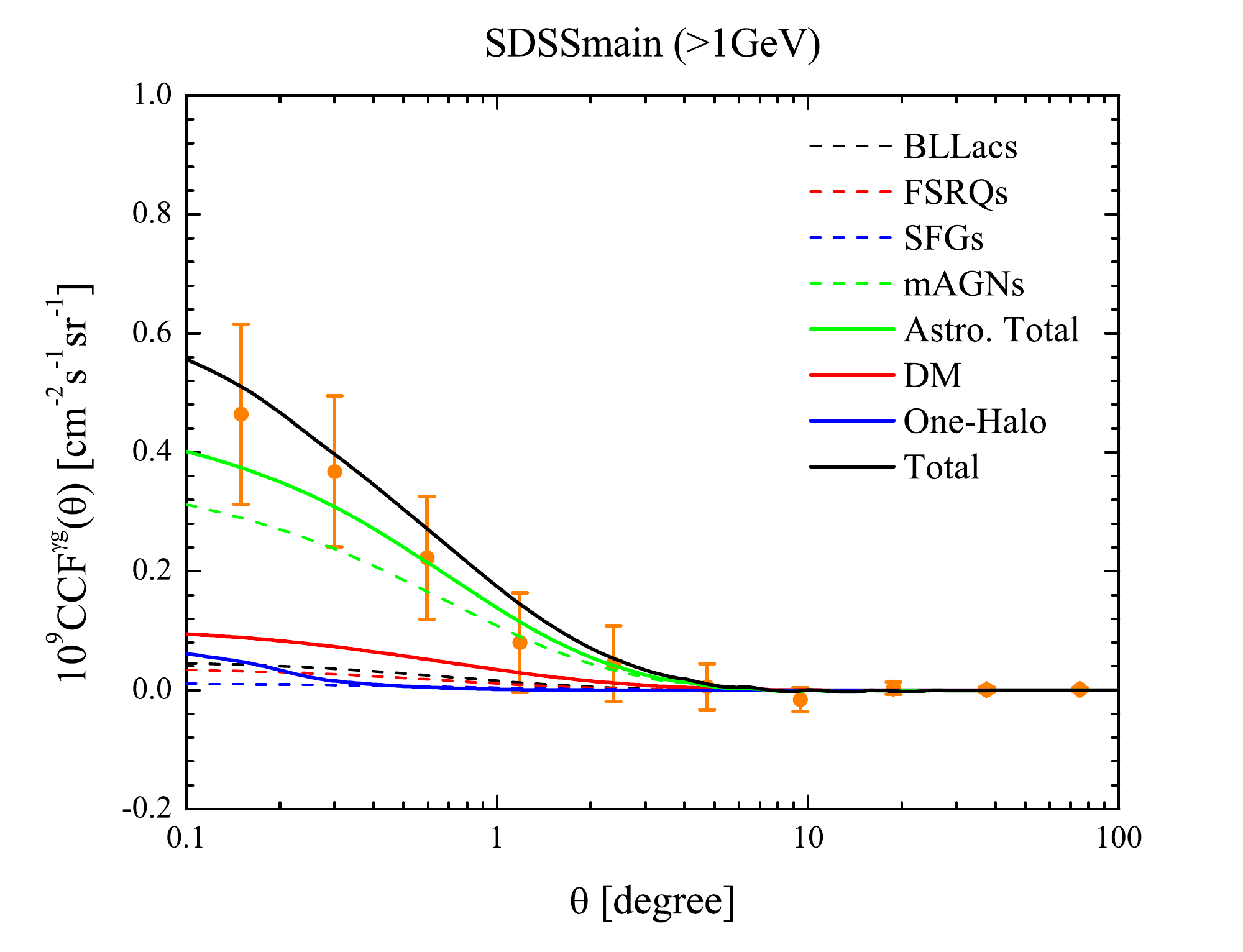}  
\hspace{-0.5cm}
\includegraphics[width=0.33 \textwidth]{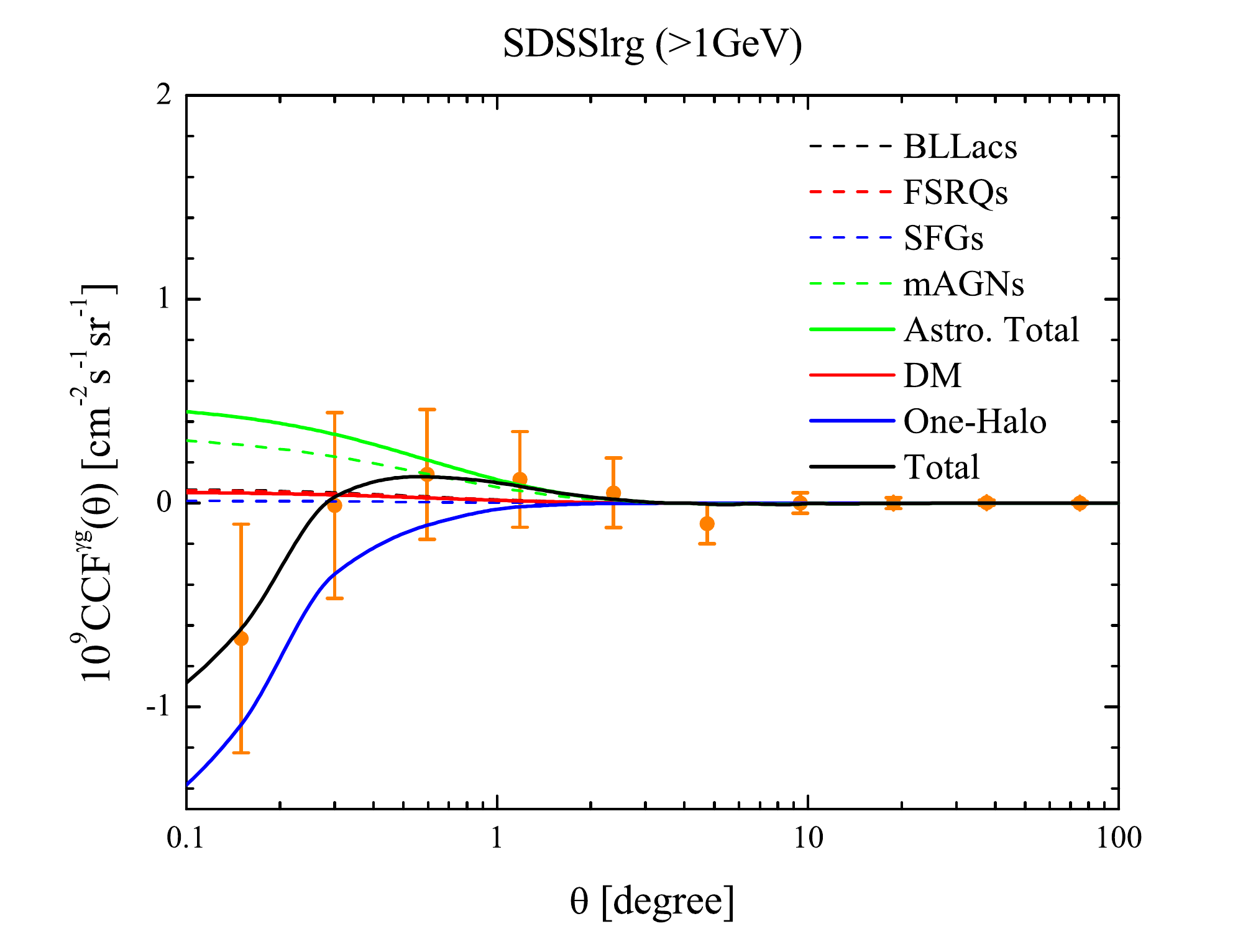} \\
%
%
\includegraphics[width=0.33 \textwidth]{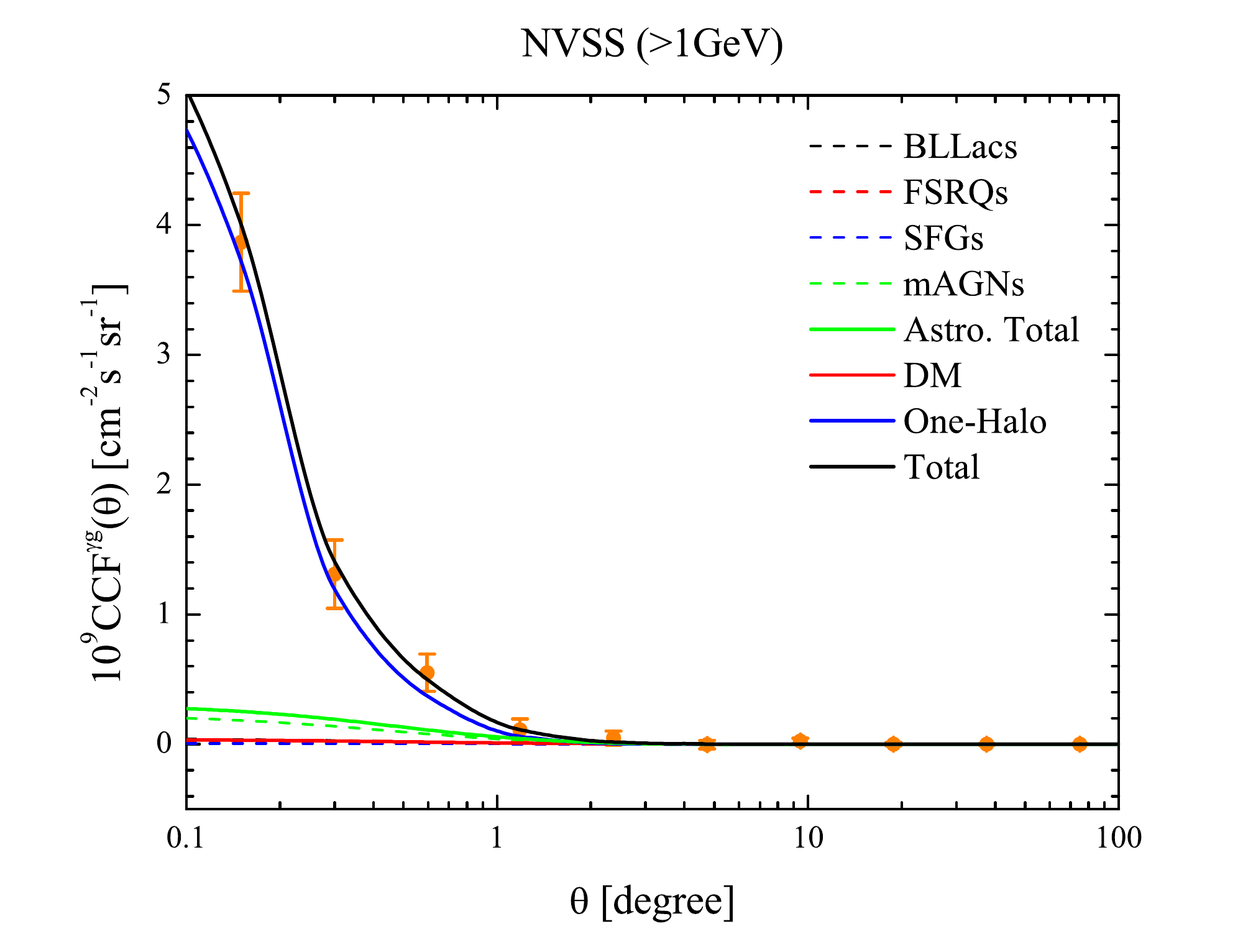} 
%
%
\includegraphics[width=0.33 \textwidth]{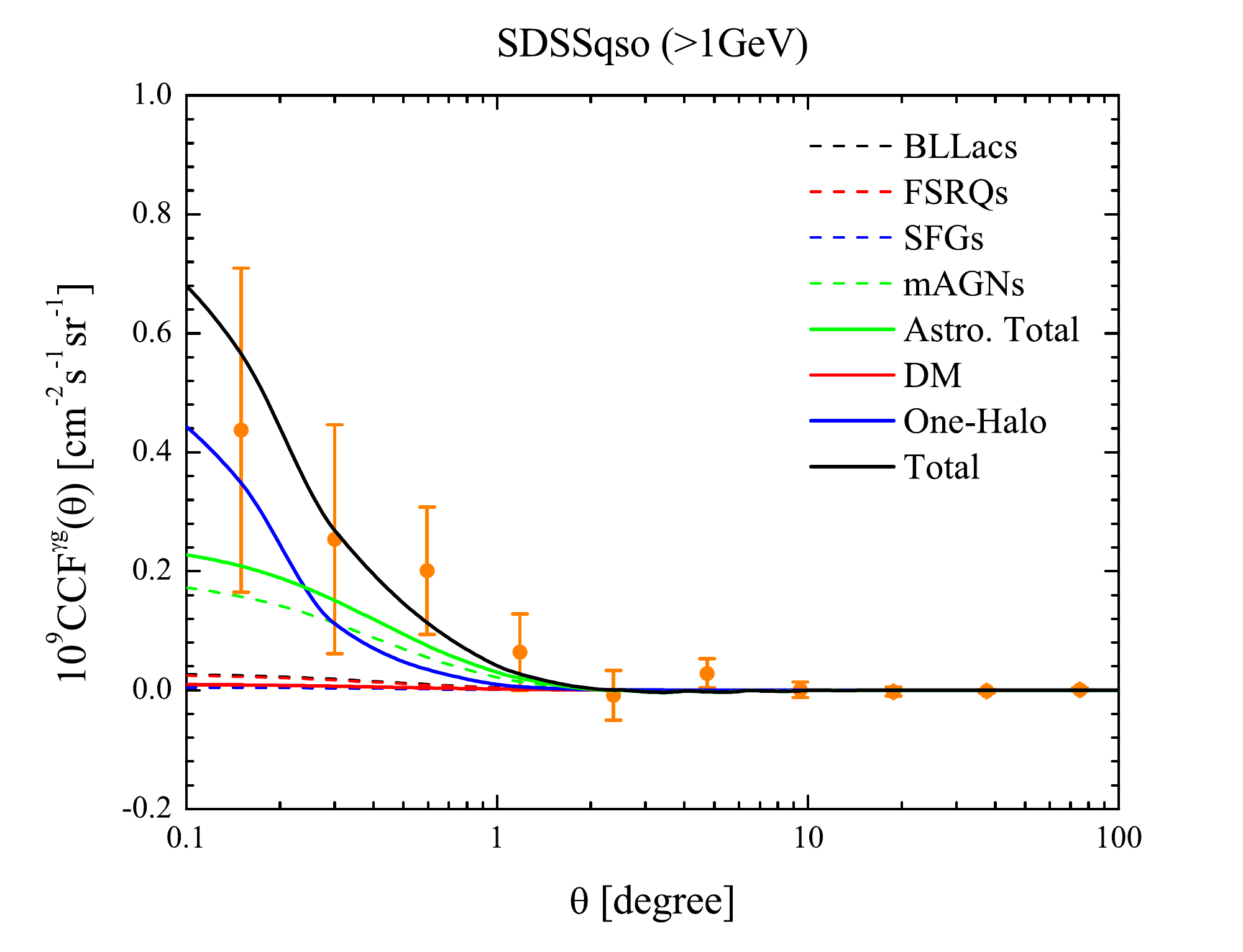} 
\end{center}
\vspace{-0.5cm}
\caption{Measured cross correlation function (CCF) \citep{Xia:2014} for $E>1$ GeV, as a function 
of the angular separation $\theta$ in the sky, compared to the best fit models of this analysis. 
The contribution to the CCF from the different astrophysical \g-rays emitters (BL Lac, mAGN, SFG, FSRQ) are shown 
by dashed colored lines, while their sum (``Astro Total") and the DM contribution are indicated by solid green and red lines, respectively. 
The {\it 1-halo correction} term is shown as a solid blue line. The total contribution to the CCF is given by the black solid line.
The analogous plots for $E>0.5$ GeV and $E>10$ GeV are shown n Appendix~\ref{otherenergies}.}
\label{fig:CCFdata} 
\end{figure*}

The triangle plot shown in  Fig.~\ref{fig:tri_astrodm} summarizes the results of our analysis
for a benchmark annihilating DM case with $b\bar{b}$ final state and \low\ substructure scheme.
The plot shows the posterior marginal distributions of the normalization parameters $A_\alpha$.
The two-dimensional plots refer to the $1\sigma$  and $2\sigma$ credible regions for each pair of parameters, 
while the diagonal shows the one-dimensional posterior distribution for each parameter. 
The parameter space is eleven-dimensional, but for clarity we show only a part
of the full triangle plot (without including here the parameters $A_{1h}^k$ and $m_{DM}$). 
The two-dimensional posterior of ($A_{DM}$, $m_{DM}$)
is shown separately in the left panel of Fig. \ref{fig:DMtree}.
The posterior probability for $m_{DM}$ is instead displayed in
the right panel of Fig. \ref{fig:DMtree},
while  the one-dimensional posteriors for the $A_{1h}^k$ parameters are shown in Fig.~\ref{fig:c1hpost}.  
Finally, Fig.~\ref{fig:CCFdata} shows the best-fit results compared with the measured cross correlation functions. 

A noticeable result from Fig.~\ref{fig:tri_astrodm} is the fact that all the $A_\alpha$ posteriors seem to peak at $A_\alpha=0$ or close to it, except (but with a low significance) for the DM and mAGN constributions. 
This does not necessarily imply that the best fit is found when the contributions from all components is zero.  Instead, it is an 
indication that strong degeneracies are present. 
A two-dimensional analogy is given by a case with only two parameters related by a simple relation  $A_1+A_2=$ const.
While the degeneracy would be clearly seen in the two-dimensional
posterior,  both the one-dimensional  $A_1$ and $A_2$ posteriors  peak at zero, although $A_1$ and $A_2$
are never both zero at the same time.
This is precisely the results we find here, although  the high dimensionality of our parameter space prevents us from 
clearly trace the parameter degeneracy  even  in the two-dimensional posteriors plots.

The degeneracy of the different astrophysical components can be 
traced to the behavior of their respective window functions $W(z)$, which possess a relatively similar evolution as a function of redshift, 
and to a similar behavior of their cross-correlation 3D power spectra.  
As can be seen in Fig. \ref{fig:kernel} in Appendix~\ref{sec:windows}, apart from the DM case for which the \g-rays emission is concentrated at low redshift 
with a fast decrease for increasing distances, astrophysical sources possess a relatively broad kernel. 
Fig. \ref{fig:ps2MASS} and Fig. \ref{fig:psall}
instead show some examples of 3D cross-spectra  between  LSS tracers and  the various astrophysical sources considered here or DM: 
we notice that, for a given LSS tracer (e.g. 2MASS in Fig.\ref{fig:ps2MASS}), the behaviors are quite similar for all astrophysical sources
 (while, instead, differences can be appreciated for the DM case).
These facts, together with  the relatively large error bars 
makes astrophysical-component separation currently difficult. On the other hand, given the somewhat different 3D cross-spectra,
perspective to separate the DM component are, perhaps, brighter.

An exception in this line of reasoning are  mAGNs and SFGs which exhibit a significant degree of degeneracy with DM in the 2D posterior contours.
The main features of the DM  signal is that it peaks at low redshift and that is mostly contributed by massive halos. To mimic such a signal an astrophysical source must then preferentially be hosted
in large halos at low $z$.
Both SFGs and mAGNs meet the redshift requirement while the blazars do not,  since their window peaks at higher $z$ . 
However, only mAGNs are believed to be hosted in large halos, while SFGs  typically populate galaxy-size halos.
Objects in large halos at low redshifts are expected to have a large bias and, more importantly,  their correlation properties
at the Mpc scale is dominated by 
a large one-halo term. 
This introduces a characteristic feature in the cross-PS that differentiates mAGNs from SFGs, 
making their contribution more similar to the DM one at $\sim$ Mpc scales (see left panel of Fig.~\ref{fig:ps1h}).
At the lowest redshift considered (namely, in the cross correlation with 2MASS), the Mpc scale corresponds to a sub-degree scale in the CCF. 
Nonetheless, given the present still large error bars, the above feature is only weakly constrained and thus a further degeneracy of both components with SFGs still remains on top of the mAGN-DM main degeneracy. Further investigation of this issue is reported later below. 
Instead, further differences between the mAGNs and and the DM cases 
are expected at smaller angles which, unfortunately, cannot be investigated given the size of the \Fermi-LAT PSF.

Difficulties in modeling the 1-halo term in the HOD framework  described in Appendix \ref{sec:HOD}
propagates into uncertainties in predicting the cross-power at small-angles. To account for this
potential source of systematic errors we introduced in Eq.~(\ref{eq:1halo}) the 1-halo correction-terms $m^{(k,n)}_{1h}$.
The 1D marginalized posteriors of the associated extra five parameters $A_{1h}^{k}$ are shown in Fig.~\ref{fig:c1hpost}
as black solid curves.
The various datasets are consistent with the case $A_{1h}^{k}=0$  with different confidence levels, except
NVSS. In this case  we find a strong and statistically significant deviation from zero.
This can also be appreciated in the fit to the observed CCF in Fig.~\ref{fig:CCFdata} where 
the presence of a prominent 1-halo correction-term is required to fit the data at small angles.
There is a likely explanation for this additional contribution: the presence in the NVSS catalog of 
 \g-ray point sources (i.e., AGN) that are just below \Fermi\  detection threshold.
These sources would add their auto-correlation signal at zero-lad that, because of the PSF, spreads out to 
$\sim 1$ deg scale.
This effect requires some fine tuning of the parameters defining the 1-halo term in Eq. (\ref{eq:PSBd1}) 
which the benchmark model fails to catch, thus requiring a large correction term.
The effect is also discussed in \cite{Xia:2014} to which we refer the
reader for further discussion.
The relevance of this term in the fit to NVSS data is expected to affect our constraints of the DM properties.
To investigate this issue we use three further fitting procedures in addition to the one adopted so far.
The four fitting procedures are as follows:
\begin{itemize}
\item {\bf NVSS-10, $A_{1h}^k\neq0$}. All the 10 NVSS data points are fitted and the {\it 1-halo-correction} terms are free parameter of the fit. This is the 
standard fitting procedure used to obtain the results shown in Fig.~\ref{fig:tri_astrodm}.
\item {\bf NVSS-10, $A_{1h}^k=0$}. All the 10 NVSS data points are fitted and all the {\it 1-halo-correction} terms are set equal to zero.
\item {\bf NVSS-6, $A_{1h}^k\neq0$}. The first 4 NVSS data points at small angles are excluded from the fit. The {\it 1-halo-correction} terms are used as free parameters in the fit.
\item {\bf NVSS-6, $A_{1h}^k=0$}. The first 4 NVSS data points are excluded from the fit. All {\it 1-halo-correction} terms are set equal to zero.
\end{itemize} 

\begin{figure*}[t]
\centering
\includegraphics[width=0.9\textwidth]{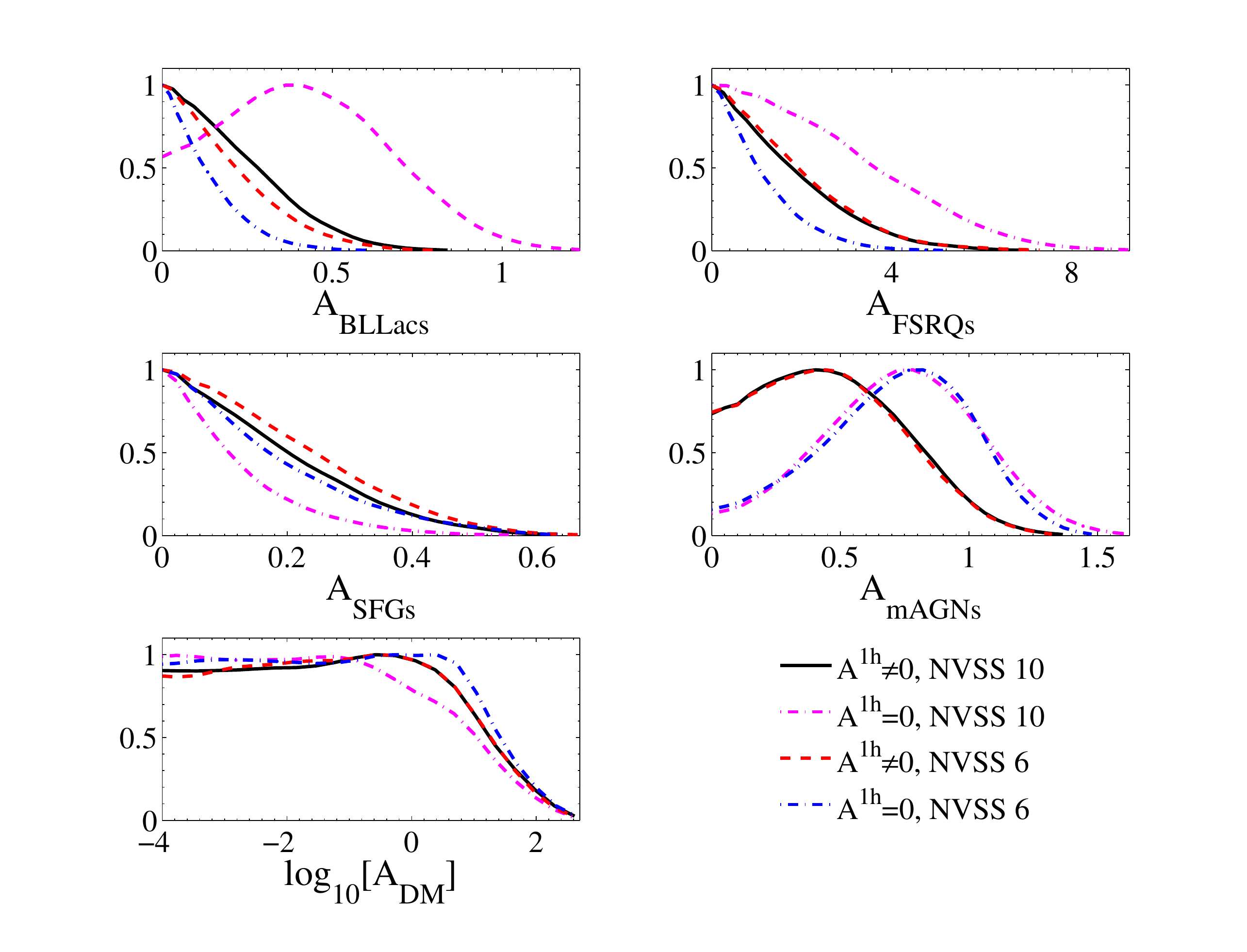}\\
\caption{Marginalized 1D posterior probabilities for the $A_{\alpha}$ terms. } 
\label{fig:Apost}
\end{figure*}

Fig.~\ref{fig:c1hpost} shows the $A_{1h}^k$ posteriors for the  NVSS-10, $A_{1h}^k\neq0$ (black solid curves) and the NVSS-6, $A_{1h}^k\neq0$ (red dashed curves) cases.
It can be seen that there are no significant differences between the two fitting schemes, except for the NVSS case
in which $A_{1h}^k$ becomes obviously unconstrained when the first four data points, where the fit is guaranteed by the {\it1-halo-correction} term, are ignored. 
Fig.~\ref{fig:Apost}  quantifies the impact of the four fitting schemes on the posterior probabilities of all $A_\alpha$ parameters.
The plots show that the fitting procedure does have an impact on some $A_\alpha$ parameter.
In particular the results obtained with the NVSS-10, $A_{1h}^k=0$ fit deviates from the others in most of the cases. 
However, this is also the scheme that provides the worst fit to the various datasets, as illustrated by the 
comparatively larger $\chi^2$ values listed in Table~\ref{tab:chi2}, so that the results from this case are likely somewhat biased.  
Instead, all the 
three remaining  schemes provide reasonably good fits to the different datasets.
The NVSS-6, $A_{1h}^k=0$ case provides a slightly worse fit to the data, particularly to the LRG sample, 
than the other schemes. 
It is interesting to notice that this fitting scheme favours a non-zero mAGN
component (see the $A_{mAGN}$ panel in Fig.~\ref{fig:c1hpost}) which is absorbed by the {\it 1-halo-correction} term 
when a fitting scheme with $A_{1h}^k\neq0$ is adopted. 
This indicates that a degeneracy between the $A_{1h}^k$ and the $A_{mAGN}$ parameter is present.
We notice that in all three cases the $\chi^2$ is lower than total number of degrees of freedom.
This is partly due to some unaccounted correlation between the three energy bins and between the different catalogues,
and partly to the fact that the error bars are probably slightly over-estimated.
Indeed, it is known that the algorithm implemented in the PolSpice software which is used in \cite{Xia:2014} is not a minimum variance estimator
 of the error bars, i.e., it does not provide the smallest error possible~\citep{Efstathiou:2003dj}.

\begin{table*}
\begin{center}
\caption{Best-fit $\chi^2_{\rm bf}$ for the four analysis described in the text, broken down into the contributions from the three energy bands  ($E_{05}$, $E_{1}$ and $E_{10}$ stand for $E>0.5$, 1, 10 GeV, respectively) and the five catalogs used. 
The number of degrees of freedom, $N_{\rm DOF}$, is expressed as the total number of data points minus the number of free parameters in the fit.
\label{tab:chi2}}
\footnotesize{
\hspace{-0.4cm}
\begin{tabular}{|c|ccc|ccc|ccc|ccc|ccc|ccc|c|c|}
\hline
$\chi^2_{\rm bf}$  &   \multicolumn{3}{|c|}{2MASS}   & \multicolumn{3}{c|}{SDSS-MG}   & \multicolumn{3}{c|}{SDSS-LRG}  & \multicolumn{3}{c|}{SDSS-QSO}   & \multicolumn{3}{c|}{NVSS}   & \multicolumn{5}{c|}{TOTAL}  \\
\cline{2-21}
   & $E_{05}$ & $E_{1}$ & $E_{10}$ & $E_{05}$ & $E_{1}$ & $E_{10}$&$E_{05}$ & $E_{1}$ & $E_{10}$ &$E_{05}$ & $E_{1}$ & $E_{10}$ & $E_{05}$ & $E_{1}$ & $E_{10}$ & $E_{05}$ & $E_{1}$ & $E_{10}$
& {\scriptsize  All E} & {\scriptsize $N_{\rm DOF}$}\\
\hline
{\footnotesize {\sc NVSS-10} $A_{1h}^k$$\neq$$0$}  & 6.5  & 8.5 & 2.5 & 4.0 & 2.5 & 6.3    & 2.4 &  2.1 & 3.0 & 16.8 & 4.2 & 6.9 & 3.8 & 3.7 & 6.6        & 33.5 & 21.1 & 25.3 & 79.9 & 150-11 \\
{\footnotesize {\sc NVSS-10} $A_{1h}^k$=$0$}      & 6.4  & 12.5 & 2.7 & 13.5 & 6.4 & 8.9 & 10.1 & 9.5 & 4.0 & 13.9 & 3.8 & 4.9 & 68.1 & 84.6 & 56.1 & 112.1 & 116.9 &  76.6 & 305.6 & 150-6\\
{\footnotesize {\sc NVSS-6  } $A_{1h}^k$$\neq$$0$}    & 6.4  & 8.8 & 2.3   & 3.3 & 2.4 & 6.8   & 2.3 &  2.1 & 2.9 & 17.4 & 4.4 & 7.1   & 1.5 & 2.1 & 2.6& 31.0 & 19.8 & 21.7 &  72.5 & 138-11\\
{\footnotesize {\sc NVSS-6  } $A_{1h}^k$=$0$}    & 6.2  & 11.3 & 2.3   & 4.8 & 2.6 & 6.8   & 6.4 &  6.3 & 2.9 & 19.0 & 4.7 & 6.2   & 1.5 & 2.0 & 2.5& 38.0 & 27.0 & 20.8 &  85.8  & 138-6\\
\hline
\end{tabular}}
\end{center}
\end{table*}

\begin{figure*}[t]
\centering
\includegraphics[width=0.9\textwidth]{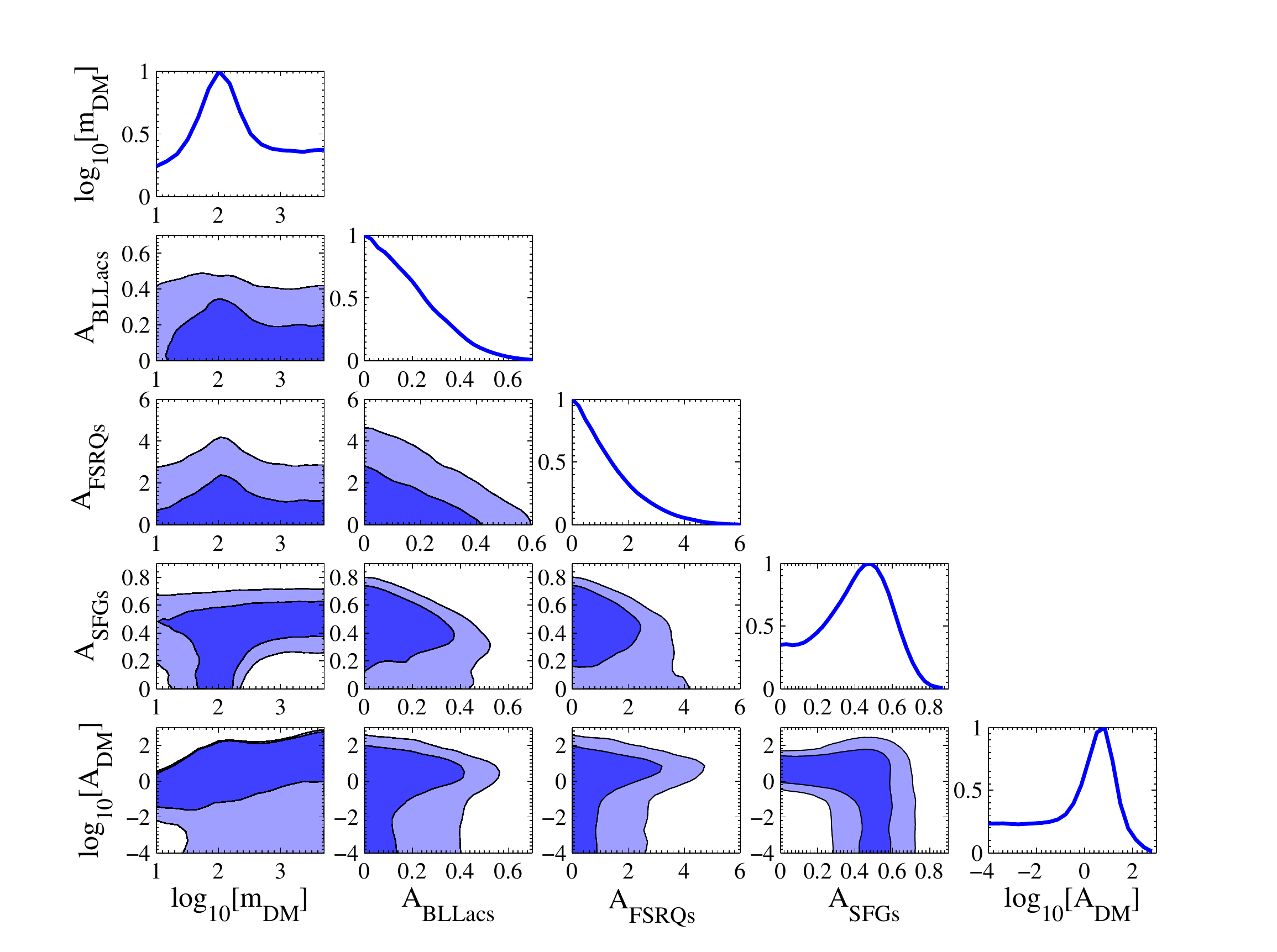}\\
\caption{Triangle plot for the NVSS-6, $A_{1h}^k=0$, $A_{mAGN}=0$ fit. The results refers to the $b\bar b$ annihilation channels and the \low\ DM substructure scheme.} 
\label{fig:trianglenoMAGN}
\end{figure*}

\begin{figure*}[tb!]
\centering
\includegraphics[width=0.49\textwidth]{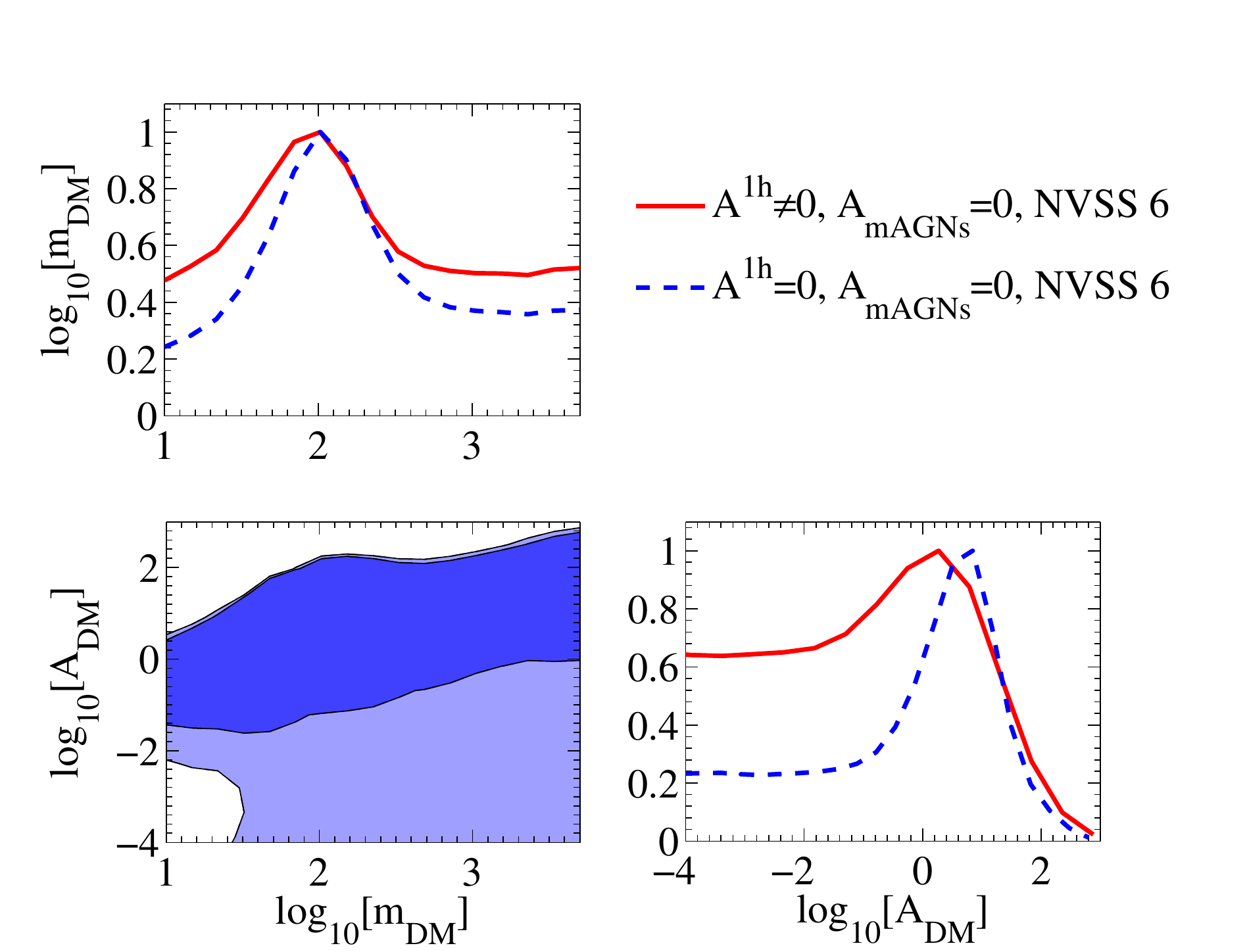}
\includegraphics[width=0.49\textwidth]{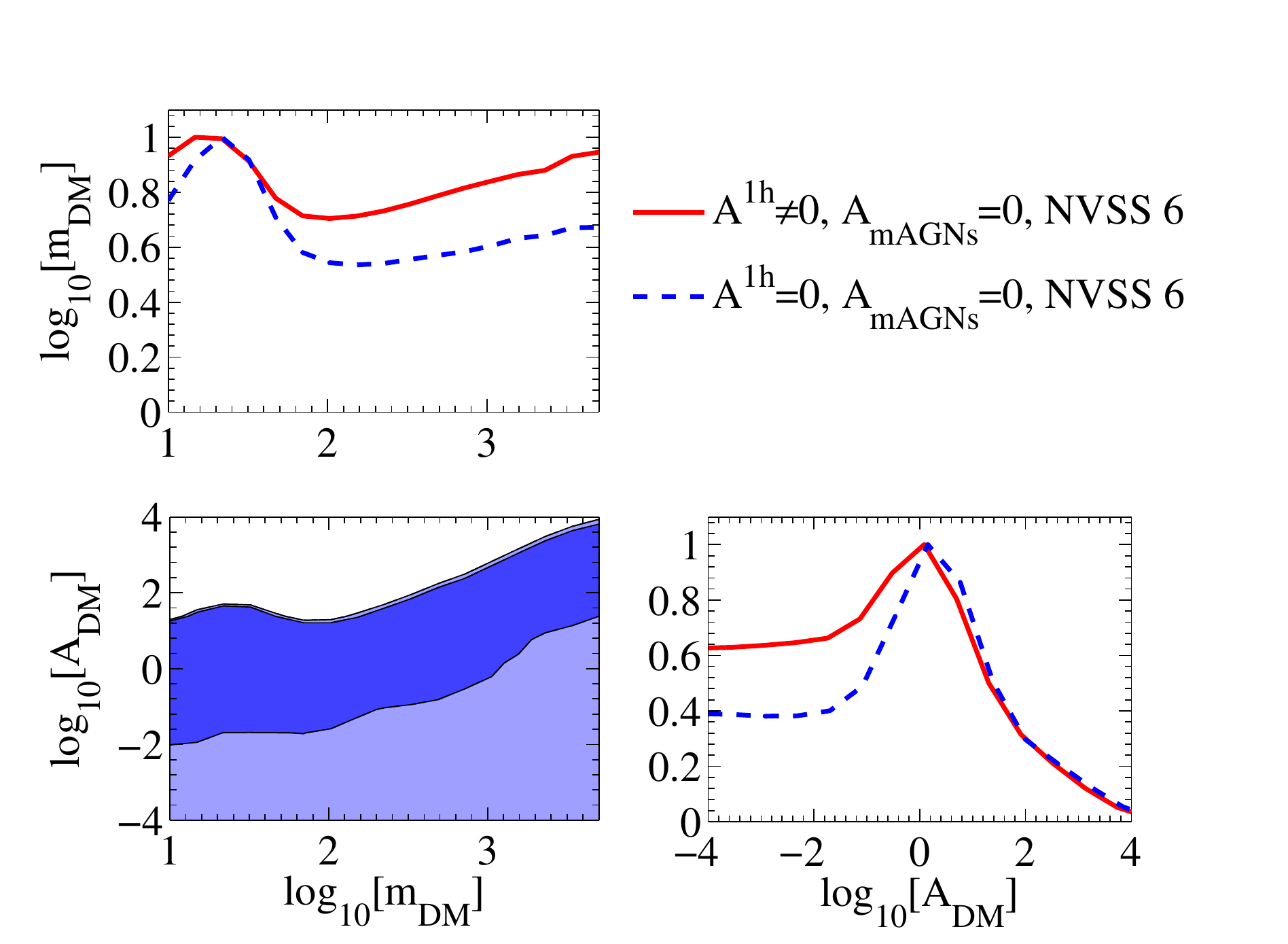} 
\caption{{\sl Left}: Detail from Fig.~\ref{fig:trianglenoMAGN} showing the DM parameters only. Furthermore, the 1D posterior panels show
both the $A_{1h}^k=0$ and $A_{1h}^k\neq 0$ cases. The plot refers to the $b\bar b$ annihilation channels and the \low\ DM substructure scheme.
{\sl Right}: The 
same as in the left panel, but for the $\tau^+\tau^-$ annihilation channel.
}
\label{fig:trianglenoMAGNonlyDM}
\end{figure*}

\begin{table*}
\begin{center}
\caption{Best fit  $\chi^2_{\rm bf}$ for the four NVSS-6,  $A_{mAGN}=0$ setup, broken down into the contributions from the three energy bands ($E_{05}$, $E_{1}$ and $E_{10}$ stand for $E>0.5$, 1, 10 GeV, respectively) and the five catalogs used.  
The number of degrees of freedom, $N_{\rm DOF}$, is expressed as the total number of data points minus the number of free parameters in the fit.
\label{tab:chi2noMAGN}}
\footnotesize{
\hspace{-0.2cm}
\begin{tabular}{|c|ccc|ccc|ccc|ccc|ccc|ccc|c|c|}
\hline
$\chi^2_{\rm bf}$  &   \multicolumn{3}{|c|}{2MASS}   & \multicolumn{3}{c|}{SDSS-MG}   & \multicolumn{3}{c|}{SDSS-LRG}  & \multicolumn{3}{c|}{SDSS-QSO}   & \multicolumn{3}{c|}{NVSS}   & \multicolumn{5}{c|}{TOTAL}\\
\cline{2-21}
   & $E_{05}$ & $E_{1}$ & $E_{10}$ & $E_{05}$ & $E_{1}$ & $E_{10}$&$E_{05}$ & $E_{1}$ & $E_{10}$ &$E_{05}$ & $E_{1}$ & $E_{10}$ & $E_{05}$ & $E_{1}$ & $E_{10}$ & $E_{05}$ & $E_{1}$ & $E_{10}$
&{\scriptsize  All E} & {\scriptsize $N_{\rm DOF}$}\\
\hline
{\footnotesize $A_{1h}^k$$\neq$$0$  $A_{DM}$$\neq$$0$}  & 7.0  & 8.0 & 2.3 & 3.1 & 2.4 & 6.3    & 2.2 &  2.0 & 3.3 & 17.5 & 4.3 & 7.1 & 1.4 & 2.0 & 2.6        & 31.3 & 18.7 & 21.6 & 71.6 & 138-10 \\
{\footnotesize  $A_{1h}^k$=$0$ $A_{DM}$$\neq$$0$}      & 6.0  & 11.3 & 2.2 & 4.0 & 2.7 & 6.6 & 6.5 & 6.4 & 2.8 & 22.2 & 5.8 & 6.9 & 1.5 & 2.0 & 2.5 & 40.2 & 28.3 &  21.0 & 89.5 & 138-5\\
{\footnotesize $A_{1h}^k$$\neq$$0$ $A_{DM}$$=$$0$}    & 6.9  & 10.7 & 3.8   & 4.7 & 2.2 & 5.8   & 2.2 &  1.9 & 3.4 & 16.8 & 4.3 & 6.9   & 1.5 & 2.0 & 2.7 & 32.1 & 21.1 & 22.7 &  75.9 & 138-8\\
{\footnotesize  $A_{1h}^k$=$0$ $A_{DM}$$=$$0$}    & 6.1  & 14.9 & 4.5   & 6.6 & 2.6 & 6.2   & 7.4 &  6.3 & 2.8 & 19.5 & 5.4 & 6.8   & 1.5 & 2.0 & 2.6 & 41.1 & 31.2 & 23.0 &  95.3 & 138-3\\
\hline
\end{tabular}}
\end{center}
\end{table*}

Let us now discuss in more details the implications for the DM component.
From the CCF plot in Fig.~\ref{fig:CCFdata} we see that  DM provides a significant contribution to the fit.
Yet, the posterior probability  of $A_{DM}$  in the 
bottom right panel of Fig.~\ref{fig:tri_astrodm} does not  provide a clear indication for a 
DM component. As discussed above, this is an indication that a DM signal may indeed be there but is
degenerate with some other component, in particular the mAGN one. In practice, the present datasets and 
our cross-correlation analysis cannot distinguish between the case of a large DM contribution with sub-dominant 
mAGN signal and that of a mAGN signal that dominates over the DM contribution.
In fact, the situation is further complicated by the aforementioned degeneracy between
mAGN and the 1-halo-correction terms.
Before discussing the degeneracy issue more in detail, it is worth pointing out that {\it i)} our results are 
robust to the choice of the fitting strategy, as shown in the right panel of Fig. \ref{fig:DMtree} and in the bottom panel of Fig.~\ref{fig:Apost},
and that  {\it ii)} despite the DM vs. mAGN degeneracy we are able to set constraints on the annihilation cross-section, as shown in the
left panel of Fig. \ref{fig:DMtree}, able to exclude the thermal value at $2\sigma$ 
for DM masses up few tens of GeV (in the \low\ substructure scheme) that, again, are robust against 
the adopted fitting scheme.
We also verified  the robustness of the DM constraints with respect to the choice of the priors for the astrophysical components.
Specifically, we considered the case of log-flat priors instead of a linear-flat ones, and  we found that the posteriors of the DM parameters are unaffected.
Some small variations are present in the constraints of the astrophysical parameters, which is expected since at the moment the significance 
of the measurement is still not very high, and in this regime some prior dependence is typically still present.

One thing to notice about the DM vs. mAGN degeneracy is that few mAGNs have been detected in \g-rays so far. As a consequence 
their model contribution to the IGRB and its anisotropies is rather uncertain. One key quantity is the relation between the 
\g-ray luminosity of these objects, $\mathcal{L}$ and the mass of their host halo $M$. Varying this relation within its
uncertainty range, which is rather large  (see e.g. \cite{Camera:2014rja}), modifies the predicted  cross-correlation signal.
Fig.~\ref{fig:ps1h} illustrates this point. In the right panel we show the cross-power spectrum mAGN-2MASS galaxies
(solid line) and how it changes when the $M(\mathcal{L})$ relation is varied within its uncertainty band (dashed curves).
Considering halo masses in the lower bound of the uncertainty strip significantly decreases the amplitude of the 1-halo term 
and reduces the amplitude of the PS on Mpc scales. As a result the PS contributed by mAGNs will be very similar to that
contributed by SFG, as shown in the left panel of Fig.~\ref{fig:ps1h}, and since their window functions are also very similar (see Fig.~\ref{fig:kernel}),
their contributions to the cross-power become fully degenerate.

We are therefore entitled to consider a scenario in which, due to this degeneracy, we set 
the mAGN contribution  equal to zero and assume that it is absorbed by the 
SFG one. To explore this situation we consider four additional fitting schemes.
In all of them we ignore the first data-points of the NVSS dataset (i.e. we use the NVSS-6 scheme)
and set $A_{mAGN}=0$. The four schemes are obtained from all possible combination of 
 $A_{DM}$ and $A_{1h}^k$ that are either set equal to zero or let free to vary. 
The four combinations are explicitly shown  in the first column of 
Table~\ref{tab:chi2noMAGN}  in which we summarize the results of the  $\chi^2$ analysis.

The inclusion of the 1-halo-correction terms improves the fit appreciably,
although the improvement is mainly driven by the LRG and QSO datasets.
The inclusion of DM with two extra-parameters 
also improves the fit decreasing the best fit $\chi^2$ by 4.3
and 5.8 for the fits with and without 1-halo-correction terms, respectively,
with improvement mainly coming from a better fit to the 2MASS data.
No scheme provides a good fit to the CCF with the QSO for $E>$500 MeV.   
This is possibly an indication of an imperfect modeling of the energy spectrum in the QSO correlation.

Fig.~\ref{fig:trianglenoMAGN} shows the triangle plot for the case $A_{DM}\neq$0 and $A_{1h}^k=$0.
When the mAGN contribution is suppressed ($A_{mAGN}=$0) a non-vanishing DM component provides quite a good fit, 
with about a $2\sigma$ deviation from zero.
Furthermore, the best fit $\chi^2$ values in Table~\ref{tab:chi2noMAGN}
are very similar to those in Table~\ref{tab:chi2} in which the mAGN component was 
included in the model (71.6 vs 72.5 and 89.5 vs 85.8 for the case with and without 1-halo-correction terms, respectively),
 a fact that corroborates the evidence for a degeneracy between DM and mAGNs.
Fig.~\ref{fig:trianglenoMAGNonlyDM} illustrates the robustness of these results to 
the inclusion of the 1-halo-correction term, $A_{1h}^k\neq$0, for two different final annihilation states:
$b\bar b$ (left set of plots) and  and $\tau^+\tau^-$ (right plots).
In all cases the best fits preference is for the presence of a DM component, even when the 
extra degree of freedom $A_{1h}^k$ is included. 
Furthermore, these plots reveal a degeneracy between $A_{DM}$ and $m_{DM}$ which is to be expected
since, as can be seen in Eq.~(\ref{eqn:window_annihilating_DM}),  the WIMP signal approximately scales with $A_{DM}/m_{DM}$. 
Indeed, $W_{\delta^2}$ contains a factor $\sv/\mdm^2$ plus an integral over the energy which introduces a contribution roughly proportional to $m_{DM}$ (being the spectrum integrated up its endpoint, which is  $m_{DM}$).
Inspection of Fig.~\ref{fig:trianglenoMAGN} also shows more clearly the 
degeneracy between DM and the SFG components, left from the mAGN-DM-SFG degeneracy
after removing the mAGN  component.
 One consequence of this is that performing a fit excluding either
 the SFG or the  mAGN component would 
 enhance the strength of the DM signal, without affecting appreciably the value of the best fit $\chi^2$. 

In conclusion, our analysis indicates that a significant DM contribution 
 to cross-correlation is entirely plausible. However, the degeneracy with other astrophysical sources, namely mAGNs and SFGs,
 largely originating from the current observational uncertainties, prevents us from drawing a definitive conclusion.
Future analyses with increased \g-ray statistics and improved angular resolution in which the cross correlation is extended 
to other catalogues of extragalactic objects will help to break the present degeneracies and to pinpoint the correct scenario.

\begin{figure*}[t]
\centering
\includegraphics[width=0.9\textwidth]{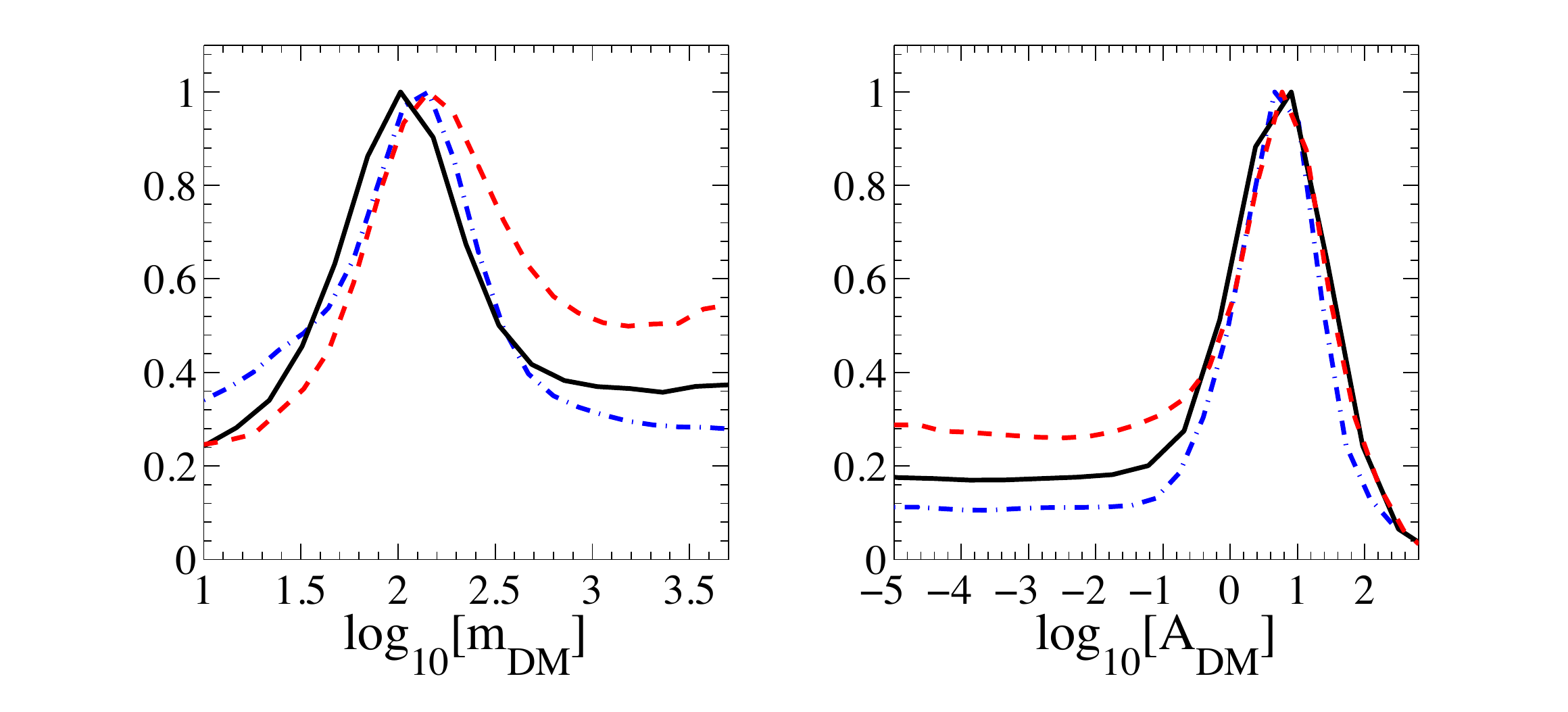}\\
\caption{ Comparison of the posteriors distributions for the DM parameters (mass $m_{\rm DM} $ (left) and annihilation rate in terms of the thermal one $A_{\rm DM}$  (right)) for the  NVSS-6, $A_{1h}^k=0$, $A_{mAGN}=0$ fit and for the 3 different SFG models described in the text.} 
\label{fig:compareXianoMAGN}
\end{figure*}

As a final remark, we also mention that, as a cross-check, we have performed the analysis employing the astrophysical models adopted in  \cite{Xia:2014}. 
The constraints on the \g-ray astrophysical contributions are different, which is expected given the different modeling. 
Regarding DM,  using the  NVSS-6, $A_{mAGN}=0$, $A_{1h}^k=0$ fit configuration,
which is the closest to the one used in \cite{Xia:2014}, 
we compare  in Fig. \ref{fig:compareXianoMAGN}  the DM posteriors  from 3 different fits using 3 different  SFG  models: the one adopted in this work (black solid curve),
the SFG1 model from   \cite{Xia:2014} (red dashed curve) and   a modified version of SFG1 (blue dot-dashed curve) with 
redshift-dependent bias equal to the the bias of the present model
(while the original SFG1 model has bias equal to 1 for all $z$).  
The SFG2 model of  \cite{Xia:2014}  is very similar to present SFG model and is not considered. 
The plot indeed shows that the DM results are not significantly dependent from the SFG model adopted.

\begin{figure*}[t]
\centering
\includegraphics[width=0.49\textwidth]{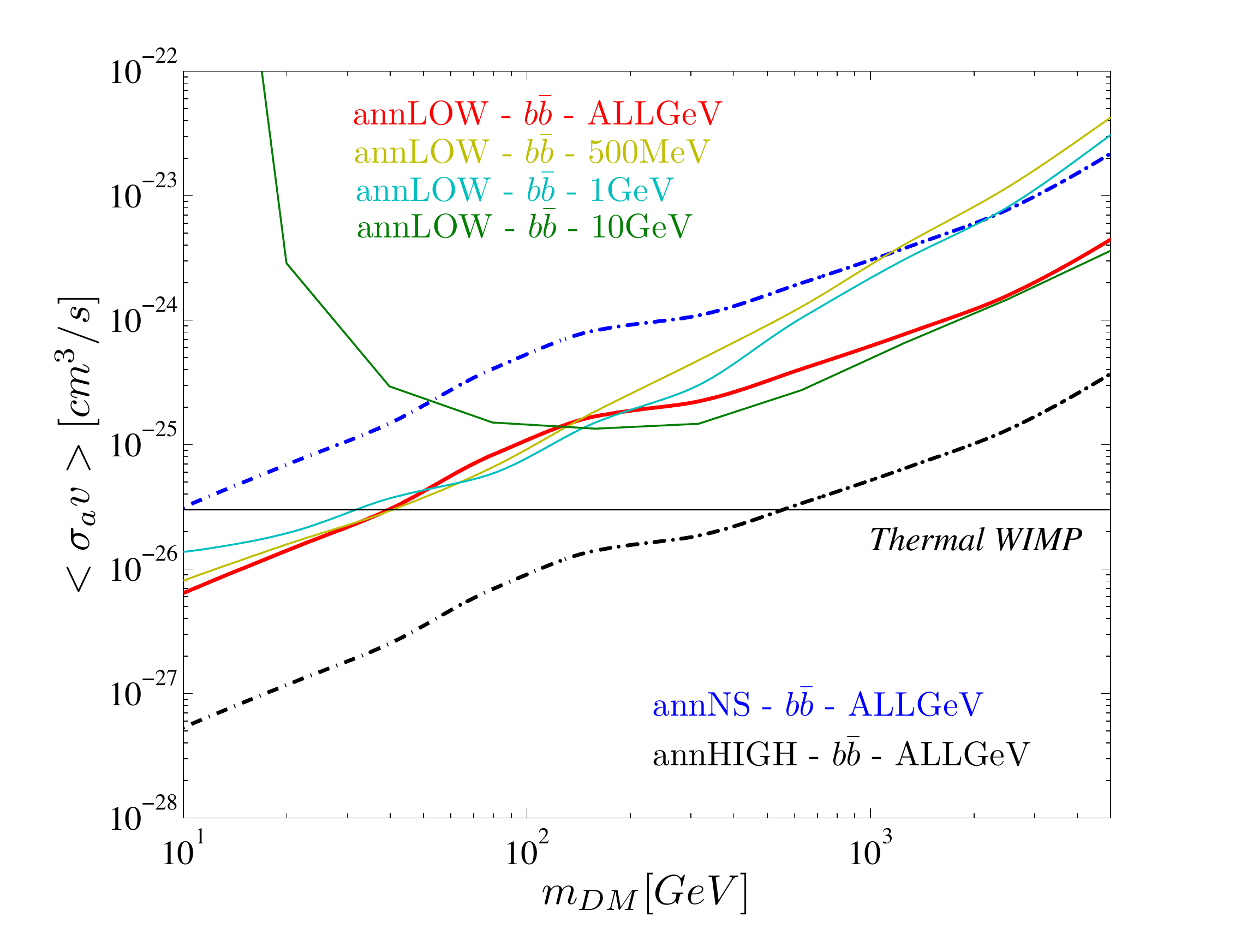}
\includegraphics[width=0.49\textwidth]{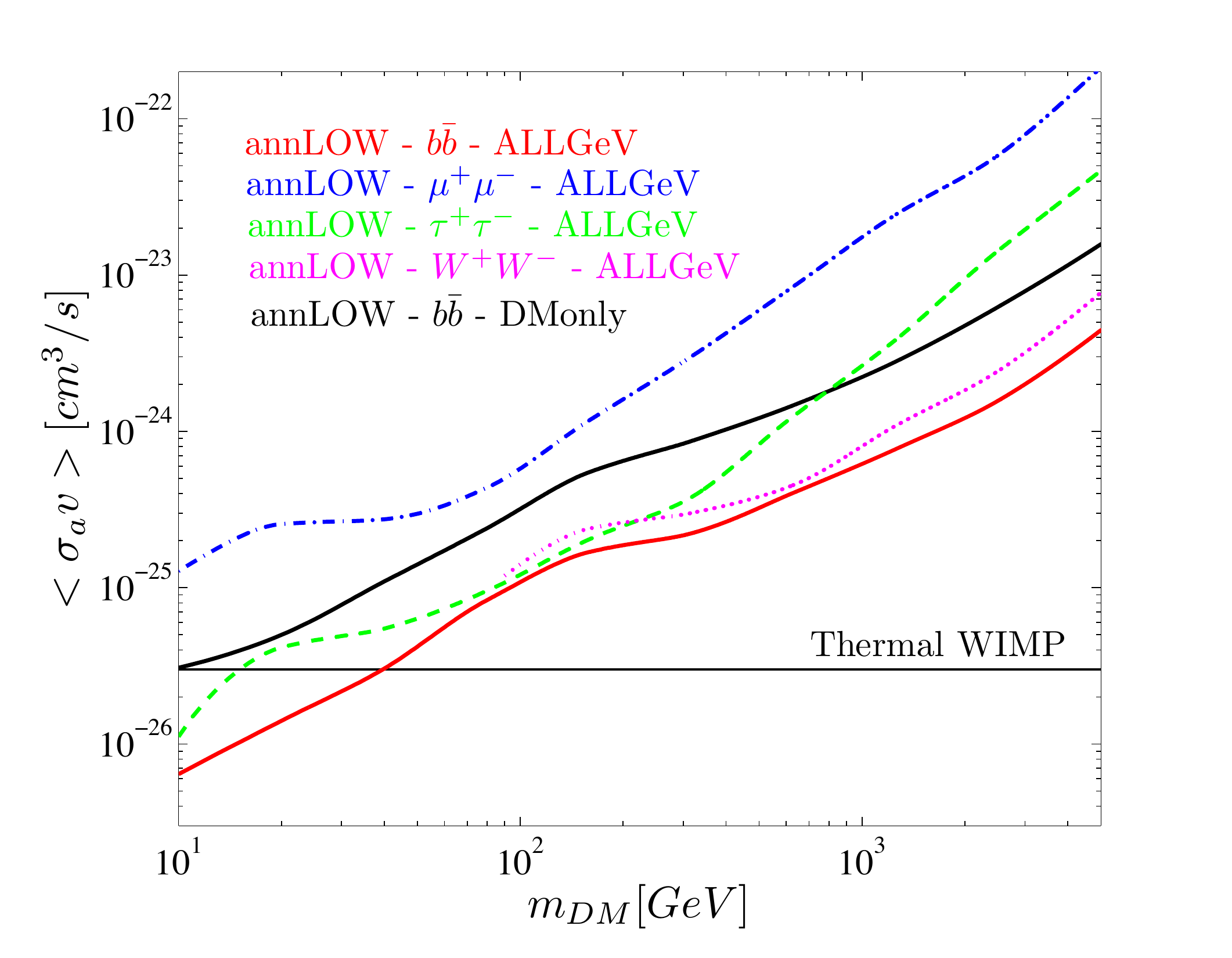} 
\caption{{\sl Left}: 95\%  upper bounds on the DM annihilation rate $\sigmav$ as a function of the DM mass, for the \low\ substructures model and 
the reference  NVSS-10 $A_{1h}^k\neq$0 fit.  
Solid lines refer to the $b \bar b$ annihilation channel: the red line refers to the analysis that combines information from all the
three energy bins under consideration ($E>0.5, 1, 10$ GeV), while the other three lines refer to the analysis performed on a single energy bin (as stated in the figure label). 
The upper dot-dashed blue line refers to the \ns\  substructure model, while the lower dot-dashed black line to the \high\ substructure model.
{\sl Right}: in addition to the $b \bar b$ case (red line) reported in the left panel, the different lines show the upper bounds for the $\mu^+\mu^-$ (blue), $\tau^+\tau^-$ (green) and $W^+W^-$ (magenta)  annihilation channels, for the  \low\ sub-structures model. 
The black line instead shows the upper bound for the $b \bar b$ case and \low\ substructure scheme, 
obtained under the assumption that the DM contribution to the 2MASS cross-correlation is the dominant one (taken from  \cite{Regis:2015zka}).
}
\label{fig:DMbounds}
\end{figure*}

\begin{figure}[t]
\centering
\includegraphics[width=0.49\textwidth]{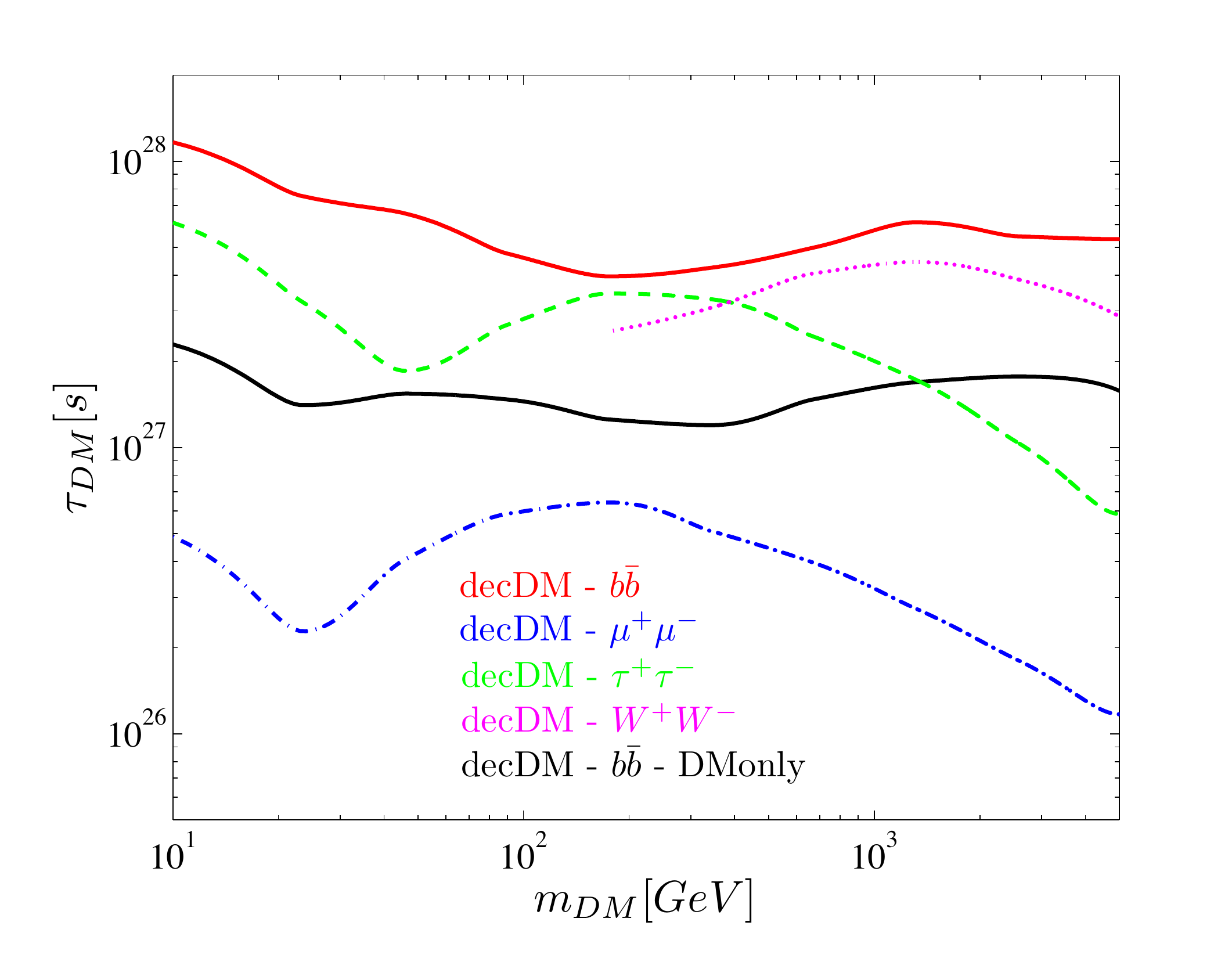}
\caption{For a decaying DM, 95\% lower limits on the DM lifetime $\tau$ as a function of its mass, for different decay channels: $b \bar b$ (red), $\mu^+\mu^-$ (blue), $\tau^+\tau^-$ (green) and $W^+W^-$ (magenta). The black line instead shows the lower bound for the $b \bar b$ case obtained under the assumption that the DM contribution to the 2MASS cross-correlation is the dominant one (taken from  \cite{Regis:2015zka})
}
\label{fig:DMdec}
\end{figure}

With no unambiguous indication for a DM components we can nevertheless set constraint on the properties of the DM candidates.
To this purpose we perform, for any given mass of the DM particle candidate, an individual 10 parameter fit 
and set the 95\%  bound on $\sv$ from the posterior distribution.
The results for the annihilating DM are summarised in  Fig.~\ref{fig:DMbounds}. In the left panel we focus on the  $b \bar b$ annihilation channel.
The solid line with different colours refer to constraints obtained from each of  the three energy band separately ($E>0.5, 1, 10$ GeV) as well as the 
ones obtained by their combination (red line). All theee results refer to the \low\ substructures model and are obtained with 
the reference  NVSS-10 $A_{1h}^k\neq$0 fitting scheme. 
The upper and lower dot-dashed curves show how the bounds change
in the the  \ns\ and  \high\ substructure model, respectively.

In the right panel we compare the results of different 
final annihilation channels  ($\mu^+\mu^-$, $\tau^+\tau^-$ and $W^+W^-$) to the original 
$b \bar b$ case (red curve) 
shown in the left plot, for the  \low\ substructures scenario and combining all energy bands. 
All the results refer to the benchmark NVSS-10 $A_{1h}^k\neq$0  case, but the other fitting schemes provide nearly indistinguishable constraints. 
The black curve is taken from \cite{Regis:2015zka} and refers to the case in which we assumed that 
all the 2MASS \g-ray correlation is produced by DM, with no
astrophysical contribution. 
As expected, including the astrophysical sources makes the constraints  stronger, of about a factor of 4. 
The gain is significant and will further improve once the DM-mAGN-SFG degeneracies discussed above
will be removed.

As expected, uncertainties on the bounds driven by the substructure model are significant.
The left 
panel of  Fig.~\ref{fig:DMbounds} shows that assuming the \high\  model would strengthen the constraints on the cross section by about 
one order of magnitude, whereas in  the \ns\ scenario, the bounds would weaken by about a factor of 5.
This implies that the thermal annihilation rate $\sv=3\cdot10^{-26} {\rm cm^3 s^{-1}}$
is excluded at the 95 \% level  up to masses of  6, 25, 250 GeV in the \ns, \low\ and \high\ scenarios, respectively. 

In Fig.~\ref{fig:DMdec} we instead show the 95\% lower bounds on  the lifetime of a decaying DM particle, for various decay final states. Bounds on DM decay, being proportional to the DM density (and not DM density squared, as instead the annihilation signal) depend on the total DM mass in structures and are not affected by the different substructure modeling. As for the annihilation case, including the astrophysical sources in the analysis 
 improves the constraints, again by about a factor of 4, with respect to those obtained by ignoring the astrophysical components~\citep{Regis:2015zka}.

Finally, to test the robustness of our DM constraints we have repeated the analysis using the same astrophysical models used in  \cite{Xia:2014}
and we found that they are  very similar to the ones obtained in the present analysis.

\begin{figure*}[t]
\centering
\includegraphics[width=0.9\textwidth]{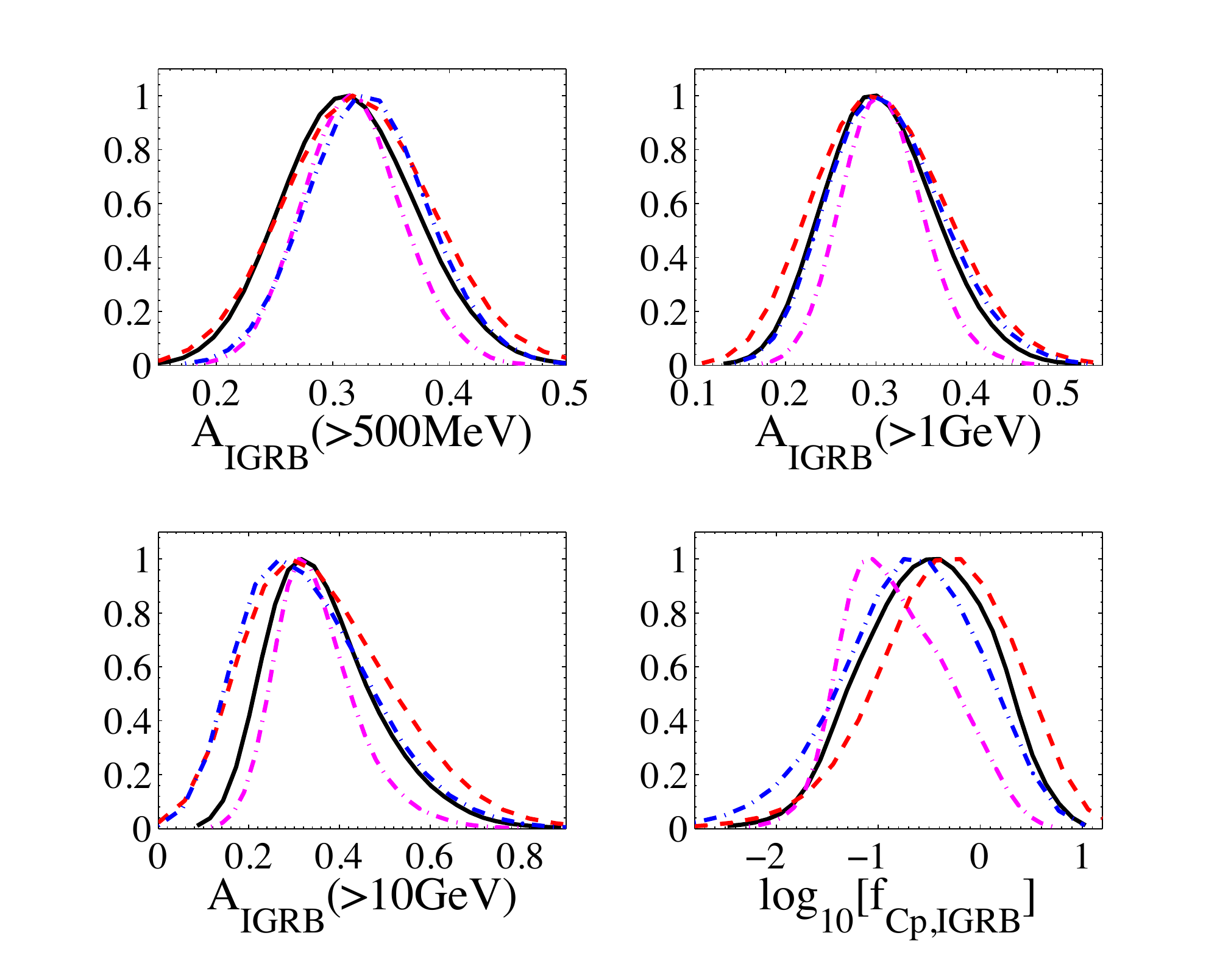}
\caption{Marginalized 1D posterior probabilities for the cumulative fractional contribution $A_{IGRB}$ 
of all \g-ray sources (BL Lac, mAGN, SFG, FSRQ and DM) to the total intensity $I_{IGRB}$ measured by \Fermi.  
$A_{IGRB}$ is expressed in terms of  $I_{IGRB} = 10^{-6},\,4\times 10^{-7},\,1.5\times 10^{-8}\,{\rm cm^{-2}\,s^{-1}\,sr^{-1}}$ 
for the energy bins $E>0.5,\,1,\,10$ GeV (to account for spectral behaviour). 
The bottom right panel show the same information for the IGRB angular auto-correlation
in the 1-2 GeV energy band.
The various lines refer to the four fits described in the text with same color and dashing conventions of Fig. \ref{fig:Apost}. 
}
\label{fig:A_EGB}
\end{figure*}

\subsection{Self consistency tests: mean intensity and auto-correlation of the IGRB }
\label{sanitycheck}

As anticipated in Section~\ref{sec:analysis} instead of including the mean IRGB intensity and 
its auto-correlation in the fit,  we use these additional observational inputs {\it a posteriori} 
as a self-consistent test for our best fitting model.

We define, $A_{IGRB}^n$, the fractional mean IGRB intensity predicted by the cross-correlation fit,
 as follows
\begin{eqnarray}\nonumber
{I_{TOT}^n}A_{IGRB}^n &=& A_{FSRQ}{I_{FSRQ}^n} + A_{BLLac}{I_{BLLac}^n} + \\
        && A_{mAGN}{I_{mAGN}^n} + A_{SFG} {I_{SFG}^n} + A_{DM}{I_{DM}^n}\;,
\label{eq:AEGB}
\end{eqnarray}
where $I_\alpha^n$ are the integrated \g-ray intensities of our {\it reference} models 
for the five \g-ray emitters considered here and shown in Fig.~\ref{fig:bench} and 
 $n=1,2,3$ identifies the energy band
The total intensity is defined as  $I_{TOT}^n\equiv\sum_\alpha I_\alpha^n$, where the sum runs over the five types of emitters.
In our model  $I_{TOT}^n$= $10^{-6},\,4\times 10^{-7},\,1.5\times 10^{-8}\,{\rm cm^{-2}\,s^{-1}\,sr^{-1}}$ for the energy ranges $E>0.5,\,1,\,10$ GeV, 
respectively, which are consistent with the {\it measured} IGRB~\citep{2015IGRB}.
We thus expect that the $A_{IGRB}^n$ have values close to unity 
to match observations.
Note that the parameters $A_{IGRB}^n$ need not to be  the same in each energy band since the total signal
$I_{TOT}^n$ and the individual contributions $I_\alpha^n$ have different scaling in energy. However,  the difference is not very large.

Similarly, we define the IGRB auto-correlation predicted from the cross-correlation fit, as a fraction  of the measured one,
in terms of the   parameter $f_{C_{P,IGRB}}$ as:
\begin{eqnarray}\nonumber
{C_{P,TOT}} f_{C_{P,IGRB}} & = & A_{FSRQ}^2{C_{P,FSRQ}} + A_{BLLac}^2{C_{P,BLLac}} + \\
     && A_{mAGN}^2{C_{P,mAGN}} + A_{SFG}^2 {C_{P,SFG}}\;,
\label{eq:CpEGB}
\end{eqnarray}
where $C_{P,FSRQ}=1.6\times 10^{-18}$, $C_{P,BLLac}=7.9\times 10^{-18}$, $C_{P,mAGN}=3.9\times 10^{-19}$
and $C_{P,SFG}=6.3\times 10^{-21}$, all of them   in units of  (cm$^{-2}$s$^{-1}$sr$^{-1})^2$sr.
are the predicted average auto-correlation signals  in the multipole range $\ell= 155-504$ and  in the energy band $1$-$2$ GeV.
We have neglected the DM contribution since it is largely subdominant with
respect to FSRQs, BLLacs and mAGNs (see Fig. \ref{fig:bench}). The SFG contribution, which is also subdominant, is considered for the sake of completeness. 
In the above equation we made the assumption that the amplitude of the auto-correlation signal 
scales with the square of the normalization parameters of the individual components. 
Unlike  the cross-correlation case we did not include 
 any 1-halo-correction term since we model astrophysical emitters as point sources 
 for which no additional small-scale power  is expected to contribute to the auto-correlation signal.
Like the mean intensity, the value $f_{C_{P,IGRB}}=1$ characterizes  a  model which saturates the measured IGRB auto-correlation.

In Fig.~\ref{fig:A_EGB} we show the posterior probabilities for  $A_{IGRB}^n$ in the three energy bands considered in our analysis, and  $f_{C_{P,IGRB}}$.
We find that the typical value $A_{IGRB}$ is between 20\% and 50\% in the two lower energy bands (upper panels)
whereas for $E>10$ GeV is in the broader range  10\% -- 80\%.
These results are robust to the details of the fitting procedure, as demonstrated by the similarity of the various curves.
They imply that the extragalactic sources considered in our model (BL Lac, mAGN, SFG, FSRQ and DM) which,
 as we have seen, provide a good match to the observed {\sl cross correlation} with LSS tracers,
also contribute to a significant fraction of the IGRB, although possibly not to the whole signal. 
This result is interesting but should also be taken with a grain of salt given the complexity of our
cross-correlation model.
For example \cite{Xia:2014}, using a different model for the SFG emission and bias, was able to account for a larger
fraction of the IGRB,  although again not 100\%.
It should be also noted that the measurement of the IGRB in \cite{2015IGRB} is affected
by systematic errors induced by the imperfect model of the foreground Galactic emission, even if the 
size of this systematic uncertainty does not seem to be large enough to saturate our models to 100\% of the total emission.
If indeed it turns out that additional \g-ray sources are required to explain the total intensity of the IGRB, 
then the results of our analysis set a rather sever constraint: their correlation with LSS tracers must be weak.
This would imply that they should be local, possibly of Galactic origin, like the millisecond pulsar or,
perhaps, diffuse inverse Compton photons from cosmic-ray electrons scattering on the optical/infrared Galactic inter-stellar radiation field.
Future analyses with newer and additional datasets will help to clarify this interesting issue.

The posterior for $f_{C_{P,IGRB}}$ is instead consistent with unity,
although its probability distribution actually spans several orders of magnitude from $10^{-2}$ to 10, meaning that
the measured auto-correlation does not provide a very stringent cross-check.

\section{Summary and Conclusions}
\label{sec:concl}

In this paper we have used the cross-correlations recently measured in \cite{Xia:2014} 
between \Fermi-LAT diffuse \g-ray  maps  and different catalogs of 
LSS-tracers  to investigate the origin of the IGRB and the nature of the various sources that may 
contribute to it, including DM annihilation or decay. This work extends that of   \cite{Regis:2015zka} which  
used only the \g-ray-2MASS correlation and considered DM as the only source of the  extragalactic \g-ray signal.
Our main results are as follows:

\begin{itemize}

\item Our theoretical models provide a good fit to the cross-correlation measured in all employed catalogs of extragalactic tracers, namely 
SDSS-DR6 quasars, 2MASS, NVSS, SDSS-DR8 LRGs and SDSS-DR8 MG. 
The quality of the fit is quantified by means of a $\chi^2$ analysis in which  we account for covariance among the errors in different 
angular bins whereas we ignore the covariance among energy bins and among the different catalogs. 
The first approximation is justified by the photon statistics, which is dominated by low energy event, making each of the energy bins 
considered in our analysis effectively independent. The second approximation is justified by the spatial distributions of the objects 
in the different catalogs that, with the partial exception of the NVSS one, do not significantly overlap with each other.

\item In our cross-correlation function (CCF) models we consider four different types of astrophysical sources (two flavours of blazars, FSRQs and BL Lacs, SFGs and mAGNs)
and, in addition, annihilating/decaying DM. 
The rationale behind the choice of these astrophysical sources is that previous analyses have shown that they are the main
 contributors to the IGRB and its  angular
auto-correlation. 
These two observational constraints are not considered in our fit. Instead, we use them {\it a posteriori} to check the consistency of our 
best fitting models which are based solely on the measured CCF. 
We find that models that provide a good match to the cross-correlation fall short of 
accounting for the mean \g-ray intensity. 
The discrepancy is not large, less than a factor of two, especially in the high energy band, and could 
be accounted for by a combination of model uncertainty and imperfect subtraction of the Galactic foreground. 
However, it may also indicate
that additional types of sources that do not cross-correlate with the LSS, like  \g-ray sources within our Galaxy, are 
required to account for the whole IGRB intensity. 

\item Including DM among the possible IGRB sources does not significantly improve the quality of the fit,
and does not indicate a preference for a particular DM mass or annihilation cross-section/decay rate.
We find that the reason for the low statistical significance on the presence  of a DM component does not
lie in the fact that the fit rejects this component, while it is rather 
due to the presence of  a model degeneracy with other types of astrophysical sources, mainly
 mAGNs and SFGs. 
 In other words, a significant DM contribution gives an equally good fit as a case with
  a negligible DM contribution and a larger mAGNs and SFGs emission.
Neglecting the mAGN component  in the fit partially breaks this degeneracy and provides a small ($\sim 2 \sigma$)
preference for DM. 
The best fit is found for a rather canonical WIMP DM candidate with $m_{DM}\sim 100$ GeV that annihilates
into $b\bar b$ at a rate which is of the order of the  thermal value for the benchmark \low\ DM clustering scenario considered.
 A candidate with a slight smaller mass
of about 30 GeV that annihilates into $\tau^+ \, \tau^-$ provides an equally good fit. 

\item Breaking this degeneracy is the main goal of future cross-correlation analyses similar to the present one. Fortunately, this is a realistic goal.
 One of the main reason for this degeneracy is the uncertainty on the  mAGN and SFG luminosity in \g-ray which, to date, has been directly measured for a handful
 of very nearby objects. However, their number is bound to increase thanks to the fact that \Fermi-LAT will keep taking data in the next few years. In addition, the 
 quality of the Fermi maps is also expected to increase both in terms of photon statistics, which will allow to better sample the energy behaviour and to improve the 
 sensitivity to characteristic DM spectral features, and angular resolution, which would allow us to push the correlation analysis to smaller angular scales where the 
 1-halo term dominates.


\item  We turn the non-detection of DM into limits on the annihilation cross-section/decay rate as function of the DM mass.
   Our derived constraints are comparable in strenght 
 to most of the current indirect detection method that exploits the $\gamma$-ray sky~\citep{Regis:2015zka}.
These constraints are rather robust to the astrophysical details of the models but, as expected, do depend on the detail of the 
DM substructure and small-scale clustering. 
For this reason and with the aim of bracketing current theoretical uncertainties, in addition to the \low\ scenario 
which represents the current, somewhat conservative, benchmark  substructure model, 
we have explored two additional, rather extreme cases: the \ns\ case in which 
we completely ignore substructures and that provides extremely conservative constraints of the DM properties, 
and the \high\ scenario in which substructures are 
more numerous and have an higher density concentration. 
In the most conservative \ns\ scenario our method excludes, at a credible level larger than 95\%, 
that DM particles with masses smaller than 10 GeV annihilating entirely into $b\bar{b}$ could have 
a thermal cross section. 
In the optimistic scenario, the same statement applies to particles lighter than  $\sim 600$ GeV.
The bounds are a factor of $\sim 4$ stronger than the most conservative case considered in  \cite{Regis:2015zka}  in which 
only DM is used in order to saturate the 2MASS cross-correlation.
Constraints on DM decay time for DM decaying into $b\bar{b}$ are $\sim 10^{28}$ s,
roughly independently  from the DM mass.

\end{itemize}

All in all we are confident that the results obtained in our analysis, which are already quite remarkable 
considering that this is the first time that a genuine cross-correlation signal 
is detected in the \Fermi-LAT \g-ray maps, will soon improve significantly. 
In this respect this work also represents a proof of concept that illustrates the potential of the cross-correlation analysis.
We base our optimism on the fact that the new  {\sc Pass8} data, with improved effective area and angular resolution,  
will soon be released by the \Fermi-LAT Collaboration and that
additional catalogs of objects al relatively low redshifts with wide, almost all-sky, 
angular coverage and well determined redshift distribution are already available \citep{Bilicki14}
and some new ones are being compiled \citep{Bilicki15}.

\section*{Acknowledgements}
This work is supported by the research grant {\sl Theoretical Astroparticle Physics} number 2012CPPYP7 under the program PRIN 2012 funded by the Ministero dell'Istruzione, Universit\`a e della Ricerca (MIUR), by the research grants {\sl TAsP (Theoretical Astroparticle Physics)} and {\sl Fermi} funded by the Istituto Nazionale di Fisica Nucleare (INFN), and by the {\sl Strategic Research Grant: Origin and Detection of Galactic and Extragalactic Cosmic Rays} funded by Torino University and Compagnia di San Paolo.
MV and EB are supported by PRIN MIUR and IS PD51 INDARK grants. MV is also supported by ERC-StG cosmoIGM, PRIN INAF
JX is supported by the National Youth Thousand Talents Program, the National Science Foundation of China under Grant No. 11422323, and the Strategic Priority Research Program, {\sl The Emergence of Cosmological Structures} of the Chinese Academy of Sciences, Grant No. XDB09000000.


\appendix

\section{Window Functions}
\label{sec:windows}

In this Appendix we discuss the modeling of the window functions adopted for the calculation of the cross-correlation angular power spectrum $C^{\gamma g}$ of Eq. (\ref{eq:clfin}), which are in turn the ingredient for the determination of the cross correlation function $CCF^{\gamma g}(\theta)$ defined in Eq. (\ref{eq:2point}).

\subsection{Dark matter}
\subsubsection{Annihilating dark matter}
\label{sec:annihilating_DM}

DM annihilations in haloes and in their substructures produce \g-ray photons.
This emission traces the DM density squared $\rho^2_{\rm DM}$: therefore the density field
responsible for the correlation signal is
$f_{\delta^2}(\chi,{\mathbf r}) = \rho^2_{\rm DM}(\chi,{\mathbf r})$.
The window function reads:
\begin{equation}
W_{\delta^2}(\chi) = \frac{(\odm \rho_c)^2}{4\pi} 
\frac{\sv}{2\mdm^2} \left[1+z(\chi)\right]^3 
\Delta^2(\chi) \int_{E_\gamma>E_{\rm min}} \de E_\gamma \, \frac{\de N_a}{\de E_\gamma} 
\left[E_\gamma(\chi) \right] e^{-\tau\left[\chi,E_\gamma(\chi)\right]},
\label{eqn:window_annihilating_DM}
\end{equation}
where $\odm$ is the cosmological abundance of DM, $\rho_c$ is the critical
density of the Universe, $\mdm$ is the mass of the DM particle, and $\sv$ denotes the velocity-averaged annihilation rate, assumed here to be the same in all haloes. 
$\de N_a / \de E_\gamma$ indicates the number of
photons produced per annihilation event, and sets the \g-ray energy spectrum. 
We will consider annihilation into $b\bar{b}$ quarks 
as representative of a typical soft annihilation spectrum (with \g-rays mostly arising from production and decay of neutral pions), and into $\mu^+\mu^-$ leptons
as representative of a hard-spectrum channel (where \g-rays mostly arising from final state radiation), with $\tau^+\tau^-$ and $W^+W^-$ final states as intermediate possibilities. 
$E_{\rm min}$ is the energy threshold of the Fermi-LAT maps considered in the analysis, namely: $E_{\rm min}=0.5,\,1\,,10$ GeV.
The factor $\exp\{-\tau[\chi,E_\gamma(\chi)]\}$ accounts for absorption due to the extra-galactic background light, and we model the optical depth $\tau$ as in \cite{Franceschini:2008tp}. 

A crucial quantity in Eq.~\eqref{eqn:window_annihilating_DM} is the so-called 
clumping factor $\Delta^2(\chi)$:
\begin{equation}
\Delta^2(z) \equiv 
\frac{\langle \rho^2_{\rm DM} \rangle}{{\bar \rho}^2_{\rm DM}} =
\int_{M_{\rm min}}^{M_{\rm max}} \de M \frac{\de n}{\de M}(M,z) \,\left[1+b_{\rm sub}(M,z)\right]
\int \de^3 \mathbf{x} \, 
\frac{\rho^2_h({\mathbf{x}|M,z)}}{{\bar \rho}^2_{\rm DM}}.
\label{eqn:clumping}
\end{equation}
The clumping factor involves the integral of the halo number density $\de n/\de M$ above the so-called minimal halo mass $M_{\rm min}$, multiplied by the total number of annihilations produced in the generic haloes of mass $M$ at redshift $z$ with density profile $\rho_h(\mathbf{x}|M,\chi)$ and with subhalos providing a ``boost'' to the emission given by $b_{\rm sub}$. 
We assume a reference value of $10^{-6} M_\odot$ for $M_{\rm min}$, which corresponds to a typical free-streaming mass in the WIMP DM scenario.
We adopt the halo mass function from \cite{Sheth:1999mn} and we assume that the halos are characterized by the so-called Navarro-Frenk-White (NFW) universal density profile \citep{Navarro:1996gj}. The profile is completely determined by the total mass of the halo and by its size. We express the latter in terms of the concentration parameter $c(M,z)$, taken from \cite{Prada:2011jf} (see also \cite{Sanchez-Conde:2013yxa} for an analytic fit of $c(M,z=0)$ of \cite{Prada:2011jf}).

Concerning the boost provided by subhaloes hosted in the main haloes, we consider three scenarios (\high, \low, and \ns) as extreme cases bracketing the effect. 
In Eq.~(\ref{eqn:clumping}) this is indeed the most uncertain quantity~\citep{Sanchez-Conde:2013yxa,Gao:2011rf,Fornasa:2012gu,Ng:2013xha}.
In the \low\ scenario, the function $b_{\rm sub}(M,z)$ is computed following \cite{Sanchez-Conde:2013yxa} (see, in particular, their Eq.~(2) assuming a subhalo mass function $dn/dM_{\rm sub}\propto M_{\rm sub}^{-2}$). The \high\ scenario stems instead from the $b_{\rm sub}(M,z)$ found in \cite{Gao:2011rf}, and assuming no redshift dependence. 
In the \ns\ case, we simply set $b_{\rm sub}=0$: this can be considered as the most conservative approach.

\begin{figure*}[t]
\vspace{-3cm}
\centering
\includegraphics[width=0.49\textwidth]{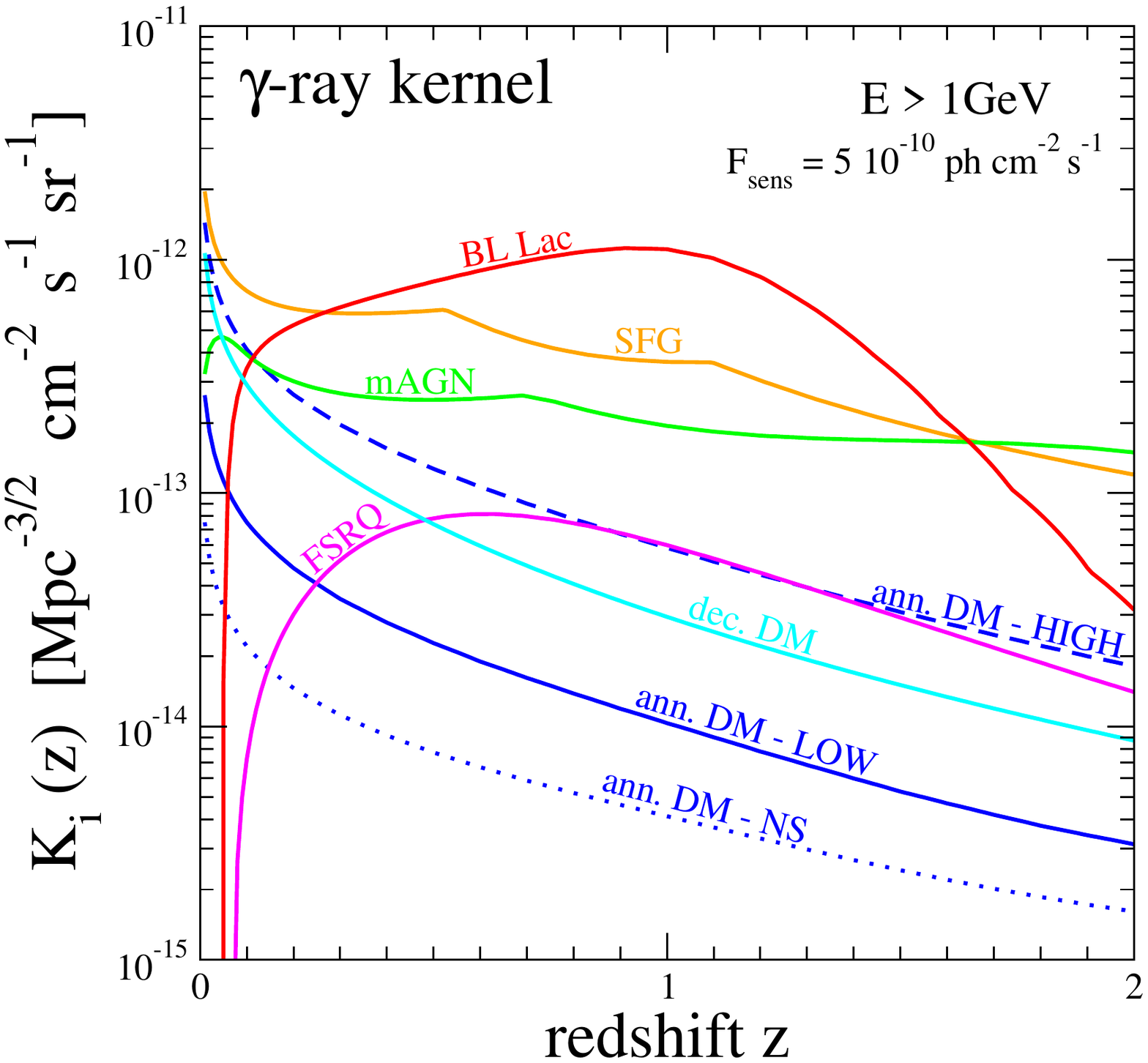}
\includegraphics[width=0.49\textwidth]{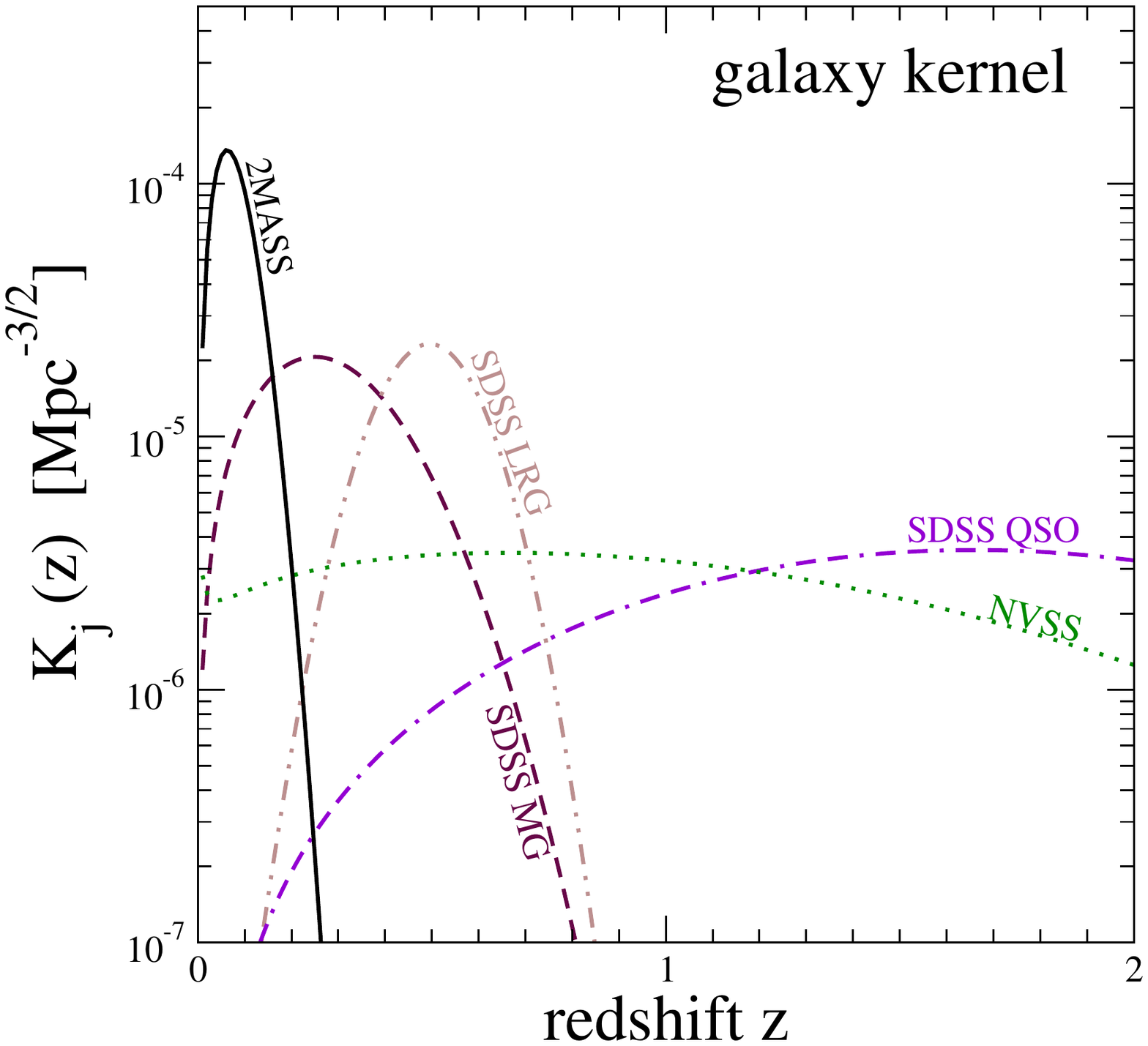}
\caption{Angular-power-spectrum kernels $K(z)$ of \g-ray emitters ({\sl left}) and galaxies ({\sl right}), shown as a function of the redshift $z$. The kernel is defined as $K_a(z)=\sqrt{c\,z/H(z)} W_a(z)/\chi(z)$, where $H(z)$ is the Hubble parameter, $W_a(z)$ denotes the window function of the objects of class $a$ and $\chi$ is the comoving distance. The \g-rays kernels are integrated for energies above 1 GeV and refer to unresolved sources fainter than $F_{\rm sens} = 5 \cdot 10^{-10}$ photons cm$^{-2}$ s$^{-1}$.
}
\label{fig:kernel}
\end{figure*}

The blue curves in Fig.~\ref{fig:kernel}a show a quantity related to the window function of Eq.~\eqref{eqn:window_annihilating_DM}, defined as the kernel of the \g-ray emission entering in the computation of the angular power spectrum discussed in Section~\ref{sec:formalism}. Specifically, we define the kernels as: 
\begin{equation}
K_a=\sqrt{c\,z/H(z)} W_a(z)/\chi(z)\, ,
\end{equation}
such that:
\begin{equation}
C_\ell^{(\gamma_i g_j)}=\int \de \ln(z) K_{\gamma_i}(z)\, K_{g_j}(z)\,P_{\gamma_i g_j}(k,z)\, .
\end{equation}
In Fig.~\ref{fig:kernel}a we choose a reference particle-physics model with $\mdm=100$ GeV, $\sv=3 \times 10^{-26}\,\mathrm{cm^3s^{-1}}$ and $b \bar{b}$ annihilation channel. The three clustering scenarios (\ns, \low\ and \high\ for dotted, solid and dashed curves, respectively) share approximately the same redshift dependence, but they correspond to different sizes of the clumping factor and consequently of the intensity of the DM-induced \g-ray flux. Note that a comparison with previous works in the literature can be non-trivial, as different groups employ different prescriptions for the ingredients of the DM clustering and, in particular, for the boost factor. Fig.~\ref{fig:kernel} can be useful also as a normalization test when confronting the results presented in the rest of the paper with other works.  

\subsubsection{Decaying dark matter}
\label{sec:decaying_DM}
If instead of being stable, the DM particles decay, while having a negligible self-annihilation rate, the produced \g-rays traces the DM density linearly, i.e., $f_{\delta}(\chi,{\mathbf r}) = \rho_{\rm DM}(\chi,{\mathbf r})$, The window function in this case reads:
\begin{equation}
W_\delta(\chi) = \frac{\odm \rho_c}{4\pi} 
\frac{\Gamma_{\rm d}}{\mdm} \int_{E_\gamma>E_{\rm min}} \de E_\gamma \, 
\frac{\de N_d}{\de E_\gamma} \left[E_\gamma(\chi) \right] 
e^{-\tau\left[\chi,E_\gamma(\chi)\right]}\;,
\label{eqn:window_decaying_DM}
\end{equation}
where $\Gamma_{\rm d}$ is the decay rate.
The photon yield, $\de N_d / \de E_\gamma$, is assumed to be the same as for annihilating DM, but with the energy of the process given by $\sqrt{s}=\mdm$ instead of $2\mdm$. In other words, $\de N_d / \de E_\gamma(E_\gamma) = \de N_a / \de E_\gamma(2E_\gamma)$ with the kinematic end-point being at $\mdm/2$. The kernel in the case of decaying DM is shown as a cyan curve in Fig.~\ref{fig:kernel}a. In the plot we report reference particle-physics model with $\mdm=200$ GeV, $\tau_{\rm d}=1/\Gamma_{\rm d} = 6 \times 10^{27}\,\mathrm{s}$ and decays into $b \bar{b}$ quarks. Note that for decaying DM, the window function does not depend on the details of the DM clustering. We notice also that DM kernels peak at low redshifts, both for annihilating and decaying DM, and have a relative fast decrease with distance.

\subsection{Astrophysical sources}
\label{sec:intastro}

For astrophysical sources, we adopt as the characterizing parameter the
source \g-ray luminosity $\mathcal L$ in the energy interval (0.1 -- 100) GeV. For a power-law energy spectrum with spectral index $\alpha$, the window function takes the form:
\begin{equation}
W_{S_i}(\chi) = \frac{(\alpha_i-2)\, \langle f_{S_i}(\chi)\rangle}{4\pi E_0^2\,\left[1+z(\chi)\right]^2}
\int_{E_\gamma>E_{\rm min}} \! \! \de E_\gamma \, 
\left( \frac{E_\gamma}{E_0} \right)^{-\alpha_i}
e^{-\tau\left[\chi,E_\gamma(\chi)\right]},
\label{eq:Wastro}
\end{equation}
where $E_0=100$ MeV is just the normalization energy, and $i$ stands for each of the \g-rays sources  adopted in our analysis: BL Lac, FSRQ, mAGN and SFG. The mean luminosity produced by an unresolved class of objects located at a distance $\chi$ from us is denoted by $\langle f_{S_i}(\chi) \rangle$ and is given by:
\begin{equation}
\langle f_{S_i}(\chi) \rangle = 
\int_{\mathcal{L}_{\rm min,i}}^{\mathcal{L}_{\rm max}(F_{\rm sens},z)} \de \mathcal{L} 
\, \mathcal{L} \, \Phi_i(\mathcal{L},z),
\label{eqn:average_g}
\end{equation}
where $\Phi_i(\mathcal{L},z)$ is the \g-ray luminosity function for the source class $i$. The upper bound, $\mathcal{L}_{\rm max}(F_{\rm sens},z)$, is the luminosity above which an object can be resolved, given the detector sensitivity $F_{\rm sens}$
for which we assume the  value $F_{\rm sens}=5 \times 10^{-10}\,\, \mathrm{photons~cm^{-2}s^{-1}}$ above 1 GeV \citep{2FGL,3FGL}.
The precise value  depends slightly on $\alpha_i$ and  on the  catalogue of resolved point sources,
although varying  $F_{\rm sens}$ within these different values has only a weak impact of the window function.
Conversely, the minimum luminosity $\mathcal{L}_{\rm min,i}$ depends on the properties of the source class under investigation. 
The four populations of astrophysical \g-ray emitters (i.e., BL Lac, FSRQ, mAGNs and SFGs) are discussed in the following. For each of them we describe the choice of $\alpha_i$ and of the \g-ray luminosity function.

\subsubsection{Blazars}
	\label{sec:blazars}
We consider BL Lacertae (BL Lacs) and flat-spectrum radio quasars (FSRQ) separately.
The \g-ray luminosity function of BL Lacs and FSRQ is taken from \cite{AjelloBLLs} and \cite{AjelloFSRQs}, respectively,
where it is derived from a parametric fit of the redshift and luminosity distributions of resolved blazars in the Fermi-LAT catalogue.
The lower limit of the integral in Eq.~\eqref{eqn:average_g} is set to $\mathcal{L}_{\rm min}=7\cdot 10^{42}\,\mathrm{erg s}^{-1}$ (BL Lac) and $\mathcal{L}_{\rm min}=4\cdot 10^{43}\,\mathrm{erg s}^{-1}$ (FSRQ). 
For the energy spectrum, we consider a simple power-law with a spectral index taken from the average spectral index in~\cite{AjelloBLLs,AjelloFSRQs}, namely, we assume $\alpha_{\rm BL Lac}=2.1$ and $\alpha_{\rm FSRQ}=2.44$.

The kernels of unresolved blazars are shown by the solid red (BL Lac) and magenta (FSRQ) lines in Fig.~\ref{fig:kernel}a. Note that they strongly decrease at low $z$ since Fermi-LAT has already detected a large number of the closest (brightest) emitters of these classes.

\subsubsection{Misaligned AGNs}
\label{sec:MAGNs}
In the case of mAGN, we follow \cite{DiMauro:2013xta}, which studied the correlation between the \g-ray luminosity and the core radio luminosity $L_{r,{\rm core}}$ at 5 GHz, and derived the GLF from the radio luminosity function. We consider their best-fit $\mathcal{L}$ vs. $L_{r,{\rm core}}$ relation and assume an average spectral index $\alpha_{\rm mAGN}$ of 2.37. The solid green line in Fig.~\ref{fig:kernel}a indicate the contribution of unresolved mAGNs.

\subsubsection{Star-forming galaxies}
As done in \cite{Fermi:2012eba}, we assume that the \g-ray and infrared (IR) luminosities are correlated in the case of SFG. We adopt the best-fit $\mathcal{L}$ vs. $L_{\rm IR}$ relation from \cite{Fermi:2012eba} while for the IR luminosity function we adopt the one from \cite{Gruppioni:2013jna} ,
(adding up spiral, starburst, and SF-AGN populations of their Table 8), as considered in \cite{Tamborra:2014xia}. 
The spectral index is taken to be $\alpha_{\rm SFG}=2.7$ for all the 3 components 
although starbursts galaxies  would require in principle a somewhat harder spectrum. 
Nonetheless, this  component is subdominant in the total SFG contribution except for high energies and at high redshift (i.e., in the ranges which are less relevant for the analyses in our work).
The above choice has thus no practical effects on our results.
The kernel associated to unresolved SFGs is the solid orange line in Fig.~\ref{fig:kernel}a. All the different single peaked sub-populations provide sizable contributions and this gives raise to different peaks.
\vspace{1cm}

The \g-ray emission produced by the four extragalactic astrophysical populations described above accounts for approximately the whole IGRB and autocorrelation angular power spectrum (see Fig.~\ref{fig:bench}). As described in the main text we however introduced a normalizing constant $A_\alpha$ for each population to be determined by the fit.
Apart from the extragalactic DM-induced emission described in Secs~\ref{sec:annihilating_DM} and \ref{sec:decaying_DM}, there may be a contribution associated with annihilations/decays in the DM halo of the Milky Way. This is not included since it does not correlate with the LSS tracers. 

\subsection{Galaxy catalogues}
\label{sec:Wgal}
For galaxies, we take the redshift distributions $dN_j/dz(\chi)$ reported in \cite{Xia:2014}. 
The associated kernels are shown in Fig.~\ref{fig:kernel}b. The 2MASS kernel peaks at low-redshift, and a comparison with the \g-rays kernels shown in the left panel of the same Fig.~\ref{fig:kernel} indicates that the 2MASS catalogue is the most suitable for investigating a DM signal in the cross-correlation analysis, followed by the SDSS Main Galaxy Sample catalogue.

\section{Halo occupation distribution of galaxies}
\label{sec:HOD}
In this work, we compute the angular cross-correlation between the unresolved \g-ray sky and the number of galaxies in specific catalogues.
In order to estimate the latter from a theoretical point of view (and since we adopt the halo model description for the structure clustering), we need to describe how galaxies populate halos. Namely, we need to model how many galaxies of a certain catalogue are present in a halo of mass $M$ and how they are spatially distributed. To this aim, we employ the halo occupation distribution (HOD) formalism.

We follow the approach described in \cite{Zheng:2004id} (for review on HOD, see also \cite{Berlind:2001xk,Cooray:2002dia}), where the HOD is parameterized by distinguishing the contributions of central and satellite galaxies, $N=N_{\rm cen}+N_{\rm sat}$ (since different formation histories typically imply different properties for galaxies residing at the centers of halos with respect to satellite galaxies). These can be modeled with the following functional forms:
\bea
\langle N_{\rm cen}(M)\rangle &=& \frac{1}{2}\left[1+{\rm erf}\left(\frac{\log M-\log M_{\rm th}}{\sigma_{\rm logM}}\right)\right] \label{eq:HOD1}\\
\langle N_{\rm sat}(M)\rangle &=& \left( \frac{M}{M_1} \right)^\alpha \exp{\left(-\frac{M_{\rm cut}}{M}\right)} \label{eq:HOD2}
\eea
With this formalism, we need five parameters for each galaxy population:
$M_{\rm th}$ denotes the approximate halo mass required to populate the halo with the considered type of galaxies, with the transition from 0 to 1 central galaxy modeled by means of Eq.~(\ref{eq:HOD1}), and set by the width $\sigma_{\rm LogM}$.
The satellite occupation is described by a power law (with index $\alpha$ and normalization set by the mass $M_1$), with an exponential cutoff $M_{\rm cut}$ at low masses.
The value of the five HOD parameters for each of the considered galaxy population is discussed in the following. For some catalogues, we will also consider similar but slightly different functional forms.

We selected those galaxy samples with available HOD which more closely resemble the catalogues considered in the cross-correlation analysis of this work.
We caution, however, that since the matching of the two samples is not perfect some differences in the associated HODs might be expected. Nevertheless, this should not affect our results in a dramatic way.

Eqs.~(\ref{eq:HOD1}) and (\ref{eq:HOD2}) provide the number of galaxies in a halo of mass $M$. Concerning the spatial distribution, we treat central and satellite galaxies separately. The former is taken as a point-source located at the center of the halo (the point-source approximation is expected to break down only for $\ell\gtrsim10^3$).
Satellite galaxies are instead described in an effective way with a spatial distribution following the host-halo profile.
In other words, we express the density field of galaxies with:
\begin{equation}
g_g(\bm x - \bm x'|M)=\langle N_{\rm cen}(M)\rangle\,\delta^3(\bm x - \bm x')+\langle N_{\rm sat}(M)\rangle\,\rho_h(\bm x - \bm x'|M)/M\, .
\end{equation}
Note that:
\begin{equation}
\int \de^3\bm x\,g_g(\bm x)=\langle N_{\rm cen}(M)\rangle+\langle N_{\rm sat}(M)\rangle=\langle N(M)\rangle\, .
\end{equation}

\subsubsection{2MASS HOD}
A determination of the HOD for the 2MASS galaxies is not present in the literature (to the knowledge of the authors).
In \cite{Zehavi:2004ii}, a sample of about 200,000 SDSS galaxies mostly residing in the redshift range $0.02 < z < 0.167$ and with r-band magnitude $14.5\leq r \lesssim17.77$ was analyzed.
Such ranges of redshift and magnitude are analogous to the ones of the adopted 2MASS catalogue~\citep{2MASS}, with the cross-identification of the latter with SDSS found to be successful for about 90\% of the sources~\citep{McIntosh:2005yx}.
We can thus exploit the HOD results of \cite{Zehavi:2004ii}. They considered a step function ($\langle N_{\rm cen}\rangle=0$ for $M< M_{\rm th}$ and $\langle N_{\rm cen}\rangle=1$ for $M\geq M_{\rm th}$) instead of Eq.~(\ref{eq:HOD1}), and set $M_{\rm cut}=0$ in Eq.~(\ref{eq:HOD2}).
The analysis was performed by splitting the sample in luminosity bins, but for our purposes we can consider the averaged best-fit parameters weighted over the number of galaxies in each bin. We found $\log M_{\rm th}=12.1$, $\alpha=1.2$, and $\log M_1=13.5$.

\subsubsection{NVSS HOD}
The NVSS sub-sample considered in this work (\cite{NVSS}) contains sources brighter than 10 mJy at 1.4 GHz. The vast majority of them is associated to bright AGNs.
To model the AGN HOD, we follow \cite{Chatterjee:2012}. For bright objects, they found $\log M_{\rm th}=13.03$, $\sigma_{\rm LogM}=0.96$, $\alpha=1.17$, $\log M_{cut}=11.5$, and $\log M_1=13.64$.

\subsubsection{SDSS-DR8 Main Galaxy Sample HOD}
In \cite{Ross:2010yf}, the clustering of more than three million photometrically selected SDSS galaxies was analyzed.
In particular, the sample was defined requiring de-reddened r-band magnitudes $r_d< 21$ and absolute magnitudes $M_r<-21.2$, in the redshift range $0.1 < z < 0.4$ and masking objects with Galactic extinction $A_r > 0.2$.
This galaxy sample is very similar to the one adopted in this work for the cross-correlation analysis~(\cite{SDSSDR8}),
 except for a more limited redshift (and magnitude) range.
Since the peak of the redshift distribution is at $z\sim0.3$, such difference is not expected to play a major role.
After averaging the best-fit values of HOD parameters over the different redshift bins considered in \cite{Ross:2010yf} (weighted for the number of galaxies in each bin), we obtain $\log M_{\rm th}=12.09$, $\sigma_{\rm LogM}=0.3$, $\alpha=1.09$, and $\log M_1=13.25$.
The functional form of the satellite HOD considered in \cite{Ross:2010yf} is:
\begin{equation}
\langle N_{sat}\rangle = \langle N_{\rm cen}\rangle \times \left( \frac{M-M_{\rm th}}{M_1} \right)^\alpha
\end{equation}
instead of Eq.~(\ref{eq:HOD2}), with $\langle N_{\rm cen}\rangle$ from Eq.~(\ref{eq:HOD1}) and $\langle N_{\rm sat}\rangle=0$ for $M<M_{\rm th}$.

\subsubsection{SDSS-DR8 Luminous Red Galaxies HOD}
A recent analysis of more than 500,00 SDSS-III CMASS galaxies~\citep{Reid:2014iaa} derived the HOD of this galaxy sample.
The satellite HOD was modeled by means of:
\begin{equation}
\langle N_{\rm sat}\rangle = \langle N_{\rm cen}\rangle \times \left( \frac{M-M_{\rm cut}}{M_1} \right)^\alpha
\end{equation}
with $\langle N_{\rm cen}\rangle$ given by Eq.~(\ref{eq:HOD1}) (and $\langle N_{\rm sat}\rangle=0$ for $M<M_{\rm cut}$).
The best-fit parameters in the fiducial model of \cite{Reid:2014iaa} are $\log M_{\rm th}=13.03$, $\sigma_{\rm LogM}=0.38$, $\alpha=0.76$, $\log M_1=14.08$, and $M_{\rm cut}=13.27$.
These results are found to be in agreement with the HOD analysis presented in \cite{White:2010ed}, which in turn was tested to reproduce the clustering of the galaxy sample of \cite{Ho:2012vy} adopted here.


\subsubsection{SDSS-DR6 Quasar HOD}
The modeling of the halo occupation distribution of SDSS quasars is taken from \cite{Richardson:2012du}.
The sample consists of 47,699 quasars in the redshift range $0.4 < z < 2.5$ with median redshift of $\bar z = 1.4$ and flux limited to $i < 19.1$.
It is very similar to the catalogue considered in this work for the cross-correlation analysis~(\cite{SDSSDR6QSO}).
The best-fit parameters entering in Eqs.~(\ref{eq:HOD1}) and (\ref{eq:HOD2}) are not provided in \cite{Richardson:2012du}, but we find that with $\log M_{\rm th}=16.5$, $\sigma_{\rm LogM}=1.65$, $\alpha=1$, $\log M_{\rm cut}=15.25$, and $\log M_1=13.1$, the best-fit curve in their Fig.~2b is well reproduced.

\begin{figure*}[t]
\vspace{-3cm}
\centering
\includegraphics[width=0.49\textwidth]{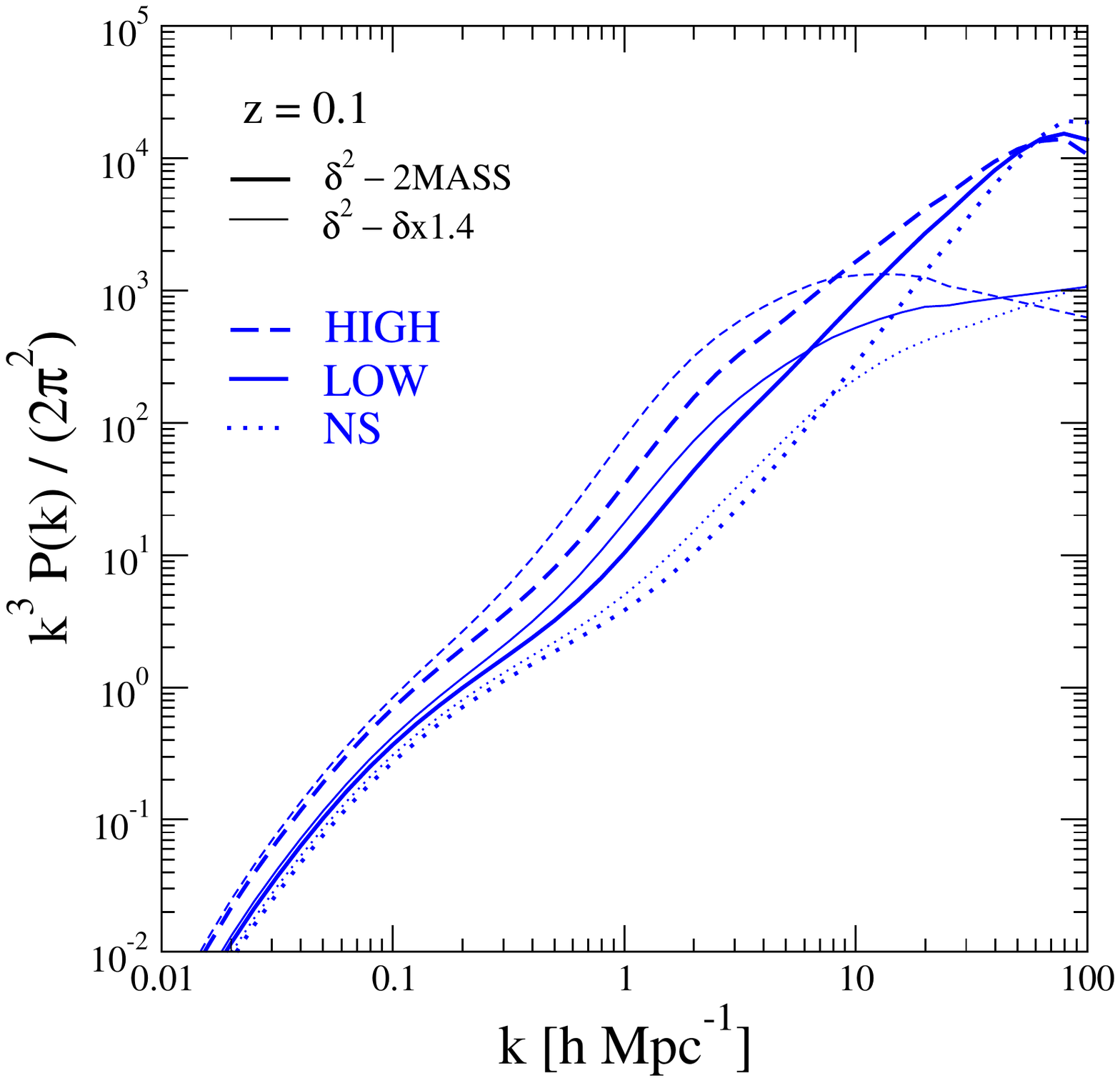}
\includegraphics[width=0.49\textwidth]{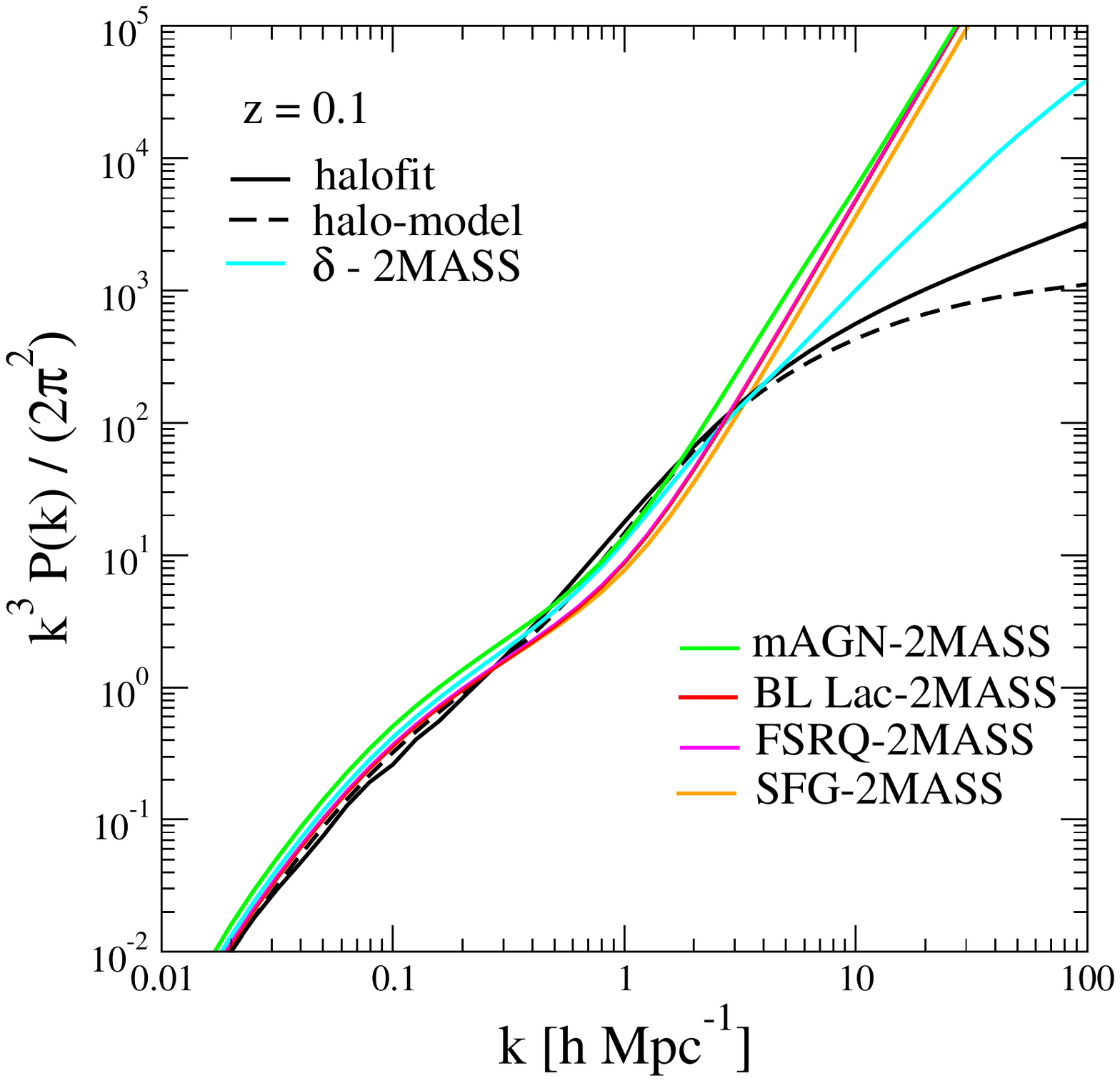}
\caption{Power spectrum (multiplied by $k^3$) of cross-correlation between \g-ray emitters and 2MASS galaxies at redshift $z=0.1$. The left panel refers to annihilating DM. The right panel shows decaying DM and astrophysical sources. The matter power spectrum obtained within the halo model employed in this work is shown with a dashed black line and is compared with the halofit results~\citep{Takahashi:2012em} derived from high-resolution N-body simulations (black solid line). }
\label{fig:ps2MASS}
\end{figure*}

\begin{figure*}[t]
\vspace{-3cm}
\centering
\includegraphics[width=0.49\textwidth]{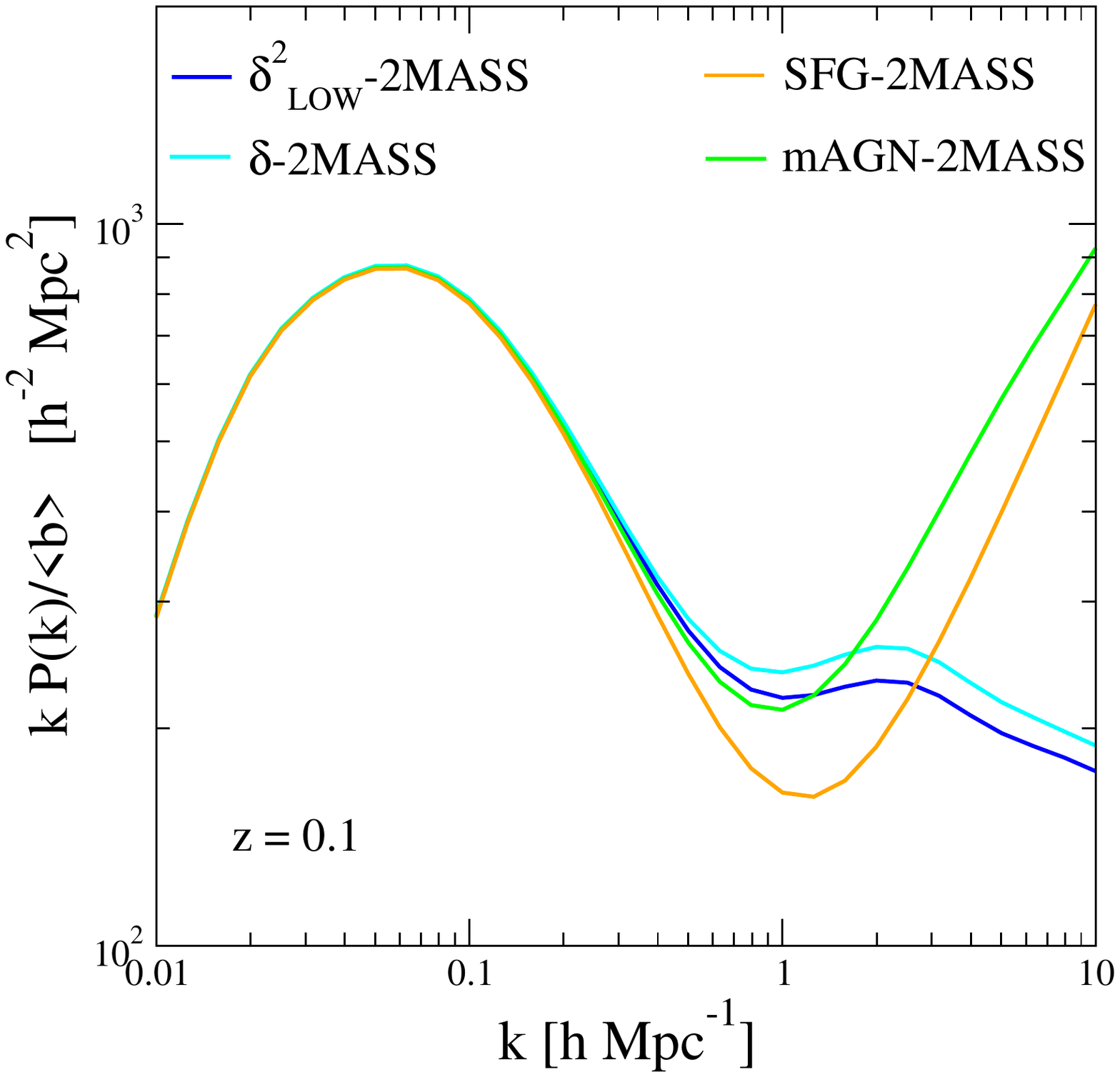}
\includegraphics[width=0.49\textwidth]{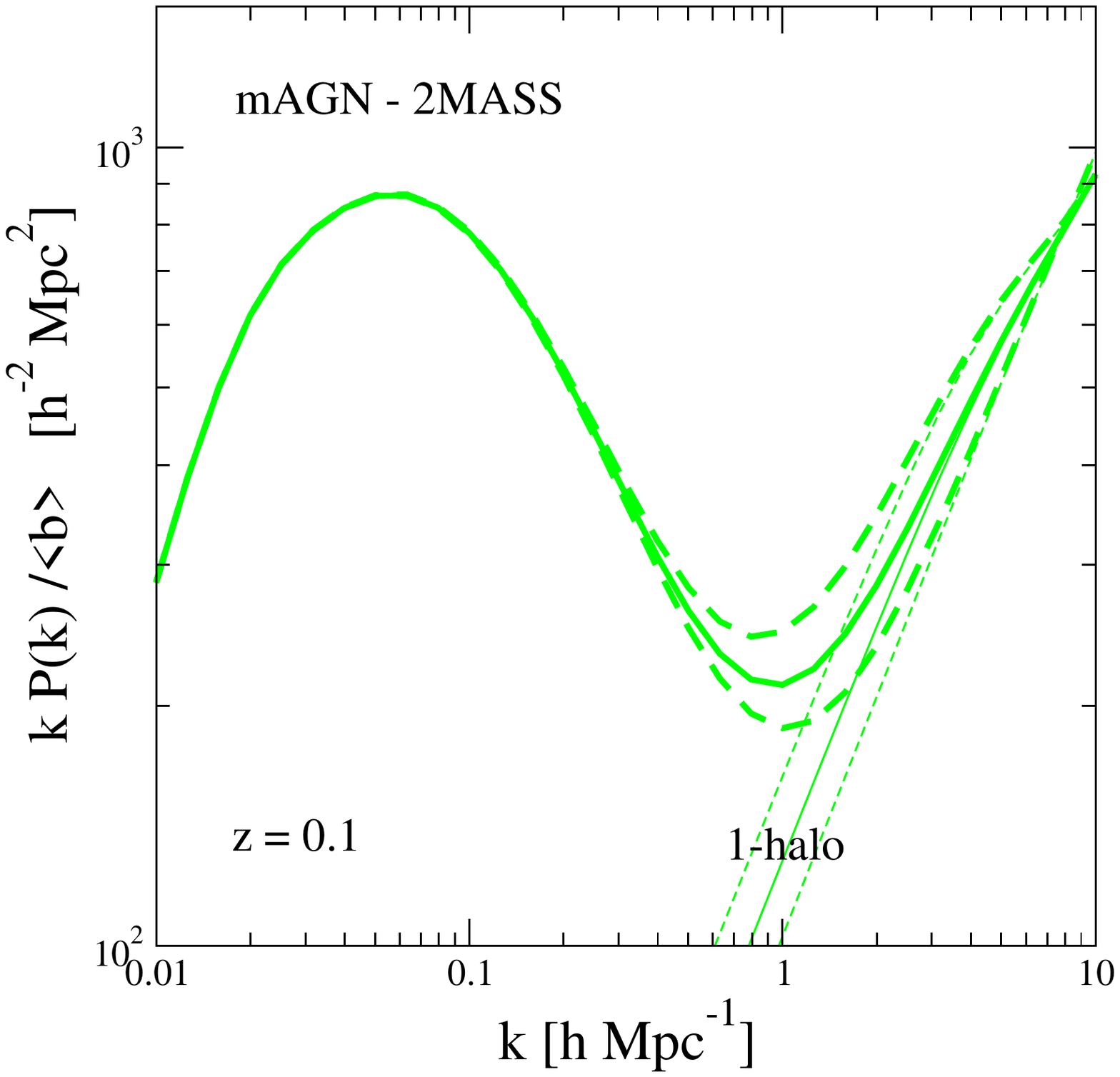}
\caption{Power spectrum of cross-correlation between \g-ray emitters and 2MASS galaxies at redshift $z=0.1$. The power spectrum is multiplied by $k$ and divided by the effective bias of \g-ray emitters $\langle b(z=0.1)\rangle$. The left panel compares predictions from annihilating (in the \low\ scenario) and decaying DM with the astrophysical models of mAGN and SFG. The right panel focuses on mAGN and reports the power spectrum with three different assumptions for the $M(\mathcal{L})$ relation (taken from \cite{Camera:2014rja}). Thin lines show the 1-halo part of the power spectrum. }
\label{fig:ps1h}
\end{figure*}

\section{3D Power Spectra}
\label{sec:3dps}
In the halo model computation of the cross-correlation power spectrum (PS), the 3D-PS is split in the one-halo ($P^{1h}$) and two-halo ($P^{2h}$) components with $P=P^{1h}+P^{2h}$.
For a derivation of the $P^{1h}$ and $P^{2h}$ discussed in the equations below, see \cite{Fornengo:2013rga}.
We remind that $S_i$ denote $\gamma$-ray astrophysical emitters (BL Lac, FSRQ, mAGN, and SFG), $g_j$ are associated to and galaxy populations (SDSS-DR6 quasars, 2MASS galaxies, NVSS radio sources, SDSS-DR8 Luminous Red Galaxies, and SDSS-DR8 ``main'' galaxies), while $\delta$ and $\delta^2$ stands for decaying and annihilating DM, respectively.
In most of the equations, the dependence on $z$ is not explicitely reported to simplify the notation.

The 3D power spectrum of cross-correlation between $\gamma$-rays from annihilating DM and galaxy catalogues is computed as:
\bea
P_{g_j,\delta^2}^{1h}(k,z) &=& \int_{M_{\rm min}}^{M_{\rm max}} dM\ \frac{dn}{dM} \frac{\langle N_{g_j}\rangle}{\bar n_{g_j}} \tilde v_g(k|M) \,\frac{\tilde u(k|M)}{\Delta^2} \label{eq:PSannLSS1} \\
P_{g_j,\delta^2}^{2h}(k,z) &=& \left[\int_{M_{\rm min}}^{M_{\rm max}} dM\,\frac{dn}{dM}b_h(M) \frac{\langle N_{g_j}\rangle}{\bar n_{g_j}} \tilde v_g(k|M) \right]\,\left[\int_{M_{\rm min}}^{M_{\rm max}} dM\,\frac{dn}{dM} b_h(M) \frac{\tilde u(k|M)}{\Delta^2} \right]\,P^{\rm lin}(k)\;.
\label{eq:PSannLSS2}
\eea
The function $\tilde u(k|M)$ is the Fourier transform of:
\begin{equation}
 u(\bm x|M)=\rho_h^2(\bm x|M)/\bar \rho_{DM}^2+b_{\rm sub}(M) \,\rho_h(\bm x|M)/M \,\int \de^3\bm x\,\rho_h^2(\bm x|M)/\bar \rho_{DM}^2
\end{equation}
where $\rho_h$ denotes the main halo profile and $b_{\rm sub}$ is the boost function associated to subhalos (introduced above). Note that $\tilde u(k=0|M)=(1+b_{\rm sub}(M,z))\,\int \de^3\bm x\,\rho^2(\bm x|M)/\bar \rho^2$.
The product $\langle N_{g_j}\rangle\,\tilde v_g(k|m)$ is instead the Fourier transform of $\langle N_{\rm cen,j}(M)\rangle\,\delta^3(\bm x )+\langle N_{\rm sat,j}(M)\rangle\,\rho_h(\bm x|M)/M$. We have $\langle N_{g_j}\rangle\,\tilde v_g(k=0|m)=\langle N_{g_j}\rangle$.
The average number of galaxies $g_j$ at a given redshift is given by $\bar n_{g_j}(z)=\int dM\,dn/dM\, \langle N_{g_j}\rangle$.
The details of the models of $\langle N_{g_j}\rangle$, $dn/dM$ and $\rho_h(\bm x|M)$ have been described in the previous Sections.

The impact of clustering assumptions on the 3D PS are illustrated in Fig.~\ref{fig:ps2MASS}a, where we consider the 2MASS catalogue and show $P_{g_j,\delta^2}(k,z=0.1)$.
The boost from substructures makes the \g-ray contributions from most massive halos to dominate the signal, and this is more pronounced in the \high\ case rather than in the \low\ scenario. In the case without substructures, low mass halos becomes more important in the total budget of the \g-ray emission. This explains the hierarchy at $k\sim1/$Mpc. For the same reasons, an opposite hierarchy occurs at very small scales ($k\gtrsim100/$Mpc).

In the case of decaying DM, the PS of cross-correlation takes the form:
\bea
 P_{g_j,\delta}^{1h}(k,z) &=& \int_{M_{\rm min}}^{M_{\rm max}} dM\ \frac{dn}{dM} \frac{\langle N_{g_j}\,\rangle}{\bar n_{g_j}}\tilde v_g(k|M) \tilde v_\delta(k|M) \label{eq:PSdecLSS1}\\
 P_{g_j,\delta}^{2h}(k,z) &=& \left[\int_{M_{\rm min}}^{M_{\rm max}} dM\,\frac{dn}{dM} b_h(M) \frac{\langle N_{g_j}\rangle}{\bar n_{g_j}} \tilde v_g(k|M) \right] \left[\int_{M_{\rm min}}^{M_{\rm max}} dM\,\frac{dn}{dM} b_h(M)\,\tilde v_\delta(k|M) \right]\,P^{\rm lin}(k)\;.
\label{eq:PSdecLSS2}
\eea
Here $\tilde v_\delta(k|M)$ is the Fourier transform of $\rho_h(\bm x|M)/\bar \rho_{DM}$.
In Fig.~\ref{fig:ps2MASS}b, we show $P_{g_j,\delta}(k,z=0.1)$ (again for the 2MASS case), together with the matter power spectrum derived within our halo model approach. The latter is compared to a revised halofit PS derived from latest high-resolution N-body simulations~\citep{Takahashi:2012em}. They agree within 20\% at $k<10/$Mpc and this supports our choices for the halo model ingredients. At larger $k$ there is a departure, with less power in the halo model, but the picture at such small scales is in any case very uncertain, also from the simulations point of view. 

\begin{figure*}[t]
\vspace{-3cm}
\centering
\includegraphics[width=0.49\textwidth]{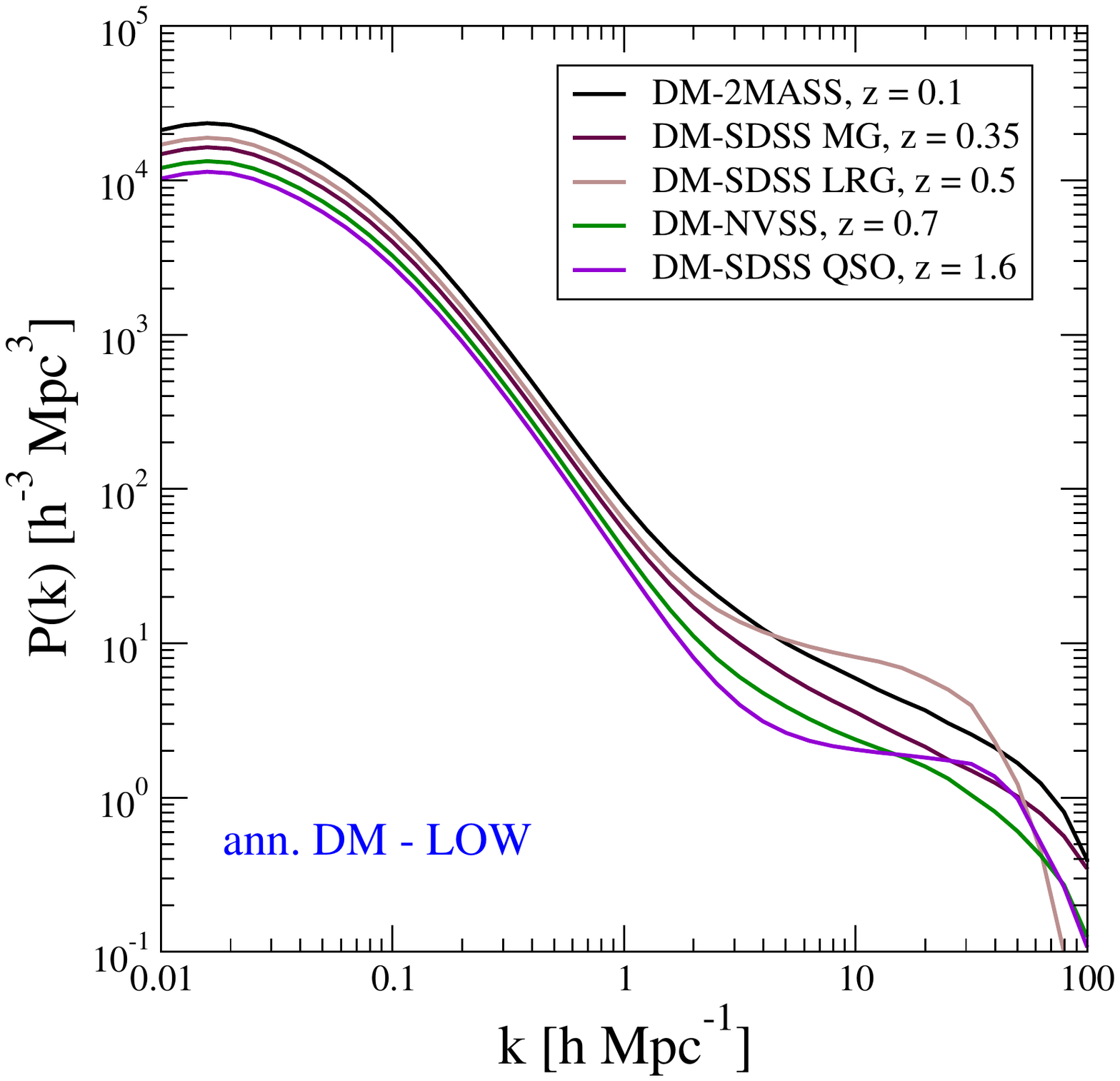}
\includegraphics[width=0.49\textwidth]{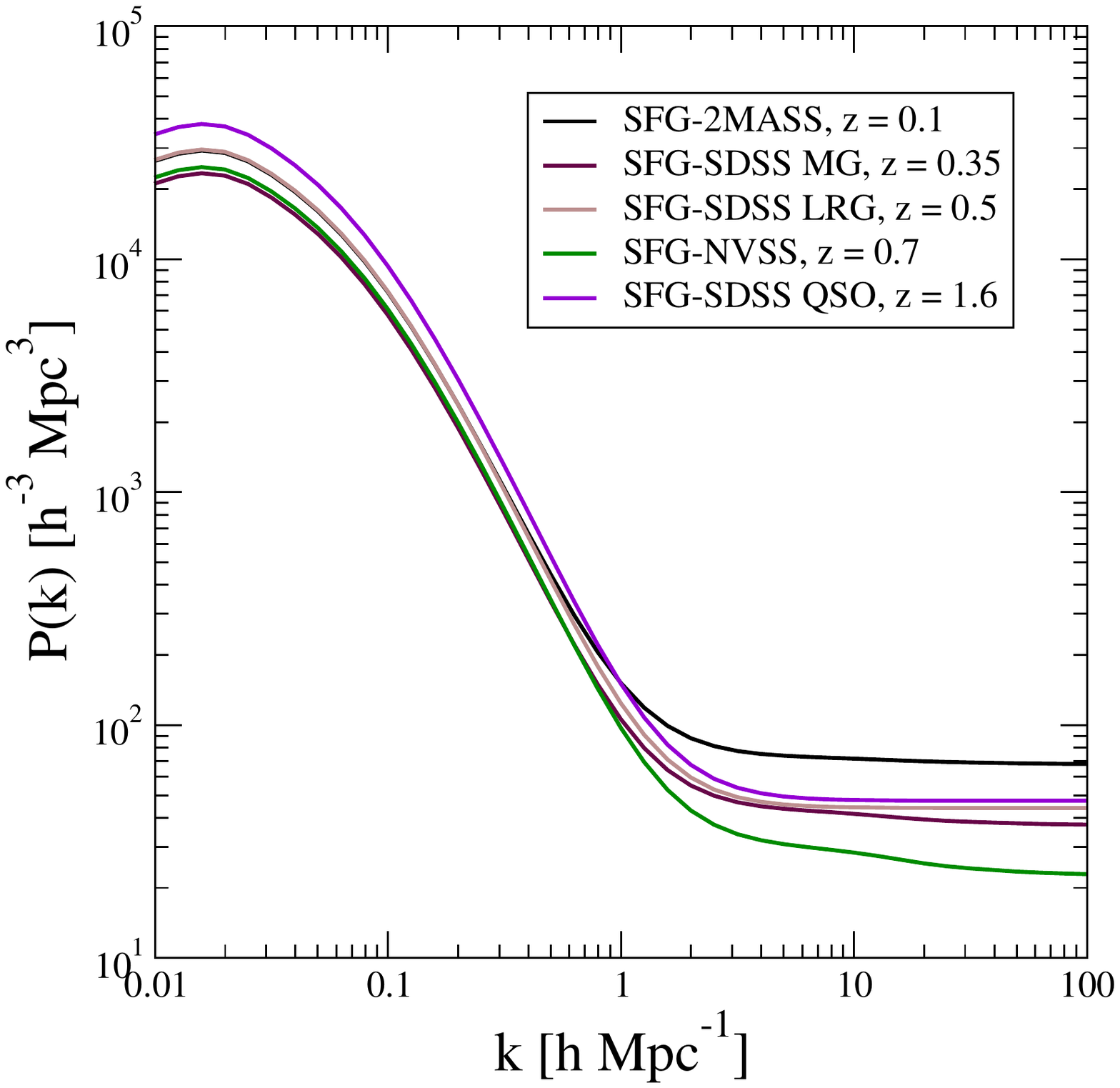}
\caption{{\sl Left}: Three dimensional power spectrum of cross-correlation between \g-rays from annihilating DM (in the \low\ scenario) and galaxies, evaluated at the redshift corresponding to the peak of the $dN_j/dz$ of each catalogue. {\sl Right}: Same as left panel, but for SFG instead of DM.}
\label{fig:psall}
\end{figure*}

We assume astrophysical \g-ray emitters to be point-like sources with the density field given by $f_{S_i}(\bm x - \bm x')=\mathcal{L}_{S_i}\,\delta^3(\bm x - \bm x')$. The 3D PS of cross-correlation with galaxy catalogues can be written as:
\bea
 P_{g_j,S_i}^{1h}(k,z) &=& \int_{\mathcal{L}_{\rm min,i}(z)}^{\mathcal{L}_{\rm max,i}(z)} d\mathcal{L}\,\Phi_i(\mathcal{L},z)\,\frac{\mathcal{L}}{\langle f_{S_i} \rangle} \,\frac{\langle N_{g_j}\!(\mathcal{L})\,\rangle}{\bar n_{g_j}}\tilde v_g(k|M(\mathcal{L})) \label{eq:PSBd1} \\
 P_{g_j,S_i}^{2h}(k,z) &=& \left[\int_{\mathcal{L}_{\rm min,i}(z)}^{\mathcal{L}_{\rm max,i}(z)} d\mathcal{L}\,\Phi_i(\mathcal{L},z)\, b_{S_i}(\mathcal{L})\,\frac{\mathcal{L}}{\langle f_{S_i} \rangle} \right]\left[\int_{M_{\rm min}}^{M_{\rm max}} dM\,\frac{dn}{dM} b_h(M)\,\frac{\langle N_{g_j}\,\rangle}{\bar n_{g_j}} \tilde v_g(k|M) \right] \,P^{\rm lin}(k)\;.\label{eq:PSBd2}
\eea
where $b_{S_i}$ is the bias of \g-ray astrophysical sources with respect to matter, for which we adopt $b_{S_i}(\mathcal{L})=b_h(M(\mathcal{L}))$.
Both Eqs.~(\ref{eq:PSBd1}) and (\ref{eq:PSBd2}) require the specification of the relation $M(\mathcal{L})$ between the mass of the host halo $M$ and the luminosity of the hosted object $\mathcal{L}$.
We will use the modeling of $M(\mathcal{L})$ derived in \cite{Camera:2014rja}, where this aspect is discussed, and to which we refer the reader for the details.
The blazar $M(\mathcal{L})$ model of \cite{Camera:2014rja} is adopted for both BL Lac and FSRQ.

\begin{figure*}[t]
\vspace{-3cm}
\centering
\includegraphics[width=0.49\textwidth]{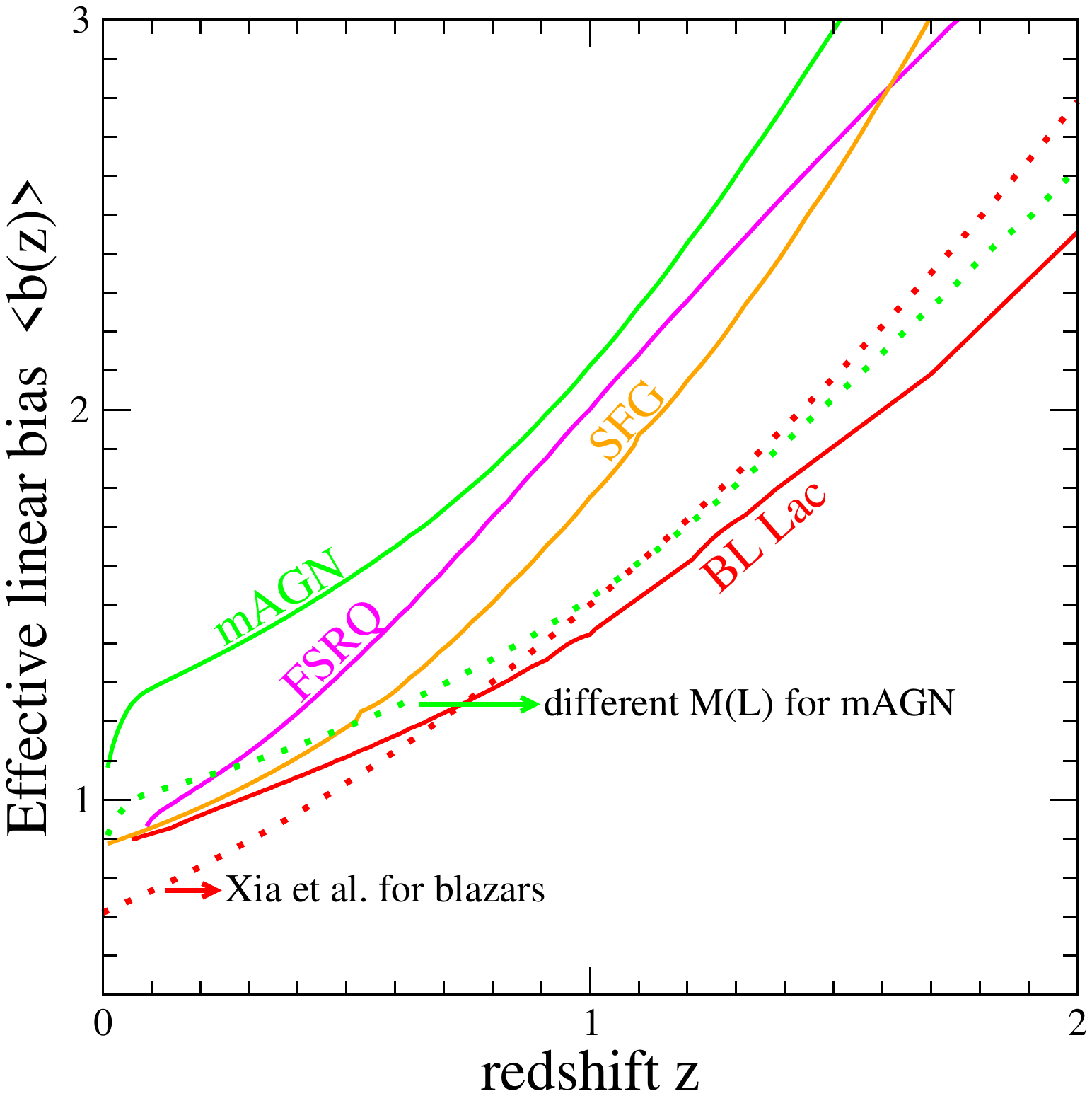}
\includegraphics[width=0.49\textwidth]{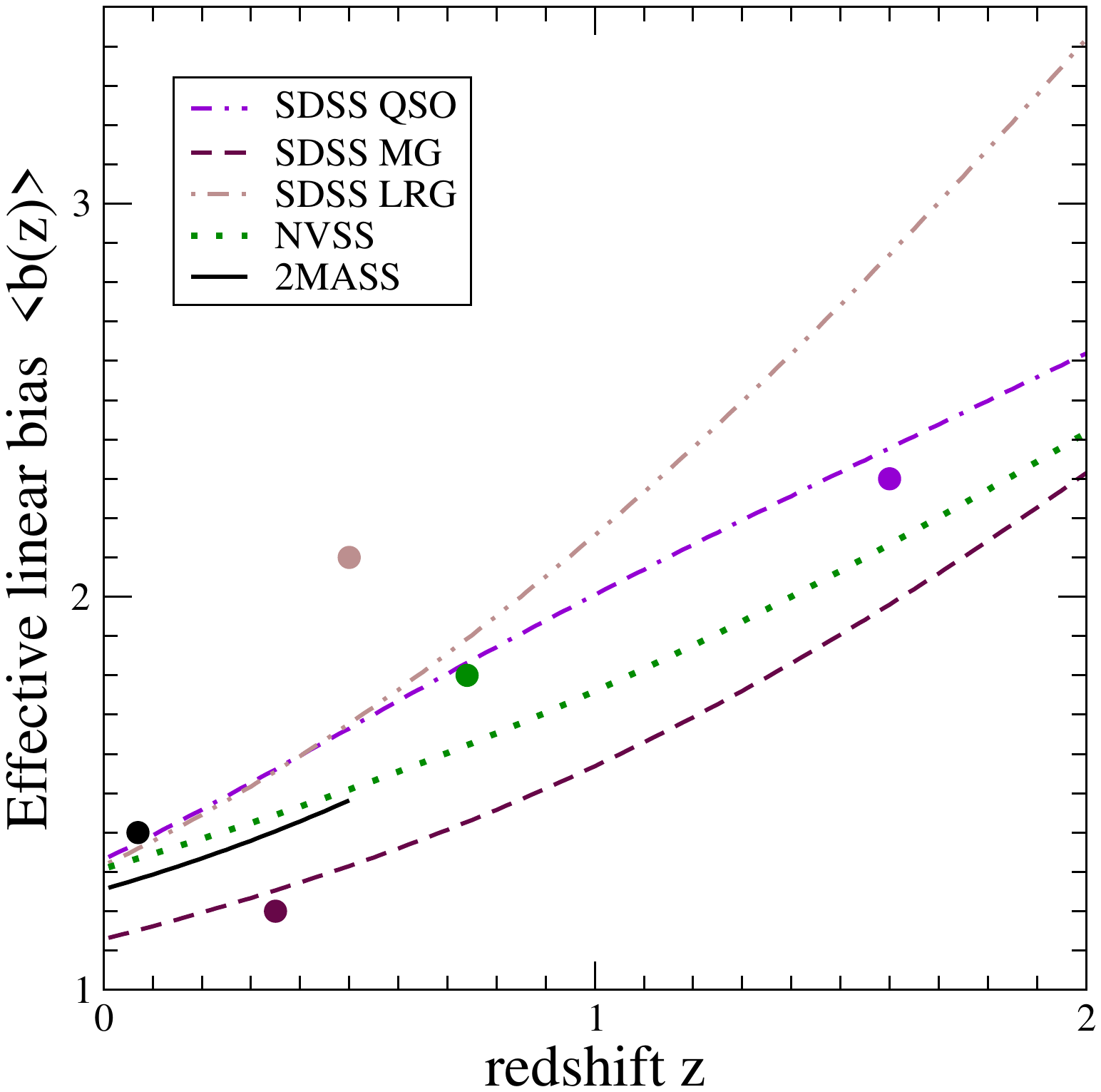}
\caption{Effective bias for \g-ray astrophysical emitters ({\sl left}) and galaxies ({\sl right}), as defined in Eqs.~(\ref{eq:biasastro}) and (\ref{eq:biasgal}), respectively. 
To illustrate the impact of the $M(\mathcal{L})$ description, we additionally show the bias of mAGN when assuming the lower limit discussed in  \cite{Camera:2014rja} for such relation (green thin line). For comparison, we also report the bias of \g-ray blazars considered in  \cite{Xia:2014} (red dotted line).
In the galaxy cases (right panel), we show with circles the value of the different bias parameters adopted in  \cite{Xia:2014}, where they were taken to be constant in redshift. The position of the dots refers to the redshift which corresponds to the peak of the galaxy distribution $dN_j/dz$.
}
\label{fig:bias}
\end{figure*}

We caution that Eq.~(\ref{eq:PSBd1}) for $P_{g_j,S_i}^{1h}$ gives only an approximate estimate of the 1-halo correlation.
Indeed, modeling the satellite galaxies as a smooth component reduces their correlation with point-like \g-ray sources.
On the other hand, we assume that a halo hosting a given \g-ray emitter also hosts the galaxies of all catalogues.
This may not be true (e.g. some catalogue is mostly formed by galactic objects which do not host an AGN), thus artificially enhancing $P_{g_j,S_i}^{1h}$.
Moreover, Eq.~(\ref{eq:PSBd1}) is based on average relations, whilst a relative small number of outliers (i.e., bright \g-ray sources in a halo with galaxies) can have a relevant impact. For all these reasons, and since $P_{g_j,S_i}^{1h}$ is approximately independent on $k$, we can include in the fit an arbitrary constant term allowing for both positive and negative corrections to Eq.~(\ref{eq:PSBd1}). We call this additional quantity {\sl one-halo correction term}, and we perform the analysis under the assumption that this term is not relevant (i.e. by setting it to vanish) and under the assumption that it is present, leaving it as a free parameter, one for each LSS tracer.

Fig.~\ref{fig:ps2MASS} (right panel) shows $P_{g_j,S_i}(k,z=0.1)$, again taking the 2MASS catalogue as illustrative. The different classes of \g-ray emitters show a similar spectrum, and have less (more) power than in the DM cases at intermediate (small) scales, as expected given their size. 

In Fig.~\ref{fig:ps1h}, we show the difference between the DM and astrophysical PS at low redshift arising from the 1-halo term.
To this aim we divide the PS by the bias in order to have the large scale PS (i.e., the two halo term) with a common normalization.
At small scales the power associated to astrophysical sources is larger than for DM.
The picture is opposite at intermediate scales, around Mpc, especially for SFG.
The adopted model of $M(\mathcal{L})$ makes the mAGN individual objects that contribute more to mAGN emission to be hosted in relatively large halos.
This implies that the mAGN PS at Mpc scales is similar to the one of DM, explaining (part of) the origin of the degeneracy between mAGN and DM mentioned in the main text.  
We investigate the impact of different $M(\mathcal{L})$ relations, taken from \cite{Camera:2014rja}, in the right panel of Fig.~\ref{fig:ps1h}.

The 3D PS of cross-correlation with the other catalogues are shown in Fig.~\ref{fig:psall}.
As illustrative examples, we selected the \low\ scenario for annihilating DM, and SFG for astrophysical \g-ray sources. The PS are computed at the redshift corresponding to the peak of the $dN_j/dz$ of each catalogue.

In Fig.~\ref{fig:bias}, we show the effective bias of astrophysical \g-ray emitters and galaxies. They are defined with $P_{g_j,S_i}^{2h}=\langle b_{S_i}\rangle\,\langle b_{g_j}\rangle \,P^{\rm lin}$ at $k=0$, so they read:
\bea
\langle b_{S_i}(z)\rangle&=&\int_{\mathcal{L}_{\rm min,i}(z)}^{\mathcal{L}_{\rm max,i}(z)} d\mathcal{L}\,\Phi_i(\mathcal{L},z)\, b_h(M(\mathcal{L}))\,\frac{\mathcal{L}}{\langle f_{S_i} \rangle} \label{eq:biasastro}\\
\langle b_{g_j}(z)\rangle&=&\int_{M_{\rm min}}^{M_{\rm max}} dM\,\frac{dn}{dM} b_h(M)\,\frac{\langle N_{g_j}(M)\rangle}{\bar n_{g_j}}\;. \label{eq:biasgal}
\eea
Eq.~(\ref{eq:biasastro}) depends on  the mass-luminosity relation $M(\mathcal{L})$, while Eq.~(\ref{eq:biasgal}) is governed by the modeling of $\langle N_{g_j}(M)\rangle$ .
The fair agreement shown by the computed bias with findings of autocorrelation studies quoted in the literature (e.g., \cite{Zehavi:2004ii,Reid:2014iaa,Ho:2012vy,White:2010ed,Ross:2010yf,Allevato:2014}) is an important check of our modeling of $M(\mathcal{L})$ and $\langle N_{g_j}(M)\rangle$.

The bias of \g-ray blazars appears systematically lower than findings in \cite{Allevato:2014}. Translating the halo bias in terms of the mean mass hosting the blazars by means of \cite{Sheth:1999mn}, their results imply halos of $M\simeq3\cdot10^{13}\,M_\odot$, while according to our results shown in Fig.~\ref{fig:bias}, FSRQs reside in halos of $M\simeq1.5\cdot10^{13}\,M_\odot$ and BL Lacs in $M\simeq5\cdot10^{12}\,M_\odot$. 
This is not surprising, if we consider that the work of \cite{Allevato:2014} focuses on resolved objects, namely on a blazar subsample given by the brightest ones, which reside in more massive halos, while on the contrary, we investigate the unresolved component, which should be hosted by less massive halos. 
Moreover, the relatively low number of known \g-ray objects prevents a firm knowledge of their clustering, and sizable uncertainties on the bias are currently present.

\section{Plots of CCF at other energies}
\label{otherenergies}

\begin{figure*}[ht!]
\begin{center}
%
\includegraphics[width=0.33\columnwidth]{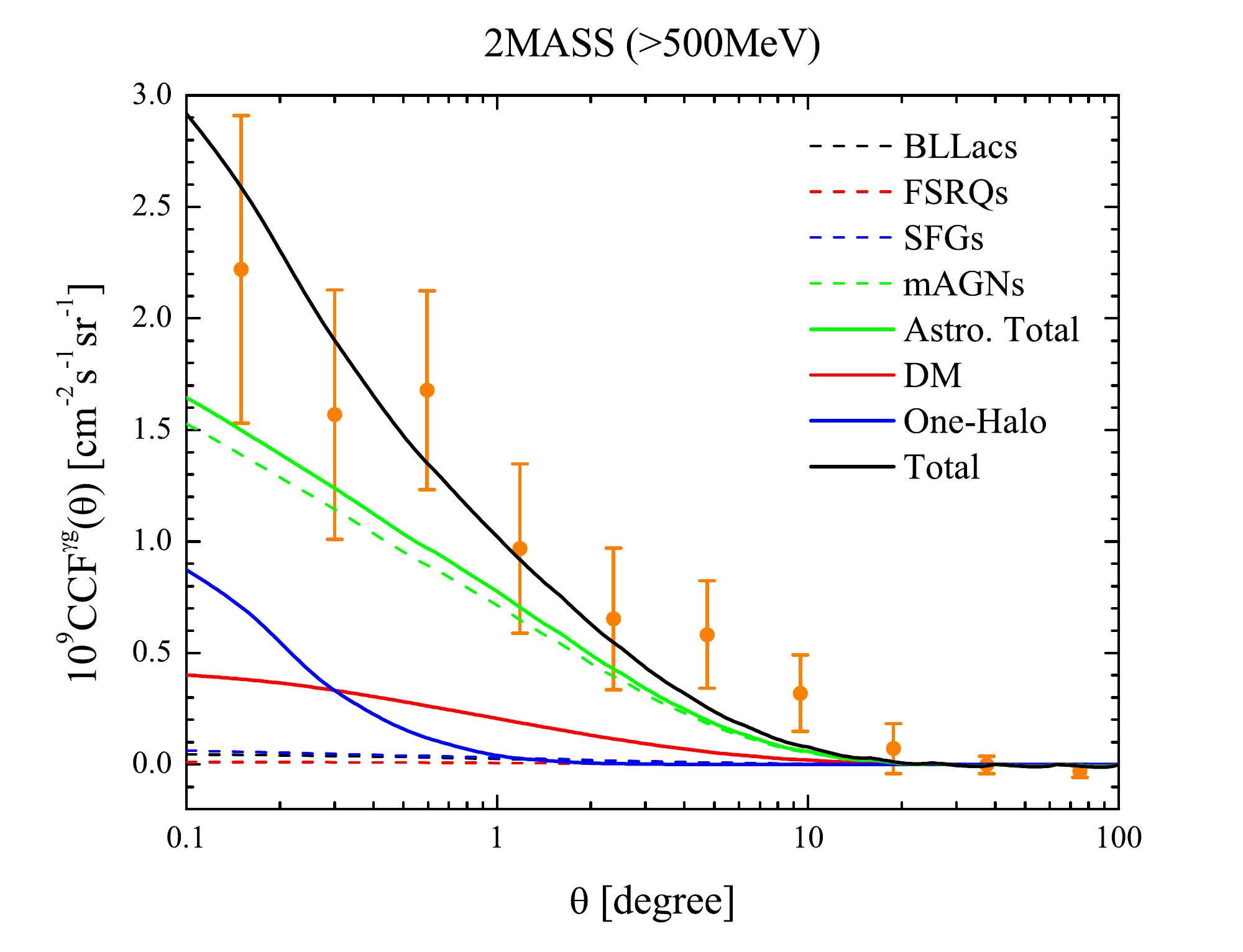} 
\hspace{-0.5cm}
\includegraphics[width=0.33\columnwidth]{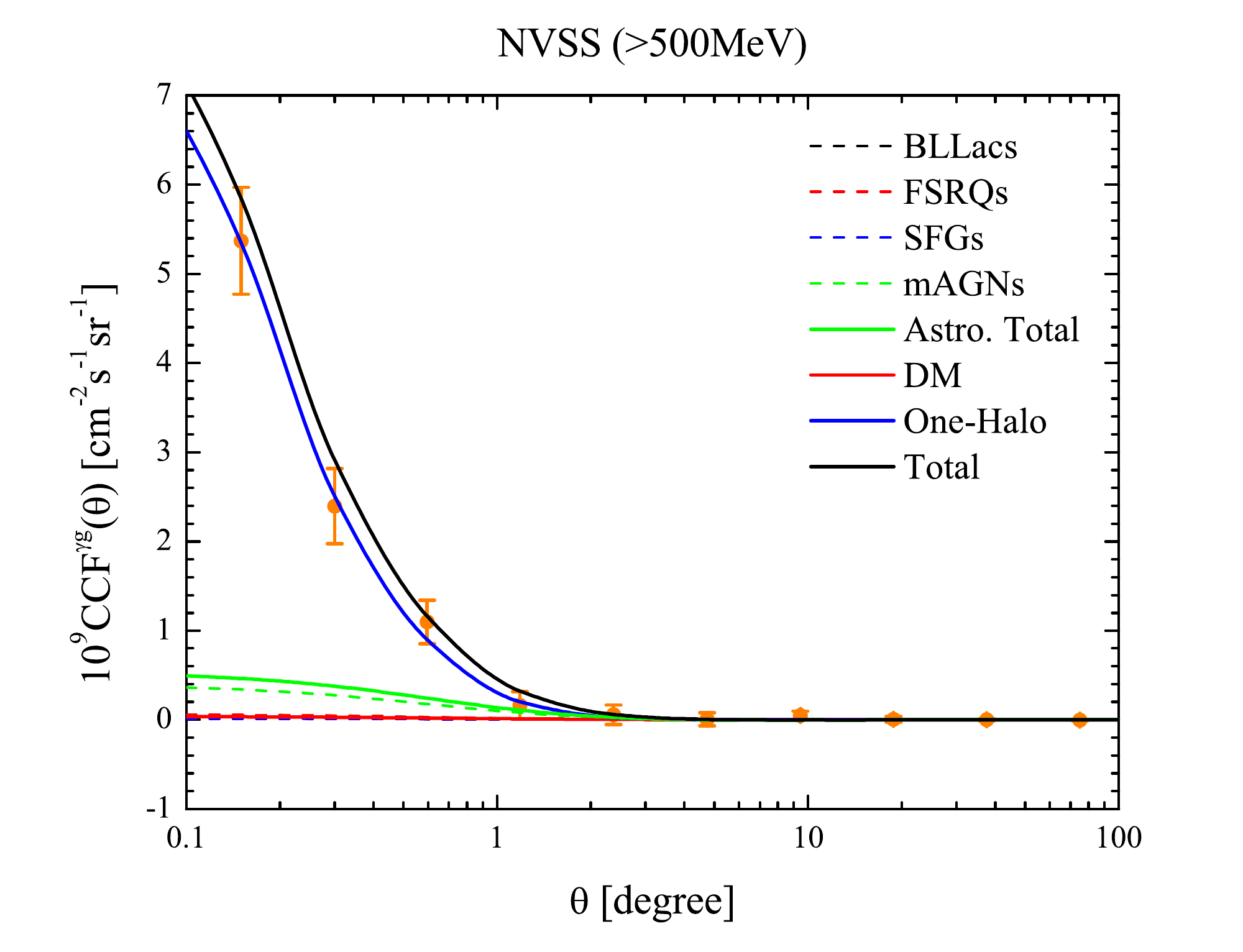} 
\hspace{-0.5cm}
\includegraphics[width=0.33\columnwidth]{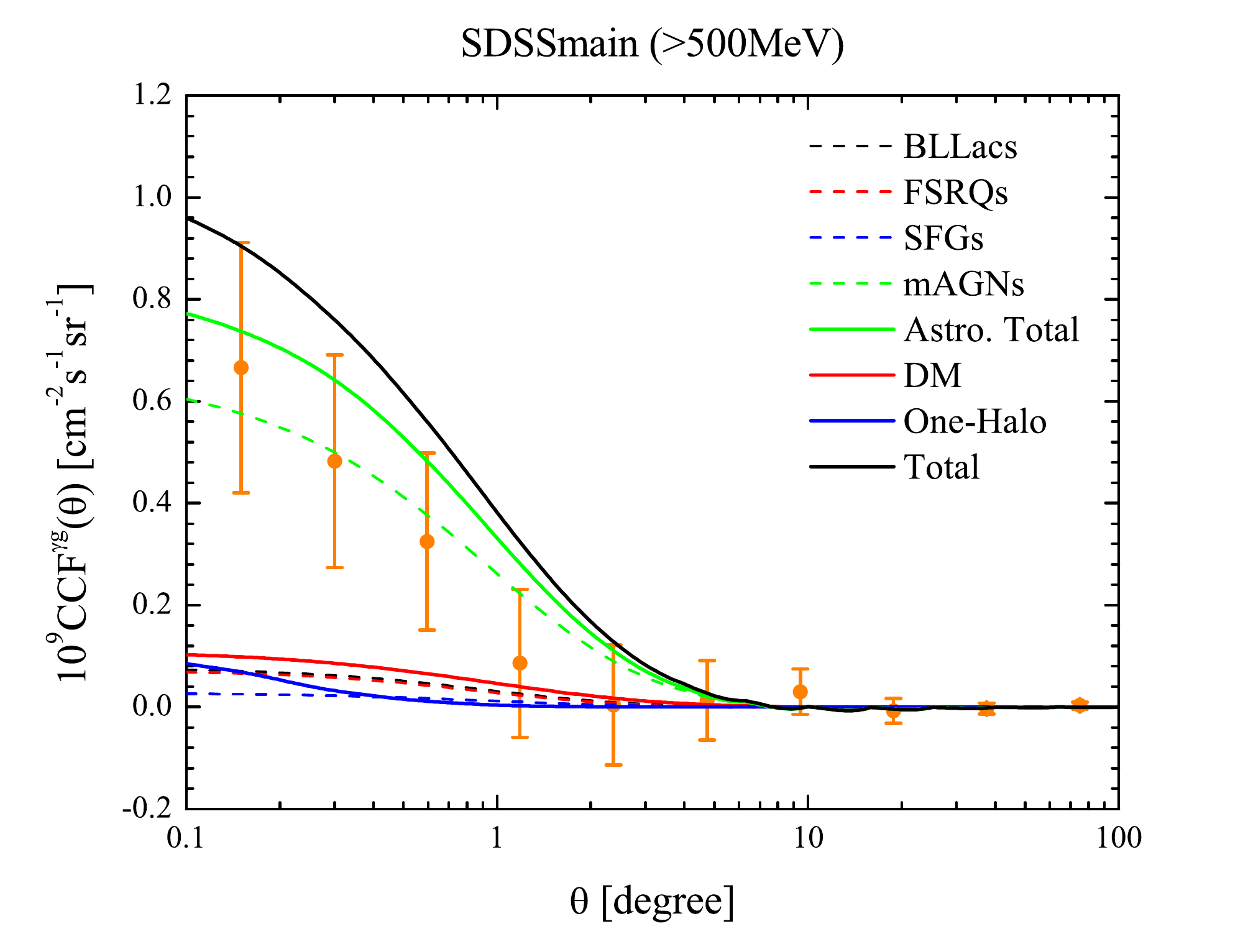}  \\
%
%
\includegraphics[width=0.33\columnwidth]{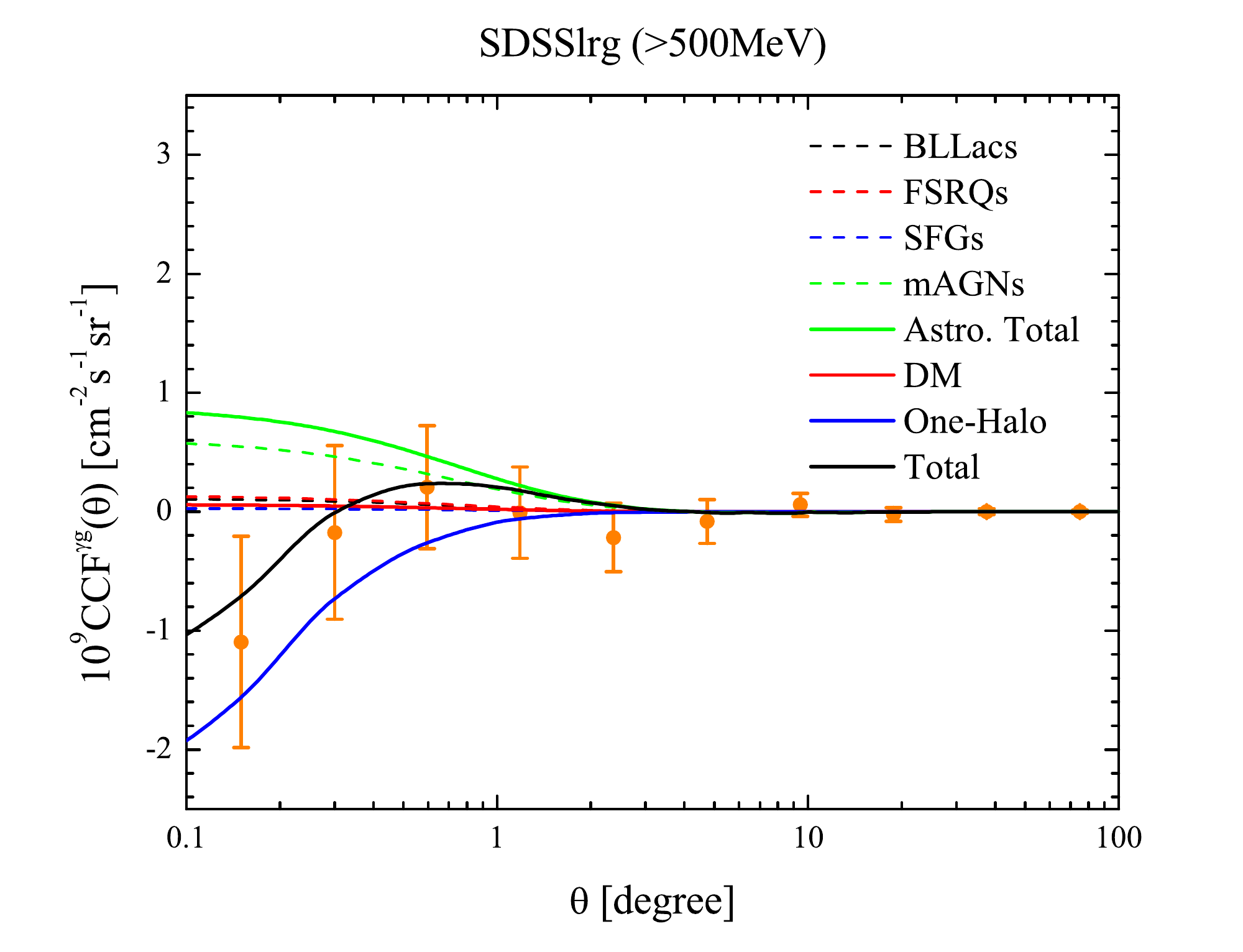} 
%
%
\includegraphics[width=0.33\columnwidth]{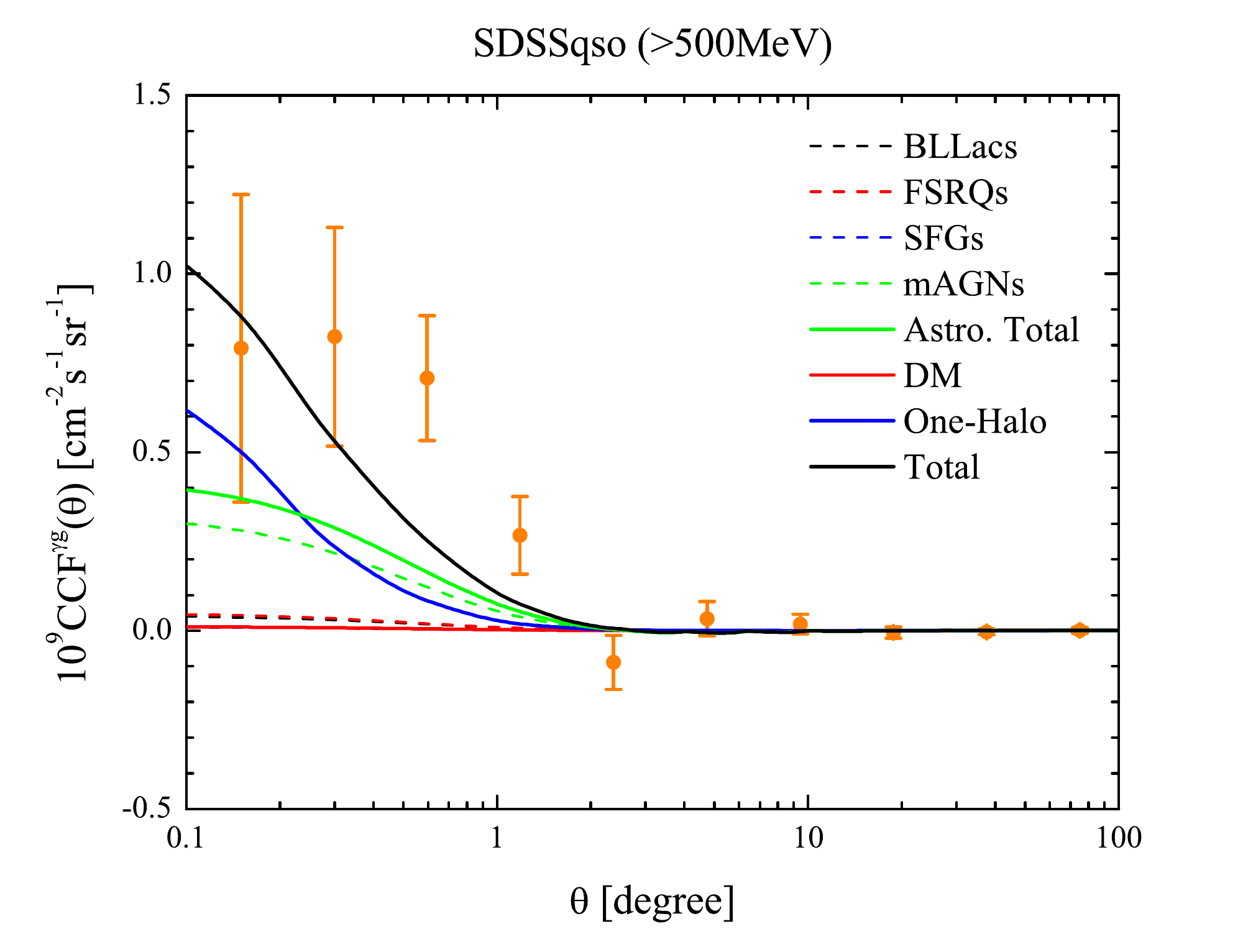} 
\end{center}
\vspace{-0.5cm}
\caption{Measured cross correlation function (CCF) \citep{Xia:2014} for $E>0.5$ GeV, as a function 
of the angular separation $\theta$ in the sky, compared to the best fit models of this analysis. 
The contribution to the CCF from the different astrophysical \g-rays emitters (BL Lac, mAGN, SFG, FSRQ) are shown 
by dashed colored lines, while their sum (``Astro Total") and the DM contribution are indicated by solid green and red lines, respectively. 
The {\it 1-halo correction} term is shown as a solid blue line. The total contribution to the CCF is given by the black solid line.}
\label{fig:CCFdata500MeV} 
\end{figure*}

\begin{figure*}[ht!]
\begin{center}
%
\includegraphics[width=0.33\columnwidth]{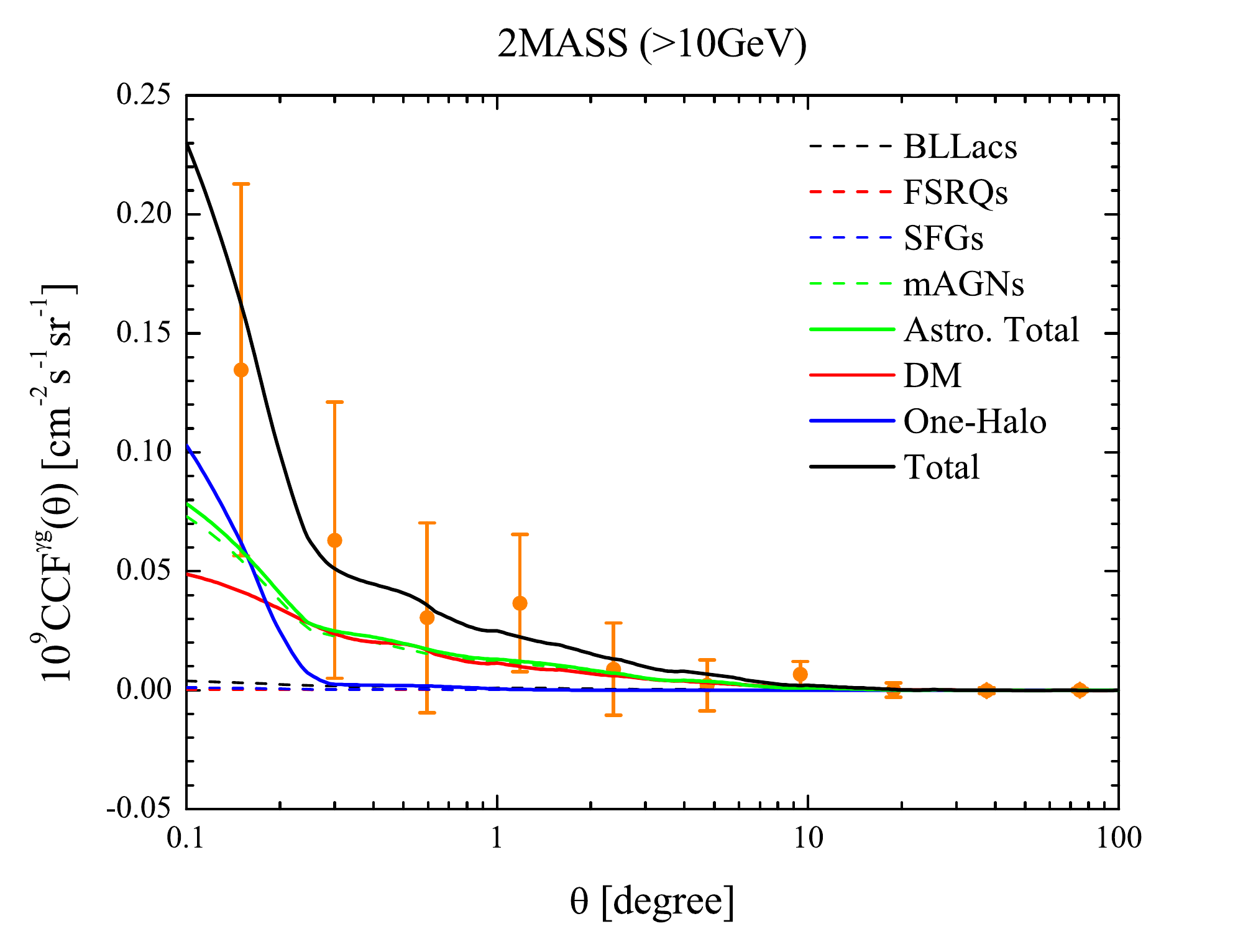} 
\hspace{-0.5cm}
\includegraphics[width=0.33\columnwidth]{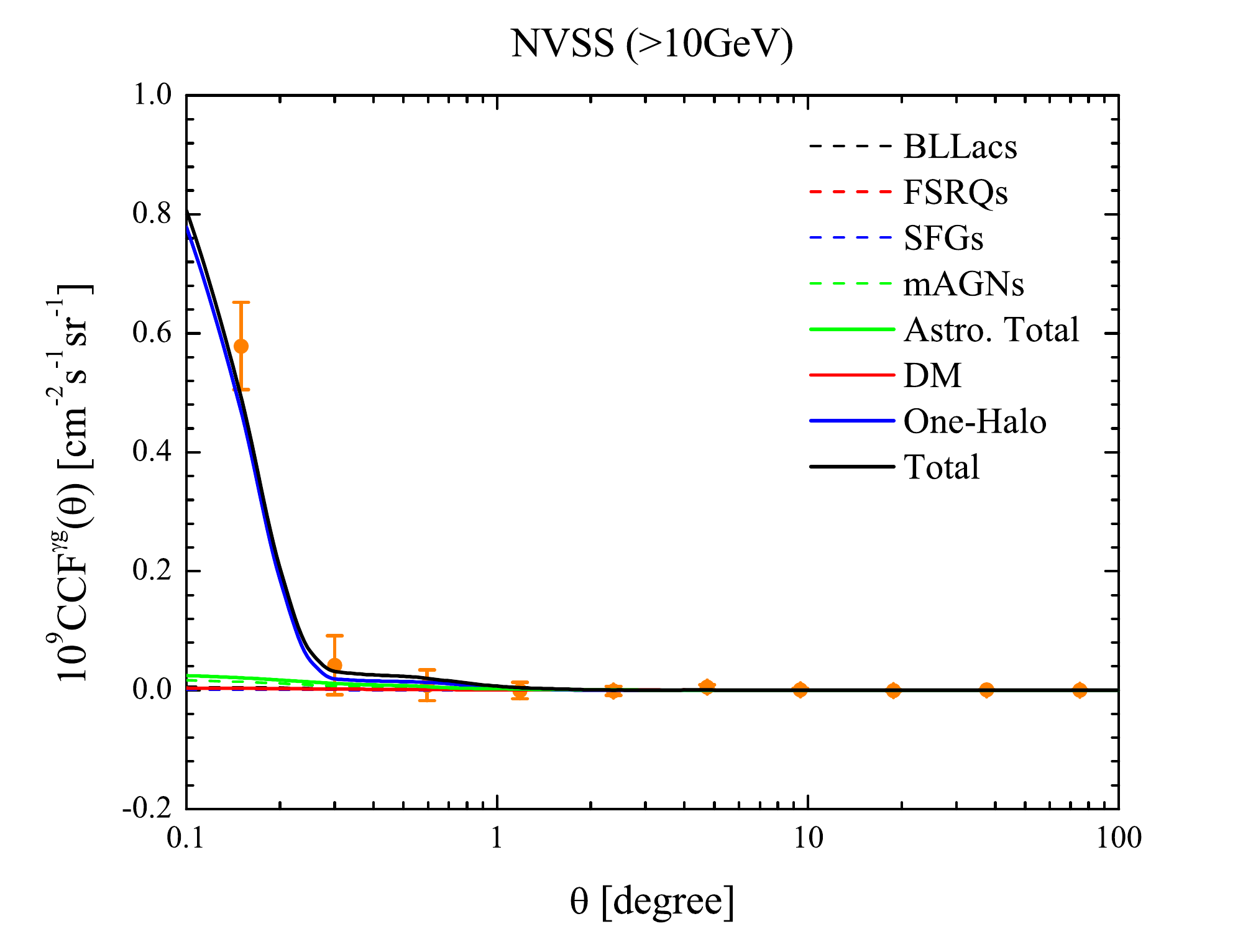} 
\hspace{-0.5cm}
\includegraphics[width=0.33\columnwidth]{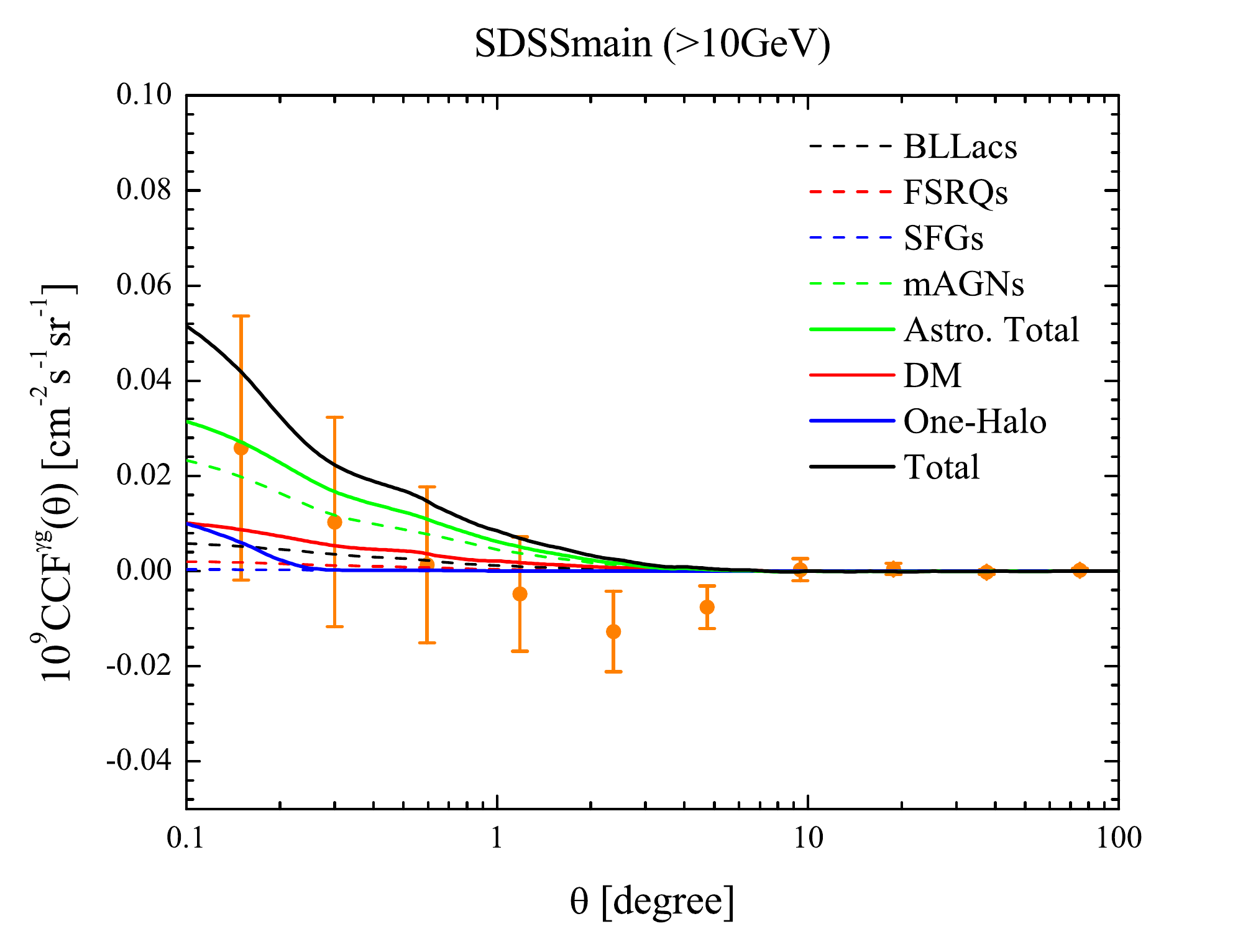}  \\
%
%
\includegraphics[width=0.33\columnwidth]{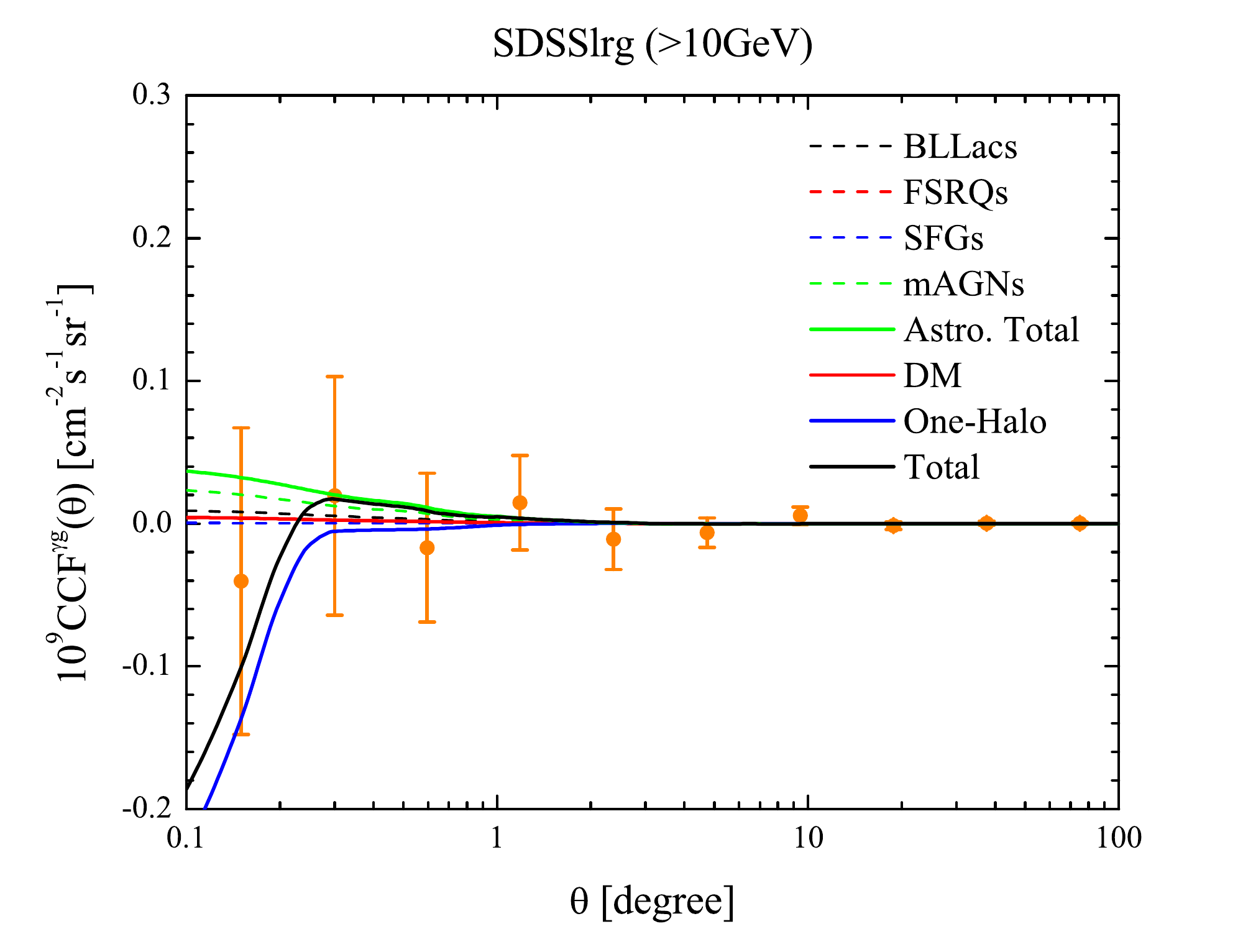} 
%
%
\includegraphics[width=0.33\columnwidth]{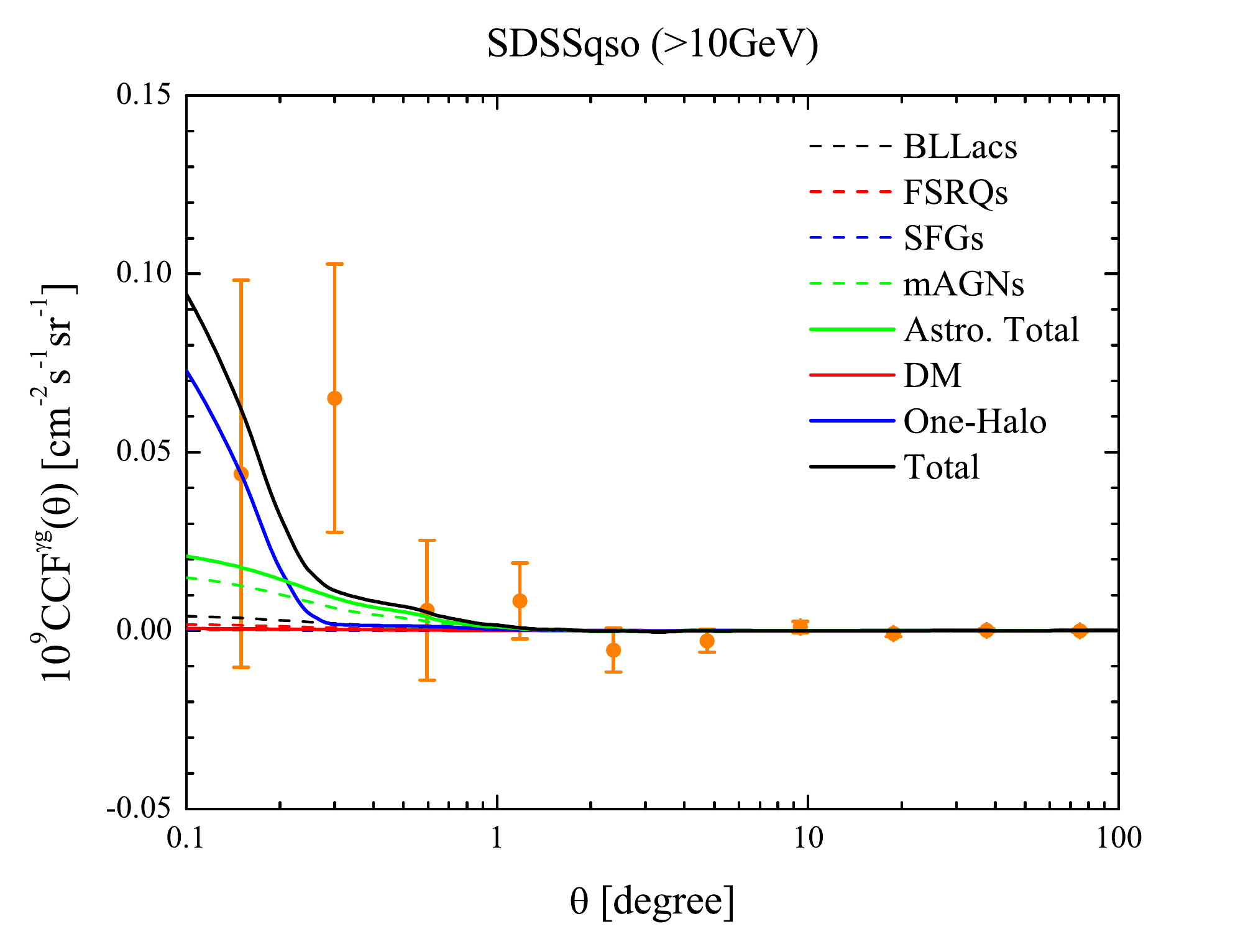} 
\end{center}
\vspace{-0.5cm}
\caption{Same as Fig.~\ref{fig:CCFdata500MeV} but for $E>10$ GeV.}
\label{fig:CCFdata10GeV} 
\end{figure*}


\bibliographystyle{apj}
\bibliography{references}

\end{document}